\documentclass[aps,prx,twocolumn,amsmath,amssymb,superscriptaddress,floatfix,longbibliography
]{revtex4-2}

\usepackage[modulo]{lineno}

\usepackage[dvipsnames]{xcolor}
\usepackage{graphicx}
\usepackage{hyperref}
\usepackage{dcolumn}
\usepackage{bm}
\usepackage{comment}
\usepackage{dashrule}
\usepackage{subfiles} 
\usepackage{gensymb}
\usepackage{accents}

\usepackage{datetime2}
\usepackage{orcidlink}
\usepackage{soul}
\usepackage[normalem]{ulem}

\usepackage[rightcaption]{sidecap}

\sidecaptionvpos{figure}{t}

\newcommand{\fil}{_{\text F}}
\newcommand{\rfil}{_{\text R}}
\newcommand{\sm}{_{\text S}}
\newcommand{\god}{_{\text T}}

\definecolor{nblue}{rgb}{0.06,0.3,0.73}
\definecolor{nblack}{rgb}{0,0,0}
\definecolor{nred}{rgb}{0.9,0.1,0.1}
\definecolor{nmagenta}{rgb}{0.7,0.0,0.3}
\definecolor{ncyan}{cmyk}{1,0,0,0}

\newcommand{\blk}{\color{nblack}}

\newcommand{\past}[1]{\overleftarrow{#1}}
\newcommand{\fut}[1]{\overrightarrow{#1}}
\newcommand{\both}[1]{\overleftrightarrow{#1}}

\newcommand{\ellipse}{\raisebox{-1pt}{\scalebox{1.3}[.4]{$\circ$}}}
\newcommand{\halo}{\accentset{\ellipse}}
\newcommand{\csm}{_{\rm cS}}
\newcommand{\tar}{_{\rm tar}}
\newcommand{\dd}{{\rm d}}
\newcommand{\bx}{\hat{\bf x}}
\newcommand{\tp}{^\top}
\newcommand{\ex}[1]{\langle #1 \rangle}
\newcommand{\Tr}{{\rm Tr}}
\newcommand{\inv}{^{-1}}
\newcommand{\erf}[1]{Eq.~(\ref{#1})}

\newcommand{\beq}{\begin{equation}}
\newcommand{\eeq}{\end{equation}}

\AtBeginDocument{\renewcommand{\Re}{\operatorname{Re}}}
\AtBeginDocument{\renewcommand{\Im}{\operatorname{Im}}}

\newcommand{\omgc}{\omega_{\text{c}}}
\newcommand{\omgL}{\omega_{\text{L}}}
\newcommand{\nc}{\bar{n}_\text{cav}}

\newcommand{\Ydet}{\hat{Y}_{\text{det}}}
\newcommand{\Yv}{\hat{Y}_{\text{v}}}

\newcommand{\effc}{\eta_\text{esc}}

\newcommand{\effwf}{\eta_\text{wf}}
\newcommand{\nth}{n_\text{th}}

\newcommand{\Pin}{P_\text{in}}
\newcommand{\Pfb}{P_\text{fb}}

\newcommand{\rmi}{\mathrm{i}}
\newcommand{\Hfb}{H_{\text{fb}}}
\newcommand{\grp}{g_{\text{RP}}}

\newcommand{\Gammafb}{\Gamma_\text{fb}}

\begin{document}

\title{Post-processed estimation of quantum state trajectories}

\author{Soroush Khademi\orcidlink{0000-0002-0871-3342}}\email{s.khademi@uq.edu.au}
\affiliation{Australian Research Council Centre of Excellence for Engineered Quantum Systems, School of Mathematics and Physics, The University of Queensland, St Lucia, Queensland 4072, Australia}

\author{Jesse J. Slim\orcidlink{0000-0003-0696-9130}}
\affiliation{Australian Research Council Centre of Excellence for Engineered Quantum Systems, School of Mathematics and Physics, The University of Queensland, St Lucia, Queensland 4072, Australia}

\author{Kiarn T. Laverick\orcidlink{0000-0002-3688-1159}}
\affiliation{Centre for Quantum Dynamics, and Australian Research Council Centre of Excellence for Quantum Computation and Communication Technology, Griffith University, Yuggera Country, Brisbane, Queensland 4111, Australia}
\affiliation{MajuLab, CNRS-UCA-SU-NUS-NTU International Joint Research Laboratory}
\affiliation{Centre for Quantum Technologies, National University of Singapore, 117543, Singapore}

\author{Jin Chang\orcidlink{0000-0003-1101-8516}}
\affiliation{Kavli Institute of Nanoscience, Department of Quantum Nanoscience, Delft University of Technology, 2628CJ Delft, The Netherlands}

\author{Jingkun Guo\orcidlink{0000-0002-3470-1433}}
\affiliation{Kavli Institute of Nanoscience, Department of Quantum Nanoscience, Delft University of Technology, 2628CJ Delft, The Netherlands}

\author{Simon Gröblacher\orcidlink{0000-0003-3932-7820}}
\affiliation{Kavli Institute of Nanoscience, Department of Quantum Nanoscience, Delft University of Technology, 2628CJ Delft, The Netherlands}

\author{Howard M. Wiseman\orcidlink{0000-0001-6815-854X}}
\affiliation{Centre for Quantum Dynamics, and Australian Research Council Centre of Excellence for Quantum Computation and Communication Technology, Griffith University, Yuggera Country, Brisbane, Queensland 4111, Australia}

\author{Warwick P. Bowen\orcidlink{0000-0001-8127-1715}}\email{w.bowen@uq.edu.au}
\affiliation{Australian Research Council Centre of Excellence for Engineered Quantum Systems, School of Mathematics and Physics, The University of Queensland, St Lucia, Queensland 4072, Australia}
\affiliation{Australian Research Council Centre of Excellence in Quantum Biotechnology, School of Mathematics and Physics, The University of Queensland, St Lucia, Queensland 4072, Australia}


\begin{abstract}
Weak quantum measurements enable real-time tracking and control of dynamical quantum systems, producing quantum trajectories --- evolutions of the quantum state of the system conditioned on measurement outcomes. For classical systems, the accuracy of trajectories can be improved by incorporating future information, a procedure known as smoothing. Here we apply this concept to quantum systems, generalising a formalism of quantum state smoothing for an observer monitoring a quantum system exposed to environmental decoherence, a scenario important for many quantum information protocols. This allows future data to be incorporated when reconstructing the trajectories of quantum states. We experimentally demonstrate that smoothing improves accuracy using a continuously measured nanomechanical resonator, showing that the method compensates for both gaps in the measurement record and inaccessible environments. We further observe the signature of a key predicted departure from classical smoothing: quantum noise renders the trajectories nondifferentiable. These results establish that future information can enhance quantum trajectory reconstruction, with potential applications across quantum computing, communication, sensing and control.
\end{abstract}

\keywords{continuous quantum measurement, cavity optomechanics, quantum state smoothing}

\maketitle

The continuous observation of dynamical systems plays a central role in physics and engineering~\cite{stengel1994optimal,wiseman2009quantum,Dong2022quantum,Carmichael1993open}.
In quantum technologies, it has been long studied and provides important capabilities for sensing~\cite{jimenez2018signal,Duan2025concurrent}, tracking and control~\cite{Sayrin2011realtime,wieczorek2015optimal,thomas2021entanglement,Magrini2021realtime}, error correction~\cite{convy2022machine,livingston2022experimental}, and the preparation of non-classical states as resources~\cite{muller2008entanglement,meng2020mechanical,thomas2021entanglement}.
By filtering the measurement outcomes up to the present time, it is possible to produce real-time estimates of the trajectory of the system. Measurement outcomes can also be \textit{post-processed} using both past and future data. This procedure is known as \textit{smoothing}~\cite{Weinert01} and allows more accurate reconstruction of trajectories.

Even though the controlled evolution of quantum states is fundamental to quantum technologies, smoothing has not yet been applied to improve quantum state trajectories themselves. It has been applied to generate trajectories of quantum weak-values~\cite{Kocsis2011observing,Bliokh2013photon}. However, their physical significance is disputed since they can exhibit negative probabilities and values that lie outside the eigenvalue spectrum~\cite{Aharonov1988how,leggett1989comment,ferrie2014result}. Beyond smoothing, future data has also been used to improve predictions of the outcomes of intermediate single-shot measurements~\cite{gammelmark2013past,rybarczyk2015forward,tan2015prediction,bao2020retrodiction}, and to retrospectively verify filtered trajectories~\cite{miao2010probing,rossi2019observing,meng2022measurement,lammers2024quantum}, but these approaches do not track the forward-in-time evolution of the quantum system.

In this work, we demonstrate smoothing of quantum state trajectories using and generalising the quantum state smoothing formalism of Ref.~\cite{guevara2015quantum} so that it can be applied to the case of a single observer monitoring an open quantum system that interacts with its environment. Our generalised formalism allows optimal smoothed quantum state trajectories to be determined for a broad range of scenarios. We use it for two: estimating the trajectories that would be obtained if all information from the monitoring channel was available from the distant past (Sec.~\ref{Sec:trans}), and if full access to the environment was available (Sec.~\ref{Sec:true}). We experimentally validate the formalism in both scenarios using a mechanical resonator. The resonator is subject to optical monitoring, which introduces significant quantum back-action. We also perform an experiment with artificially injected noise and show that smoothing can partially compensate for the noise-induced degradation in the accuracy of state trajectories. Finally, we experimentally substantiate a key prediction of quantum state smoothing (Sec.~\ref{Sec:traj}): while classical smoothing typically results in differentiable trajectories (hence the name), the stochasticity of quantum noise cannot, in general, be entirely eradicated, resulting in the formal non-differentiability of smoothed quantum trajectories~\cite{laverick2021quantum}. 

The ability to more accurately infer the time evolution of a quantum system using future measurements could advance quantum computing, communication, sensing and control protocols that incorporate post-processing stages. For instance, it could be used to better post-select computational sequences and identify errors in quantum computing~\cite{campbell2020universal,convy2022machine,livingston2022experimental}; it could improve the fidelity of optical quantum teleportation through more accurate selection of the required displacement operations~\cite{furusawa1998unconditional,bowen2003experimental}; and it can enhance prediction of the actions of machines that employ quantum sensors. It could also provide insight into foundational questions such as the nature of quantum state trajectories~\cite{wiseman1996quantum}, conditional entropy production~\cite{rossi2020experimental}, quantum clocks~\cite{he2023effect}, quantum decoherence~\cite{wiseman2001inequivalence}, and time symmetry in quantum mechanics~\cite{aharonov1964time}.

\section{Quantum trajectories of the long-time-limit filtered state} \label{Sec:trans}

\begin{figure*}[t!]
    \centering
    \includegraphics[width=\linewidth]{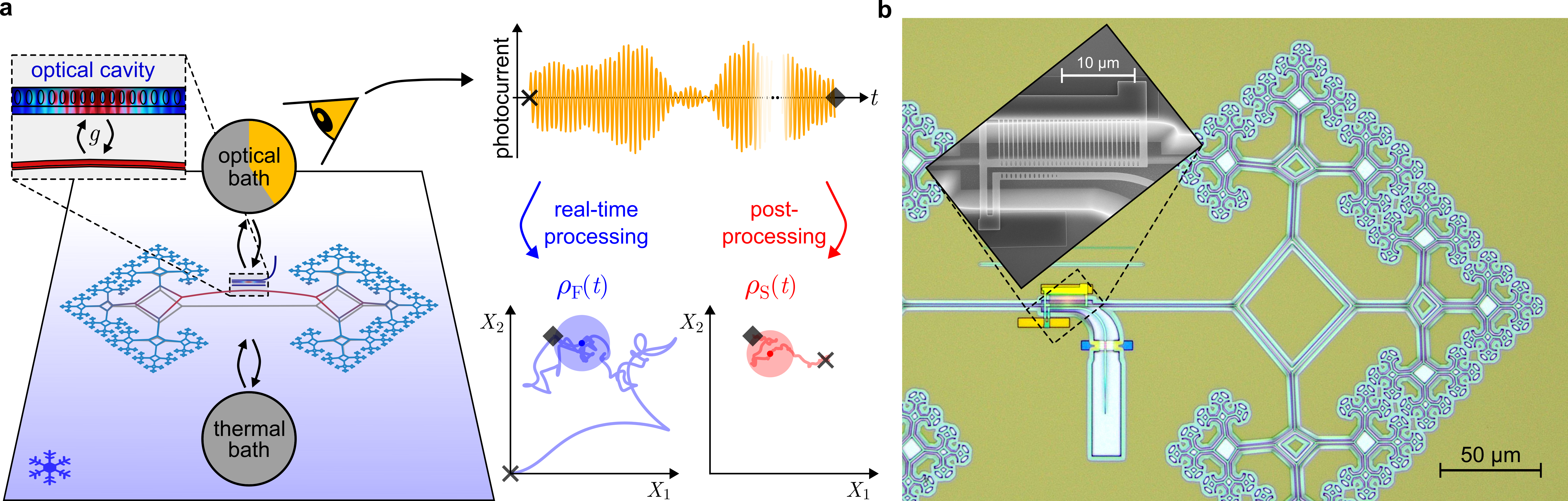}
    \caption{\textbf{Post-processed quantum trajectories in experiment. a,} A high-quality mechanical resonator, formed by the out-of-plane mode of a hierarchical arrangement of stressed silicon nitride strings, is in contact with a cryogenic thermal bath. It is also in contact with an optical bath, via a photonic crystal cavity coupled with rate $g$. A fraction $\eta$ (yellow) of the optical bath is observed, and the acquired photocurrent is processed to reconstruct the resonator's quantum trajectory. Trajectories obtained from real-time (blue) and post-processed (red) analysis of a sample measurement record diffuse in the phase space spanned by (rotating) mechanical quadratures $X_{1,2}$. Their means start at distant points (crosses), while their final values coincide (diamonds). At intermediate time $t$, variances of the inferred Gaussian quantum states are indicated (circles), illustrating that the post-processed state $\rho_{\sm}(t)$ is purer, with smaller variance, than the real-time filtered state $\rho_{\fil}(t)$. \textbf{b,} Optical image of the large optomechanical device, with vertically stacked optical cavity and tapered waveguide for fibre coupling. A scanning electron micrograph (inset) shows the cavity's and waveguide mirror's photonic crystal design.}
    \label{FigSetup}
\end{figure*}

We consider a quantum system that is continuously monitored by an observer. The observer monitors the system via measurements on a bath that it is coupled to (see FIG.~\ref{FigSetup}{\bf a}). This formulation is quite general. It could, for example, apply to a superconducting qubit monitored via a microwave field~\cite{he2023effect}, an atomic system via light~\cite{jimenez2018signal}, or a variety of other monitored quantum systems~\cite{thomas2021entanglement}. As detailed later, in our experiment the system is a mechanical resonator which interacts with a bath that has both optical and thermal components, with monitoring performed via a part of the optical component.

In the ideal case that the measurement record extends far into the distant past, the time evolution of the system is described in real time by the long-time-limit (LTL) filtered state trajectory $\rho\fil^\text{LTL}(t)$ (Methods~\ref{sec:GeneralACDK}). 
However, the observer might only have access to measurement data for a finite time into the past --- for instance, due to memory constraints or simply because they began recording recently. They then obtain the filtered state trajectory $\rho\fil(t)$ in real time, which differs from $\rho\fil^\text{LTL}(t)$, especially in the transient period shortly after onset of monitoring. To determine how close this trajectory is to $\rho\fil^\text{LTL}(t)$, it is necessary to define a measure. The standard conditioning process provides $\rho\fil(t)$ by implicitly minimising the ensemble-averaged value of the squared Hilbert--Schmidt distance, $\Delta^2_\text{HS}[\rho_\text{F}^\text{LTL},\rho_\text{F}]=\text{Tr}[(\rho_\text{F}^\text{LTL}-\rho_\text{F})^2]$, or, equivalently, the relative entropy (Methods~\ref{sec:QSSforLTL}). We therefore take these as the measures of closeness. Trajectories closer to $\rho_\text{F}^\text{LTL}(t)$ can be obtained, and early-time information recovered, by post-processing the measurement record and employing the future data.

To do this optimally, we adapt the quantum state smoothing formalism originally developed for a different scenario --- where two observers, Alice and Bob, simultaneously measure distinct baths interacting with the quantum system~\cite{guevara2015quantum,CGW19} --- and tested very recently in that context~\cite{Shota25}. We reformulate the problem to a sequential scenario so it can be applied to estimate the trajectory of the LTL filtered state. Specifically, the two observers measure the same bath in the same way, with Bob monitoring from the distant past until time $t=0$ and Alice from $t=0$ onwards, while Bob retains access to Alice's record. The task is for Alice to optimally estimate, using only her own record, Bob’s inference, $\rho_{\rm F}^{\rm LTL}(t)$, of the system’s quantum state.

In Methods~\ref{sec:QSSforLTL}, we explain how to apply the above idea to general quantum systems, but for our experiment we can use the formalism of linear-Gaussian quantum (LGQ) systems~\cite{laverick2019quantum}. An LGQ system is characterised by the first and second moments of its phase-space operators. Here we generalise the derivation in Ref.~\cite{laverick2021linear}, minimising the squared Hilbert--Schmidt distance to obtain the equations for these moments of our smoothed estimate of the LTL filtered state (Methods~\ref{MT:LQG}):
\begin{align}
\begin{aligned}
    \langle\hat{\mathbf{x}}\rangle\sm(t) & = \big(\mathbf{V}\sm(t)-\mathbf{V}\fil^\text{ss}\big)\biggl[
    \big(\mathbf{V}\fil(t)-\mathbf{V}\fil^\text{ss}\big)^{-1} \langle\hat{\mathbf{x}}\rangle\fil(t) \\
    &\hspace{5em}+ \big(\mathbf{V}\rfil(t)+\mathbf{V}\fil^\text{ss}\big)^{-1} \langle\hat{\mathbf{x}}\rangle\rfil(t)\biggr]\,, \\
    \mathbf{V}\sm(t) & = \biggl[\big(\mathbf{V}\fil(t)-\mathbf{V}\fil^\text{ss}\big)^{-1}+\big(\mathbf{V}\rfil(t)+\mathbf{V}\fil^\text{ss}\big)^{-1}\biggr]^{-1}\\
    &+\mathbf{V}\fil^\text{ss}\, .
    \label{QSSforLTLfil}
\end{aligned}
\end{align} 
In this equation, the vector $\hat{\mathbf{x}} = (\hat{q}_1,\hat{p}_1,...,\hat{q}_N,\hat{p}_N)^\top$ contains the canonical operators for position $\hat{q}_j$ and momentum $\hat{p}_j$ of the $N$ system modes, while $\langle\hat{\mathbf{x}}\rangle_\text{C}$ and $\mathbf{V}_\text{C}$ are the mean and covariance matrix of Gaussian Wigner functions inferred from the measurement record, with subscript $\text{C}$ specifying types of conditioning: $\text{F}$ for filtered state (inferred from past measurement data); $\text{R}$ for retrofiltered effect operator (inferred from future data); and $\text{S}$ for smoothed quantum state (inferred from the whole record). Equations for the filtered and retrofiltered moments, derived from the system's stochastic master equation~\cite{laverick2021linear}, are provided in Methods~\ref{sec:GeneralACDK}. In Eq.~\eqref{QSSforLTLfil}, $\mathbf{V}\fil^\text{ss}$ is the deterministic covariance matrix of the LTL filtered state, which equals $\mathbf{V}\fil(t)$ in the steady state ($t \to \infty$). This equation indicates that the smoothed trajectory converges asymptotically to the filtered trajectory after the transient phase, i.e., once $\mathbf{V}\fil(t)\to\mathbf{V}\fil^\text{ss}$. We finally note that the same equation is obtained by choosing the relative entropy as our estimation cost function (Methods~\ref{sec:QSSforLTL}).

\subsection*{Strong optical monitoring of a  nanomechanical resonator}

The benefits of quantum state smoothing are most pronounced when a substantial fraction of the information leaking from the system into its environment is collected. Cavity optomechanical systems --- where an optical probe enables quantum-limited readout of mechanical motion --- provide a natural platform for realising this condition~\cite{aspelmeyer2014cavity,mason2019continuous,rossi2019observing,bowen2015quantum}, allowing the time evolution of the mechanical resonator to be inferred with a conditional variance approaching the ground-state size. Significant quantum back-action is, in such settings, associated with the measured bath. Beyond foundational interest, tracking the trajectory of a macroscopic resonator underpins applications in quantum control~\cite{guo2019feedback}, gravitational wave and dark matter detection~\cite{muller2008entanglement,hirschel2024superfluid,baker2024optomechanical}, and proposed tests of quantum gravity~\cite{girdhar2020testing}.

To validate Eq.~\eqref{QSSforLTLfil} and demonstrate the utility of quantum state smoothing in experiment, 
we track the quantum trajectory of the fundamental out-of-plane resonance of the on-chip mechanical structure shown in FIG.~\ref{FigSetup}, which has a frequency of $\Omega/2\pi = 1.04$~MHz. We monitor this resonance via dispersive coupling to a telecom-wavelength, broad-band photonic crystal cavity with linewidth $\kappa/2\pi = 11.5$~GHz.
We operate the optomechanical system well inside the fast-cavity regime $\kappa \gg \Omega, g\,$, so that the cavity dynamics can be adiabatically eliminated. Here, $g = g_0\sqrt{\bar{n}_\text{cav}}$ is the linearised optomechanical coupling rate, with $g_0$ the single-photon coupling rate and $\bar{n}_\text{cav}$ the mean intracavity photon number. In this regime, the mechanical displacement is directly imprinted on the phase of a resonant laser probe reflected off the cavity~\cite{bowen2015quantum}, which we detect with efficiency $\eta$ using shot-noise-limited homodyne measurement (Methods~\ref{SecHomoSetup}).

As illustrated in FIG.~\ref{FigSetup}\textbf{a}, the cavity mediates an interaction between the resonator and an optical bath. The rate at which information leaks into the optical bath is given by $\gamma_\text{opt} = 4 g^2/\kappa$~\cite{bowen2015quantum,meng2020mechanical}. At the same time, information leaks into the resonator's mechanical thermal bath at a rate $\gamma_\text{th}$. In the limit that the resonator temperature $T \gg \hbar \Omega/k_\text{B}$, relevant to our experiments, this is approximated as $\gamma_\text{th} \approx \Gamma n_\text{th}$~\cite{aspelmeyer2014cavity}, where $\Gamma$ is the resonator's energy decay rate and $n_\text{th}\approx k_\text{B} T/(\hbar \Omega)$ its thermal phonon occupancy. With these bath interactions and the homodyne measurement, the resonator constitutes a single-mode linear Gaussian system.

As only the detected part of the optical bath can be accessed, to effectively monitor the system one must maximize both the optical detection efficiency $\eta$ and
the ratio $\gamma_\text{opt}/\gamma_\text{th} = \mathcal{C} / n_\text{th}$, where $\mathcal{C} = 4g^2 / (\kappa \Gamma)$ is the optomechanical cooperativity.
Our optomechanical device is carefully engineered to do this~\cite{guo2022integrated,guo2023active}.
Acoustic clamping losses are minimised by the resonator's hierarchical fractal structure~\cite{fedorov2020fractal}, resulting in an exceedingly low decay rate of $\Gamma/2\pi = 11.5$~mHz at cryogenic temperatures. Simultaneously, vertical integration of the optical cavity allows for a large single-photon optomechanical coupling rate of $g_0/2\pi = 159~$kHz. The full device characterisation is elaborated in Methods~\ref{sec:methods:characterisation}. Further details on its structure and fabrication are discussed in Ref.~\cite{guo2022integrated}.

The device is placed in a dilution refrigerator to reduce $n_\text{th}$ (Methods~\ref{sec:methods:refrigeration}). We drive the cavity to a mean photon number of $\bar{n}_\text{cav} = 41.4$ (resulting in $g/2\pi = 1.02$~MHz and $\mathcal{C} = 3.16 \times 10^4$), limited by higher-frequency mechanical resonances that introduce coloured noise at larger values of $\bar{n}_\text{cav}$. Optical absorption raises resonator temperature to $T=12.1$~K (Methods~\ref{sec:methods:mode_temperature}), so that $n_\text{th} = 2.45 \times 10^5$ and $\mathcal{C}/n_\text{th} = \gamma_\text{opt}/\gamma_\text{th} \approx 0.13$ --- an appreciable fraction of information thus enters the optical bath.

At the mean cavity photon number used, we find that the mechanical resonator intermittently switches to a self-oscillating regime.
We suppress this instability by applying weak, phase-tuned, band-limited optical feedback (Methods~\ref{sec:methods:setup_feedback} and \ref{sec:methods:char_feedback}), which is effectively Markovian and increases the effective mechanical decay rate to $\Gamma_\text{fb}/2\pi = 85$~Hz. 
While the feedback can be modelled explicitly within the system dynamics, in our case it suffices to just do the replacements $\Gamma\to\Gamma_\text{fb}$ and $n_\text{th}\to n_\text{th}\times\Gamma/\Gamma_\text{fb}$ in the analysis (Methods~\ref{MethodsSec:Analysis}) because the feedback gain is about 30 dB below the squashing limit (Methods~\ref{sec:methods:char_feedback} and \ref{SecFeedbackModelling}). 
These replacements leave the ratio $\gamma_\text{opt}/\gamma_\text{th}$ unchanged.

To maximise $\eta$, the cavity is evanescently coupled to an on-chip waveguide with efficiency $\eta_\text{esc} = 0.82$, and the waveguide's shape is engineered to maximise transfer to a
complementarily tapered optical fibre. 
After cool-down, we carefully optimise the fibre–waveguide alignment inside the dilution refrigerator to achieve a notably high transfer efficiency of $\eta_\text{wf} = 0.77$ (Methods~\ref{sec:methods:optical_coupling}). Minimising inefficiencies in the optical fibre path and employing high-efficiency, low-noise photodiodes, we achieve a total efficiency of $\eta = 0.38$ (Methods~\ref{sec:methods:total_losses}), verified independently using a new method that exploits the mechanical instability (Methods~\ref{sec:methods:eta_check}). 
Together with our high ratio of optical to thermal bath coupling, this allows us to track the resonator's evolution with a real-time conditional variance $4.7$ times the ground state variance, indicating that quantum back-action plays a significant role in the system dynamics. 
Validating this indication, we demonstrate ponderomotive squeezing of light by quantum back-action~\cite{bowen2015quantum} (Methods~\ref{sec:methods:mode_temperature}).

\subsection*{Experimental estimation of the long-time-limit filtered state}

To determine the resonator’s filtered and smoothed trajectories ($\rho_\text{F}$ and $\rho_\text{S}$), and statistically compare them with its LTL filtered trajectory ($\rho_\text{F}^\text{LTL}$), we monitor the nanomechanical resonator for several seconds and record the homodyne photocurrent. The photocurrent is normalized to the shot noise level and demodulated at mechanical frequency $\Omega$. This neglects the fast-oscillating dynamics of the resonator while retaining the effective interaction-frame dynamics~\cite{doherty2012quantum}. The demodulation produces two measurement currents corresponding to heterodyne detection of the resonator’s quadratures, $\hat X_{1}=\cos(\Omega t)\,\hat q-\sin(\Omega t)\,\hat p$ and $\hat X_{2}=\sin(\Omega t)\,\hat q+\cos(\Omega t)\,\hat p$, where the mechanical position $\hat{q}$ and momentum $\hat{p}$ are normalized to have ground-state variance of one. We divide these quadrature currents into 16,653 750-$\mu$s-long measurement records. 

The measurement records are processed using the filtering and retrofiltering equations given in Methods (Eqs.~\eqref{equ:rFT} and \eqref{equ:rRT}) to obtain trajectories of the mean values $\langle\hat{X}_j\rangle_\text{F}$ and $\langle\hat{X}_j\rangle_\text{R}$, where $j \in \{1,2\}$. The trajectories of the covariances, $\mathbf{V}_\text{F}$ and $\mathbf{V}_\text{R}$, are deterministic and set by the experimental parameters $\Gamma$, $\mathcal{C}$, $n_\text{th}$, and $\eta$ (Eqs.~\eqref{VFT} and \eqref{VRT}). Together, $\langle\hat{X}_j\rangle_\text{F}$ and $\mathbf{V}_\text{F}$ provide the filtered quantum state trajectory. The smoothed quantum state trajectory is obtained by introducing $\hat{\mathbf{x}}=(\hat{X}_1,\hat{X}_2)^\top$ with the moments calculated by filtering and retrofiltering
into Eq.~\eqref{QSSforLTLfil}. Finally, the LTL filtered trajectory is determined in a manner similar to the filtered trajectory, but with the filtering interval extended over a few preceding records --- this ensures that the conditional covariance matrix converges to $\mathbf{V}_\text{F}^\text{ss}$ before time zero, as the transient phase lasts for about half a record.
\begin{SCfigure*}
    \centering 
    \includegraphics[width=\linewidth]{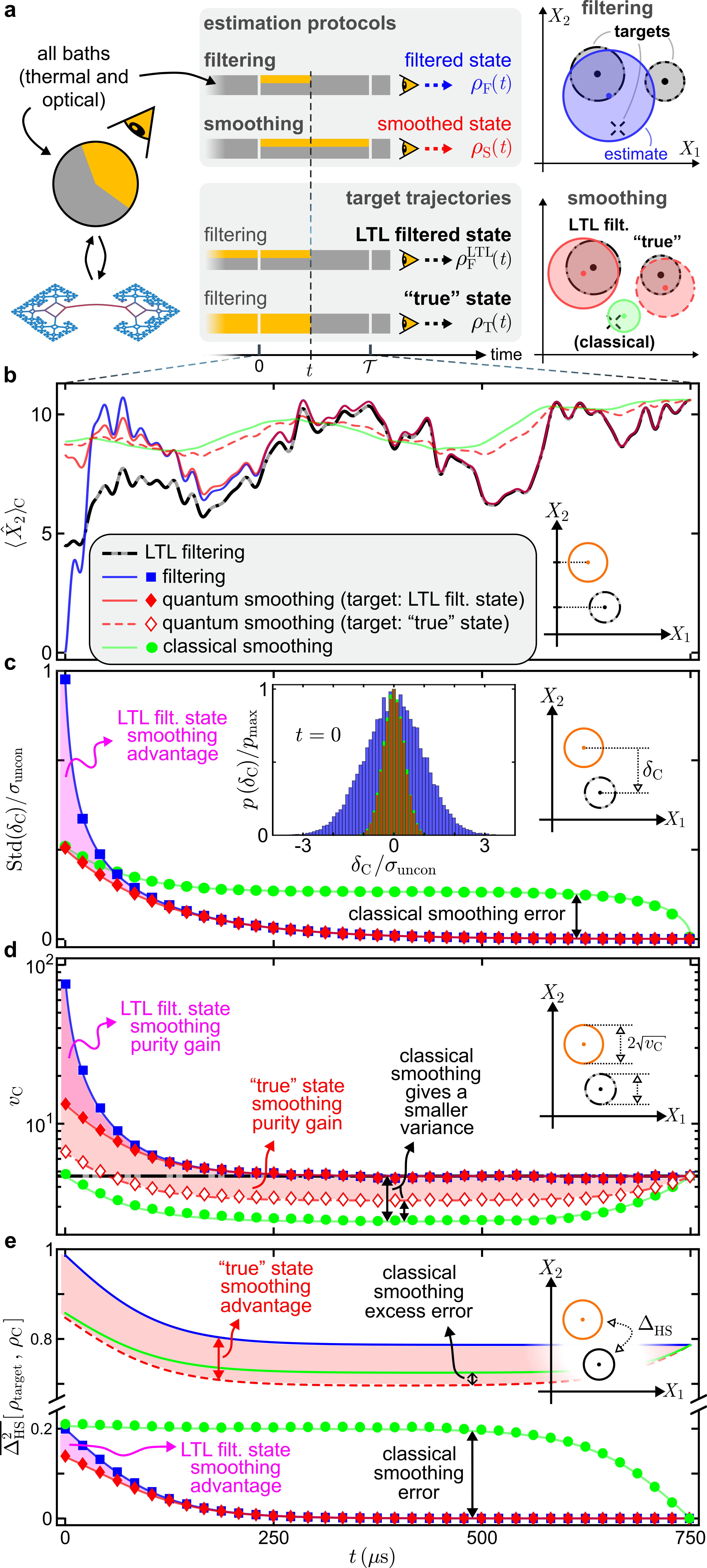}
    \caption{\textbf{Inferred trajectories of the mechanical resonator. a,} Conceptual illustration of the inference process. At left, the pie chart represents the total environment interacting with the resonator, partitioned into the measured fraction (yellow, enlarged for visual clarity) and the remaining unmeasured baths (grey). At centre, each bar represents the continuous interaction of this combined bath with the resonator. Yellow portions indicate the extent to which it is measured over time, with the outputs used to infer a quantum state for the resonator at record time $t$ (dashed line). Estimator states are inferred from the actual measurement record --- either from $0$ to $t$ (filtered state) or from $0$ to the record end $\mathcal{T}$ (smoothed state) --- whereas target trajectories assume access to a long-time-limit (LTL) measurement history and to all baths (``true"). At right, the phase space diagrams illustrate how both target trajectories share a single real-time estimator ($\rho_{\fil}$, blue) yet differ in their optimal post-processed estimators ($\rho_{\sm}\text{'s}$, red). The filtered state $\rho_{\fil}$ also provides the optimal estimate of the fictitious true value (cross) of the resonator’s classical counterpart, which has its own post-processed estimator (green).
    \textbf{b--e,} Comparing the estimation protocols. Insets on the right indicate the quantity plotted in each panel, as computed from estimator (orange) and target (black) states.
    \textbf{b,} Stochastic traces of the quadrature mean $\langle\hat{X}_2\rangle(t)$ for a sample measurement record.
    \textbf{c,} Standard deviation of distance $\delta_\text{C}(t)=\langle\hat{X}_j\rangle_\text{F}^\text{LTL}(t)-\langle\hat{X}_j\rangle_\text{C}(t),\,j=1,2$ between the LTL target and its estimators (C = F, S) or the classical smoother (C = cS), normalised to the unconditional state size $\sigma_\text{uncon}$. Statistics derived from an ensemble of 16,653 records (data points) correspond well to theory (lines, Supplementary~\ref{supp:delta_C}). Inset histogram shows the distributions of $\delta_\text{C}(0)$ across the ensemble, scaled to have a maximum of unity.
    \textbf{d,} Deterministic variances $v_\text{C}$ of the estimator states, of the classical smoother and of the LTL target. Estimators' variances agree well with the corresponding value of $\sigma^2_\text{uncon}-\mathbb{V}\mathrm{ar}_{\text{ens}}[\langle\hat{X}_j\rangle_\text{C}(t)]$, calculated across the ensemble of records (data points, Supplementary~\ref{SecConsistency} and \ref{sec:ErrorBars_forSC}).
    \textbf{e,} Distinguishability of the estimator states from their corresponding targets, and of the classical smoother from both quantum targets. The measure is the estimation cost function, given by the ensemble-averaged Hilbert--Schmidt distance squared, $\Delta^2_\text{HS}[\,\rho_\text{target}(t)\,,\,\rho_\text{C}(t)]=\text{Tr}[(\rho_\text{target}(t)-\rho_\text{C}(t))^2]$. In the lower part, data points and lines show the experimental and theoretical values for $\rho_\text{target}(t)=\rho_\text{F}^\text{LTL}(t)$, respectively. In the upper part, the lines show the theoretical values for $\rho_\text{target}(t)=\rho_\text{T}(t)$.}
    \label{FigData1}
\end{SCfigure*}

The outputs of the analysis are shown in FIG.~\ref{FigData1}. The schematic in FIG.~\ref{FigData1}\textbf{a} summarizes the two estimation protocols and their target, $\rho_\text{F}^\text{LTL}(t)$. Filtering uses the record data from time zero until the present time $t$. Smoothing uses the entire record, combining filtering of past data (between zero and $t$) with retrofiltering of future data (between $t$ and $750~\mu$s). FIGs.~\ref{FigData1}\textbf{b–e} compare the trajectories, each defined by the mean values $\langle \hat{X}_1\rangle$ and $\langle \hat{X}_2\rangle$ (for brevity, only $\langle \hat{X}_2\rangle$ is shown; $\langle \hat{X}_1\rangle$ is presented in Supplementary~\ref{sec:DataForX1}), together with a single parameter $v_\text{C}$ that fully specifies the corresponding conditional covariance matrix ($\mathbf{V}_\text{C}=v_\text{C}\,\mathbf{I}$, with $\mathbf{I}$ the $2\times 2$ identity matrix). This single-parameter description of $\mathbf{V}_\text{C}$ is possible due to the dynamical symmetry of $\hat X_{1}$ and $\hat X_{2}$ when $\Omega\gg\gamma_{\rm opt},\gamma_{\rm th}$~\cite{doherty2012quantum}.

FIG.~\ref{FigData1}\textbf{b} shows $\langle\hat{X}_2\rangle$ for a sample measurement record. For filtering (blue), it starts at zero --- the unconditional mean --- well separated from the LTL target value (dashed black). For smoothing (solid red), it begins closer to the target, as smoothing incorporates future data. The two estimates converge in the second half of the record interval. To statistically assess the accuracy of these estimates, we calculate the distance $\delta_\text{C}(t)=\langle\hat{X}_j\rangle_\text{F}^\text{LTL}(t)-\langle\hat{X}_j\rangle_\text{C}(t)$ for filtering ($\text{C}=\text{F}$) and smoothing ($\text{C}=\text{S}$) across the ensemble of records. The standard deviation of $\delta_\text{C}$ is shown in FIG.~\ref{FigData1}\textbf{c}. At $t=0$, it is smaller for smoothing (red) than for filtering (blue) by a factor of three, and it remains smaller throughout the transient phase, demonstrating the advantage of smoothing.

FIG.~\ref{FigData1}\textbf{d} shows the deterministic conditional variances. In the transient phase, $v_\text{S}$ (solid red) is smaller than $v_\text{F}$ (blue), indicating that the smoothed state is purer than the filtered state as purity is given by $1/\sqrt{\det[\mathbf{V}_\text{C}]} = 1/v_\text{C}$~\cite{serafini2017quantum}. Specifically, the smoothed state is about six times purer at $t=0$. While these variances cannot be directly measured --- a set of projective measurements evaluating $v_\text{F}$ would disturb the resonator’s dynamics --- they are involved in the inference of $\langle\hat{X}_j\rangle_\text{C}$ (Methods~\ref{SME-ME-&Analysis}) and are thus linked to the experimental mean-value traces. As shown in Supplementary~\ref{SecConsistency} and \ref{sec:ErrorBars_forSC}, $v_\text{C}(t)$ is expected to equal the variance difference $\sigma^2_\text{uncon}-\mathbb{V}\mathrm{ar}_{\text{ens}}[\langle\hat{X}_j\rangle_\text{C}(t)]$, where $\sigma^2_\text{uncon}$ is the unconditional variance and the last term is evaluated over a sufficiently large ensemble of measurement records. We test this inference consistency by calculating the variance difference from the collected ensemble (solid red diamonds and blue squares in FIG.~\ref{FigData1}\textbf{d}). The agreement with $v_\text{C}(t)$ indicates the reliability of our analysis.

Finally, the accuracy of the estimator state trajectories is assessed by calculating the ensemble-averaged squared Hilbert–Schmidt distance. The data points in FIG.~\ref{FigData1}\textbf{e} show the experimentally measured values, in close agreement with the theoretical curves (Supplementary~\ref{supp:HSc-Dis-Squ}). At $t=0$, the measured value for smoothing (solid red diamond) is roughly two-thirds of that for filtering (blue square), and it remains smaller for smoothing throughout the transient phase. The smoothed state is thus demonstrated to be closer to the target than the filtered state, yielding a more accurate trajectory for the resonator.

\section{Quantum trajectories of the ``true" state} \label{Sec:true}

Because all real quantum systems interact with baths that are not measured, the LTL filtered state is never pure in practice. Were one able to measure the
unobserved baths, or access the outputs of naturally occurring measurements on them, one could produce a more accurate trajectory of the system’s evolution. We refer to the system’s state, conditioned on measurements on all baths far into the past, as the ``true" state $\rho_\text{T}(t)$ --- without committing to any ontic or epistemic interpretation of quantum states. Ideally, this state is pure and evolves according to a stochastic Schrödinger equation. The trajectory of the true state, as the
best possible description of the system’s evolution, is our next estimation target.

As with the LTL filtered state, the system’s filtered state $\rho\fil(t)$ is the optimal real-time estimate of $\rho_\text{T}(t)$, with the squared Hilbert--Schmidt distance or the relative entropy as the cost function~\cite{guevara2015quantum}. We now show how to obtain a better estimate using future data. In previous studies of the quantum state smoothing formalism, Bob and Alice measure their respective baths from time $t=0$ assuming the same initial state \cite{guevara2015quantum,GueWis20,laverick2021quantum,laverick2021linear,CGLW21,LGW21}, with most taking that to be a pure state. Here we generalize this to an impure initial state for Alice by imagining that Bob monitors all baths from the distant past until $t=0$, at which point he hands one bath (or part of a bath) to Alice, who begins measuring it while Bob continues observing the remaining baths as well as having access to Alice's record. For a Gaussian true state, the smoothed estimate is characterised by Eq.~\eqref{QSSforLTLfil}, with $\mathbf{V}_\text{F}^\text{ss}$ replaced by the deterministic covariance matrix of the true state (that is, Bob's state), \blk $\mathbf{V}_\text{T}^\text{ss}$ --- see Methods~\ref{sec:QSSforLTL}, where the case of general quantum systems is also explained.

The set of possible true states conditioned, in part, on naturally occurring measurements (on the unobserved baths) depends on the types of interactions between the unobserved baths and the larger environment which acts as a measurement apparatus for those baths. Since these interactions are generally unknown, this set is also not known in general. However, in our case, there is a unique \textit{unravelling} (type of measurement) which satisfies the following desiderata for naturalness: (i) the resulting stochastic master equation for the system should be invariant under transformations that preserve the master equation (see Methods~\ref{sec:methods_Nat_Unrav} and Refs.~\cite{gisin1992quantum,gisin1992wave});
and (ii) the measurement should be Markovian, just as the baths are. Under these criteria, the natural unravelling of the resonator’s symmetric master equation is unique and corresponds to heterodyne detection~\cite{WisDio01}. This unravelling on the unobserved baths of course matches our unravelling on the observed bath, which is another reason it is natural for our system. Considering the total unravelling, the true state of the system is a stochastic coherent state (that is, the ground state with a stochastic phase-space displacement) (Methods~\ref{SecFullHete}).

\subsection*{Experimental ``true" state estimation}

We now estimate the resonator's ``true" trajectory (FIG.~\ref{FigData1}\textbf{a}). Its real-time estimate is the filtered trajectory already characterised by $\langle\hat{X}_{j}\rangle_\text{F}$ and $v_\text{F}$. Its post-processed estimate is given by Eq.~\eqref{QSSforLTLfil}, with all $\mathbf{V}\fil^\text{ss}$ terms replaced by $\mathbf{V}\god^\text{ss}=\mathbf{I}$, the covariance matrix of a displaced ground state, and with $\hat{\mathbf{x}}=(\hat{X}_1,\hat{X}_2)^\top$, considering the previously calculated values of $\langle\hat{X}_{j}\rangle_\text{F}$, $\langle\hat{X}_{j}\rangle_\text{R}$, $v_\text{F}$ and $v_\text{R}$. The results are shown in FIG.~\ref{FigData1}. Until the very end of the measurement record, the smoothed estimate differs from both the filtered estimate and from the smoothed estimate of the LTL filtered trajectory. For example, the red dashed line in FIG.~\ref{FigData1}\textbf{b} shows the markedly different value of $\langle\hat{X}_2\rangle_\text{S}$ for the sample record. As the smoothing here targets the true state rather than the LTL filtered state, some dynamics uncorrelated with the LTL state (dashed black) are evident.

The variance of the smoothed quantum state $v{\sm}$ is shown by the red dashed curve in FIG.~\ref{FigData1}\textbf{d}. It begins below the filtering variance $v{\fil}$ (blue curve) at $t=0$ and remains lower as both gradually decrease until they reach their steady-state values. Near the end of the record, where less future data is available for smoothing, $v{\sm}$ begins to increase towards $v{\fil}$. As $v{\sm}$ is smaller than $v{\fil}$, the smoothed state is purer than the filtered state. Similar to LTL estimation, there is a purity gain in the initial transient phase, exceeding an order of magnitude at $t=0$. In contrast to LTL estimation, the purity gain persists into the steady state, where both $v_\text{S}$ and $v_\text{F}$ have steady values, leaving the smoothed state 43\% purer. The consistency of the inference of this smoothed state is tested by calculating $\sigma^2_\text{uncon}-\mathbb{V}\text{ar}_{\text{ens}} [\langle\hat{X}_j\rangle_\text{S}(t)]$  (hollow red diamonds in FIG.~\ref{FigData1}\textbf{d}), which matches $v{\sm}(t)$ as predicted analytically (Supplementary~\ref{SecConsistency}).

The curves in the upper part of FIG. \ref{FigData1}\textbf{e} show the average value of the estimation cost function $\Delta^2_\text{HS}[\rho_\text{T},\rho_\text{C}]=\text{Tr}[(\rho_\text{T}-\rho_\text{C})^2]$ --- red dashed for $\text{C}=\text{S}$ and blue for $\text{C}=\text{F}$ (Supplementary~\ref{supp:HSc-Dis-Squ}). Throughout the entire record, the smoothed quantum state better estimates the target and thus its trajectory captures the resonator's evolution more accurately. These curves are not accompanied by experimental data because the mean values of the pure true state are inaccessible to our experiment; obtaining them would require access to the unobserved baths or to the information leaking from them into the rest of the environment. Consequently, this analysis remains contingent on our assumptions about the unravellings of the unobserved baths.
\begin{SCfigure*}
    \centering
    \includegraphics[width=\linewidth]{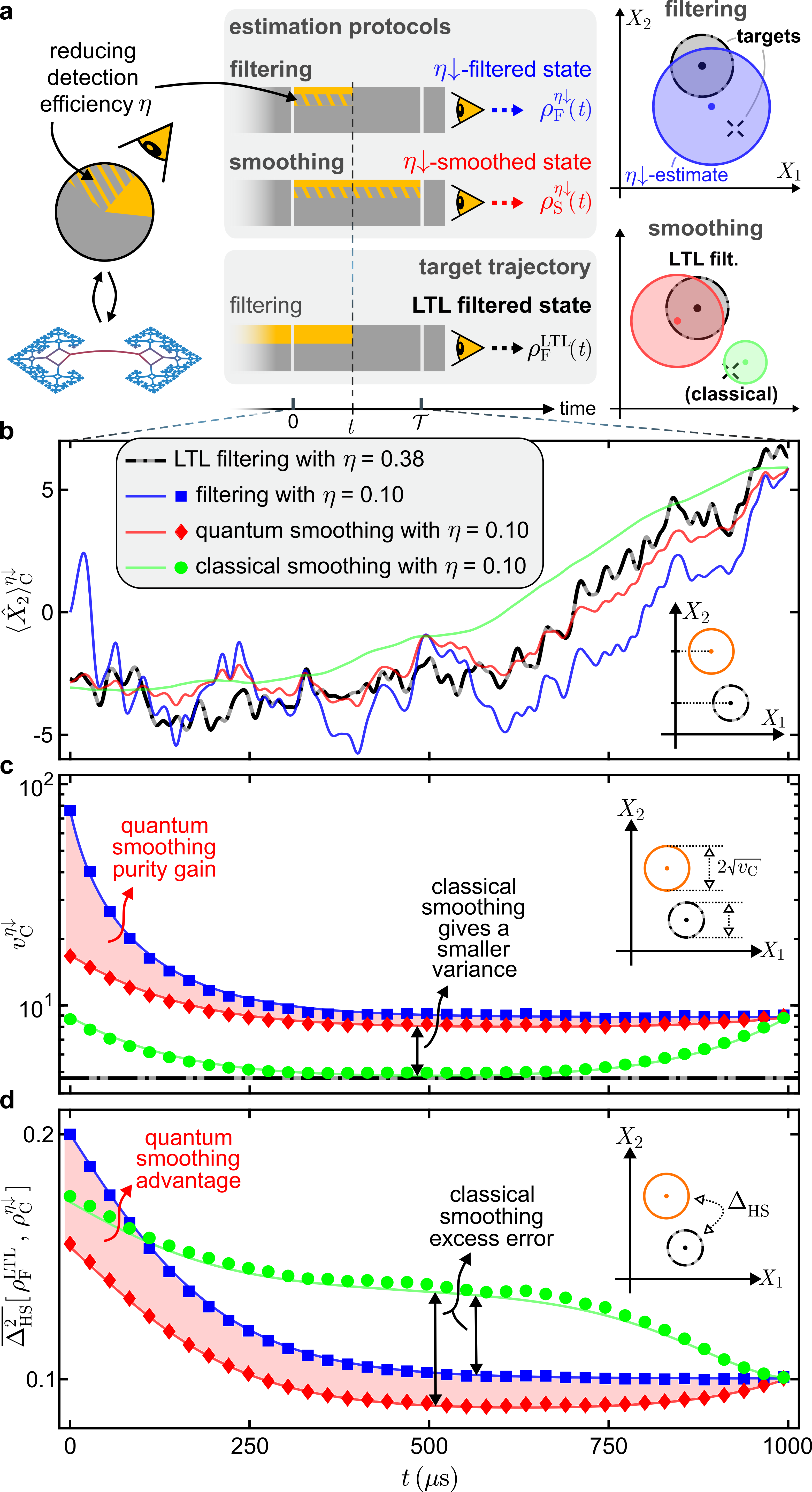}
    \caption{
\textbf{Trajectories after noise injection.} \textbf{a,} Conceptual illustration. As in Fig.~\ref{FigData1}a, the pie chart at left represents all the resonator's baths, partitioned into measured (yellow) and unmeasured (grey) parts. At centre, each bar depicts the continuous resonator-bath interaction, with yellow portions denoting the measured fraction used to infer a quantum state for the resonator at record time $t$ (dashed line). The grey shadings over the yellow parts indicate an artificial reduction of the detection efficiency $\eta$, implemented by adding white noise to the photocurrent. This noisier current is used to estimate the trajectory of the primary LTL filtered state, a target effectively conditioned on a partially unobserved bath (similar to the ``true" target in FIG.~\ref{FigData1}). The filtered estimate (blue) relies on real-time data, whereas the smoothed estimate (red) incorporates the entire data record in post-processing. As shown in phase space (right), the filtered state also yields the optimal estimate of the fictitious true value (cross) of the resonator’s classical counterpart, which has its own smoothed estimator (green). \textbf{b--d,} Estimation analysis. The right insets show the quantity plotted, obtained from estimator (orange) and target (black) states. 
\textbf{b,} Stochastic quadrature means $\langle\hat{X}_2\rangle_\text{F}^\text{LTL}(t)$ and $\langle\hat{X}_2\rangle_\text{C}^{\eta\downarrow}(t)$, with C = F (filtering), S (smoothing) or cS (classical smoothing), shown for a sample measurement record. 
\textbf{c,} Deterministic variances of the target, estimator states, and classical smoother. The values of $v_\text{C}^{\eta\downarrow}$ agree well with $\sigma^2_\text{uncon}-\mathbb{V}\text{ar}_{\text{ens}}[\langle\hat{X}_j\rangle^{\eta\downarrow}_\text{C}(t)]$ computed over an ensemble of 12,489 records (data points, Supplementary~\ref{SecConsistency} and \ref{sec:ErrorBars_forSC}). The similarity in variance between the target and classical smoother is coincidental. \textbf{d,} Comparison of the estimator states and the classical smoother with the target. The data points represent the ensemble average of the Hilbert–Schmidt distance squared, $\Delta^2_\text{HS}[\rho_\text{F}^\text{LTL}(t),\rho^{\eta\downarrow}_\text{C}(t)]=\text{Tr}[(\rho_\text{F}^\text{LTL}(t)-\rho^{\eta\downarrow}_\text{C}(t))^2]$. Theoretical predictions are shown as lines.}
    \label{FigData3}
\end{SCfigure*}

To provide a full demonstration of the performance of our true state estimators, we conduct another experiment in which the LTL filtered state trajectory serves as a true-state target. We add white noise to the homodyne photocurrent, thereby effectively creating an additional unobserved bath whose measurement outcomes are already incorporated into $\rho_\text{F}^\text{LTL}(t)$. For the noise-added data, $\rho_\text{F}^\text{LTL}(t)$ serves as a true-state target because it is partly conditioned on the newly created unobserved bath (FIG.~\ref{FigData3}\textbf{a}). Since this true-state target is experimentally accessible, the noise injection enables direct validation of our true state estimation techniques.

The noise-added photocurrent is analysed in a similar manner to previous experiments (Methods~\ref{sec:NI_analysis}). The noise-added smoothed mean values typically track their target state more closely than the noise-added filtered values (FIG.~\ref{FigData3}\textbf{b}), and the smoothing variance is smaller than the filtering variance (FIG.~\ref{FigData3}\textbf{c}). To quantify the performance of these trajectories in estimating their target, we calculate their squared Hilbert–Schmidt distance to the LTL filtered state, averaged over all available records. As shown in FIG.~\ref{FigData3}\textbf{d}, it is 23\% smaller for smoothing (red) than for filtering (blue) at the record start ($t=0$) and 13\% smaller in the steady state ($t \approx 500\,\mu$s). This result validates the advantage of smoothing for estimating a true state. Significantly, it demonstrates that post-processing can partially compensate for noise-induced degradation in the accuracy of quantum trajectories. Since the effect of adding uncorrelated white electronic noise is quantitatively equivalent to increasing the inefficiency of linear detection --- in our case lowering $\eta$ from 0.38 to 0.10 (Methods~\ref{sec:NI_analysis}) --- this result also indicates that post-processing can partially compensate for inefficiency in measuring the system's baths.

\section{Contrasting quantum and classical smoothing} \label{Sec:traj}

The filtering equations that apply to linear-Gaussian quantum systems are the same as those of classical dynamical systems, as, in the Heisenberg picture, the system operators commute with the past observables of the bath~\cite{wiseman2009quantum}. However, they generally do not commute with the future observables, and thus quantum state smoothing cannot be regarded as a direct extension of classical smoothing~\cite{guevara2015quantum}. This leads to characteristic differences.

One key difference is that the quantum-smoothed mean values are predicted to generally exhibit stochastic character (with special exceptions identified in Ref.~\cite{laverick2021quantum}). By contrast, classical smoothing yields mean value traces that are --- as the name suggests --- smooth, with a first derivative that is continuous in time (except in the presence of correlated measurement and process noise~\cite{laverick2021quantum}). To validate these theoretical predictions, we apply the classical smoothing equations, given by Eq.~\eqref{QSSforLTLfil} with all $\mathbf{V}\fil^\text{ss}$ terms replaced by $\mathbf{0}$~\cite{fraser1969optimum}, to our measurement records. As the weak feedback-induced correlation between the measurement and process noise is negligible (Methods~\ref{sec:methods:char_feedback} and \ref{sec:NI_analysis}), the trajectories of the classical mean values are indeed smooth, with no abrupt changes in the time derivative (green lines in FIGs.~\ref{FigData1}\textbf{b} and~\ref{FigData3}\textbf{b}). The stochastic behaviour of the quantum traces should, in principle, manifest as sharp features with an ill-defined time derivative. While the finite acquisition speed of any real experiment prevents such sharp features from being directly observed, signatures of the stochastic behaviour are evident in our experiment as enhanced fluctuations of the mean values, i.e., faster changes in the time derivative of all red curves in FIGs.~\ref{FigData1}\textbf{b} and \ref{FigData3}\textbf{b}. 

\begin{figure}
    \centering
    \includegraphics[width=\linewidth]{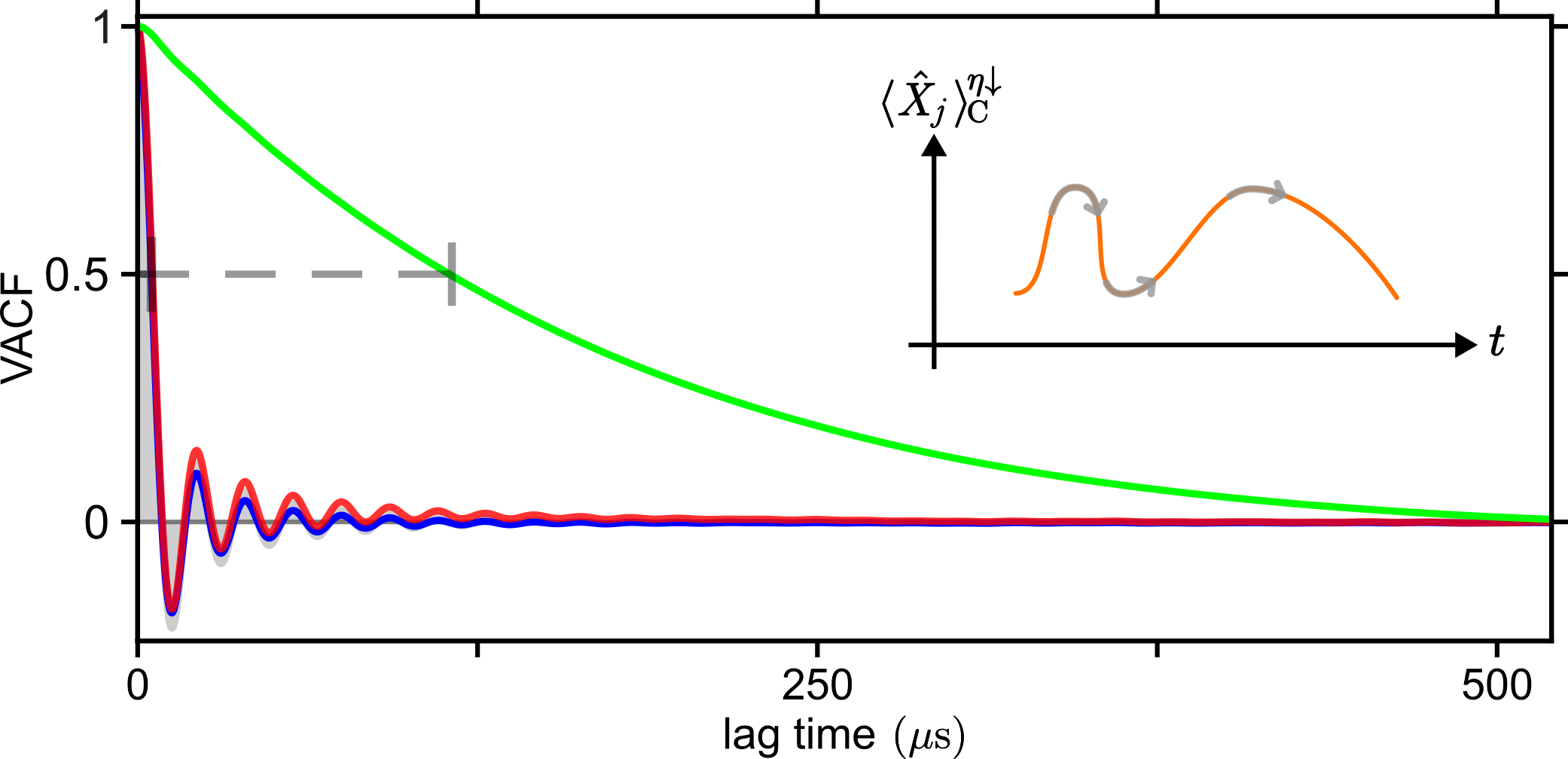}
    \caption{\textbf{Fluctuations of the mean value traces.} The ``quadrature velocity" $(\dd\langle\hat{X}_j\rangle_\text{C}^{\eta\downarrow}/\dd t)$ autocorrelation function (VACF) for filtering ($\text{C}=\text{F}$, blue), quantum state smoothing ($\text{C}=\text{S}$, red) and classical smoothing ($\text{C}=\text{cS}$, green), using the noise-injection experiment ($\eta\mathord{\downarrow}$) data. The autocorrelation function is first averaged across the ensemble of measurement records, and then normalised to its maximum value. Its decay time is a measure of how rapidly the slope of $\langle\hat{X}_j\rangle_\text{C}^{\eta\downarrow}(t)$  changes (see inset). As indicated by the dashed line, the decay is markedly slower for classical smoothing. The rippling of the blue and red curves is consistent with the autocorrelation of the impulse response of the demodulation low-pass filter (grey shading).}
    \label{FigData4}
\end{figure}

To compare the fluctuations quantitatively, we compute the autocorrelation of the time derivative of the mean values, or ``quadrature velocities". The autocorrelation functions for the noise-injection experiment are shown in FIG.~\ref{FigData4}. As expected, the quadrature velocities for classical smoothing exhibit sustained correlations, with a 115-$\mu$s decorrelation time. By contrast, both quantum smoothing and filtering show rapid decorrelation, about 21 times faster than classical smoothing and consistent with the time scale of the experimentally analysed noise. Thus, our experiment verifies one of the key expected differences between classical and quantum smoothing.

A second key distinction is that the classical smoothing equations --- when applied to a quantum system --- can produce unphysical quantum states. In fact, the classical smoothing variance can fall below the minimum value set by the Heisenberg uncertainty principle (see Discussion and Ref.~\cite{LCW25}). Our experiments show precursors of this apparent violation of the uncertainty principle in FIGs.~\ref{FigData1}\textbf{d} and~\ref{FigData3}\textbf{c} where the classical smoothing variances (green curves) are consistently smaller than both quantum smoothing and filtering. Additionally, the classical mean values (green lines in FIGs.~\ref{FigData1}\textbf{b} and~\ref{FigData3}\textbf{b}) --- while giving the \textit{weak values} of the quadratures~\cite{tsang2009optimal,LCW25} --- are not optimal estimates of the mean values of the target states. For example, the green circles in FIG.~\ref{FigData1}\textbf{c} show the statistical deviation of the mean values when targeting the LTL filtered state, demonstrating their inferior accuracy compared to quantum state smoothing. Finally, the larger squared Hilbert--Schmidt distances for classical smoothing (green curves in FIGs.~\ref{FigData1}\textbf{e} and~\ref{FigData3}\textbf{d}) confirm that the trajectories produced by the classical equations are more distinguishable from the targets than the quantum smoothed trajectories. These differences, which scale predominantly with the ratio of the target state variance to the filtered state variance, highlight the fundamental inadequacy of classical smoothing for reconstructing quantum state trajectories.

\section{Discussion}

Our demonstration that future information allows more accurate quantum state trajectories could enable a range of new applications in quantum computing, communication, sensing and control~\cite{convy2022machine,livingston2022experimental,campbell2020universal,Duan2025concurrent, Sayrin2011realtime,Magrini2021realtime}. In quantum computing, it may benefit protocols that employ post-processing~\cite{convy2022machine,livingston2022experimental,campbell2020universal}. For instance, it has recently been proposed that continuous measurement of leakage signals can be used to identify excursions of qubit systems out of the logical subspace, and that this could be used for error correction~\cite{campbell2020universal}. Our quantum state smoothing techniques could improve the accuracy with which these excursions can be identified (see Supplementary~\ref{supp:applications}.1). More generally, information must be stored to employ post-processing in quantum error correction~\cite{convy2022machine,livingston2022experimental}, presenting resource constraints for large-scale quantum computers. Here, trading off some stored past data for future data may allow a net reduction in storage requirements.

In quantum communication, quantum teleportation is typically implemented by performing a joint measurement on the input state and one mode of a two-mode entangled state, and then applying measurement-dependent operations to the other entangled mode~\cite{wiseman2009quantum}. In continuous-in-time quantum teleportation, the second entangled mode could be stored allowing the use of future data by our post-processing techniques. This could improve the accuracy of the operations performed and thereby increase fidelity (see Supplementary~\ref{supp:applications}.2).

In quantum sensing and control, our post-processing techniques may allow improved precision. We validate this, in spirit, in the noise-injection experiment, where future information is used to mitigate noise, improving the precision with which the quadrature mean values are known. Moreover, these techniques can allow better prediction of the actions of machines that use quantum sensors to respond to classical stimuli from their environment (see Extended Data FIG.~\ref{SuppFig:SensingConcept}). Assuming access to partial signals that leak from the machines, the stimuli can be optimally estimated by classical smoothing equations~\cite{tsang2009optimal}. Here, however, it is the machines' knowledge of the stimuli that matters, rather than the stimuli themselves. As we show in Supplementary~\ref{supp:applications}.3, 
our quantum state smoothing techniques allow optimal estimation of this.

Our approach is also relevant to protocols that exploit the transient conditional dynamics of quantum systems shortly after monitoring begins. For example, continuous weak measurements are widely used to generate non-classical states of quantum systems~\cite{muller2008entanglement,meng2020mechanical,thomas2021entanglement}. Employing future information could allow useful states to be identified earlier in the transient phase and improve the fidelity of the final states that are generated. Similar transient-phase physics has been shown to play an important role in studies of conditional entropy production under weak continuous measurement~\cite{rossi2020experimental}. Since quantum smoothing yields purer quantum states, access to future information would appear to modify the rate of entropy production, offering a new perspective on these studies.

As discussed earlier, while our formalism allows improved estimation of the true state of quantum systems that are interacting with an environment, there is usually an ambiguity about the type of the true state. This raises interesting philosophical and practical questions about decoherence in quantum mechanics. In particular, different assumptions about the nature of the unobserved environment lead to different sets of possible true states, even when the observed measurement record is identical. For instance, while an environment that performed position detection would drive the resonator towards a coherent true state~\cite{bowen2015quantum}, phonon-number resolving measurements would instead drive it towards a Fock state~\cite{brawley2016nonlinear,deleglise2008reconstruction}. This highlights that, when adapted to the case of a single observer, quantum state smoothing is not necessarily a reconstruction of objective reality but rather a conditional description that depends on the physical model of the interactions between system, bath and environment.

Our work also highlights the incompatibility of classical smoothing with quantum state trajectories, in particular by showing precursors of the apparent violation of the Heisenberg uncertainty principle by classical smoothing. As derived in Supplementary~\ref{sec:SSvariances}, such a violation would require $\eta>0.25$ and $\mathcal{C}/n_\text{th} > 1/(4\eta-1)$. Our experiment satisfies the first condition ($\eta=0.38$) but not the second, the latter being limited by the maximum optical power that can be used before mode competition degrades performance (see Supplementary~\ref{supp:PSDs} and Ref.~\cite{khademi2024optomechanical}).

\section*{Data availability}

\noindent
The data presented in this study is available from the Zenodo repository.

\section*{Acknowledgements}

\noindent
We thank A. C. Doherty, W. W. Wasserman, A. G. White, I. Marinković, G. I. Harris and J. S. Bennett for useful discussions. This research was supported primarily by the Australian Research Council Centre of Excellence for Engineered Quantum Systems (EQUS, CE170100009). It was also supported by the Air Force Office of Scientific Research under award number FA9550-20-1-039, the Australian Research Council Centre of Excellence CE170100012 (Centre for Quantum Computation and Communication Technology), and the Australian Research Council Centre of Excellence in Quantum Biotechnology (QUBIC, CE230100021). K.T.L. acknowledges the Plan France 2030 through the project NISQ2LSQ (Grant ANR-22-PETQ-0006) and the project OQuLus (Grant ANR-22-PETQ-0013).

\section*{Author contributions}
\noindent
S.K. conceived the concepts of LTL filtered state estimation and the noise-injection experiment. H.M.W., S.K. and K.T.L., with input from W.P.B., developed the ideas of natural measurements and estimation of a pure state. K.T.L., H.M.W. and S.K. carried out the novel theoretical derivations. S.K. applied the framework to the experiment. S.K. and J.J.S., with input from W.P.B., built the setup, characterised the optomechanical device and devised the experimental protocols. The device was fabricated by J.C., J.G. and S.G. Final data was collected by S.K. Data analysis was carried out by S.K., with additional input from K.T.L., W.P.B., J.J.S. and H.M.W. The application examples were worked out by S.K., with critical feedback from W.P.B. and H.M.W. Figures were prepared by S.K., J.J.S. and W.P.B., with contributions from J.C., J.G. and S.G. The manuscript was written by S.K., W.P.B., J.J.S., K.T.L. and H.M.W., with feedback from all authors. W.P.B. and H.M.W. supervised the project.

\section*{Competing interests}
\noindent
The authors declare no competing interests.

\renewcommand{\figurename}{Extended Data FIG.}\setcounter{figure}{0}
\makeatletter
\renewcommand{\theHfigure}{ED.\arabic{figure}}
\makeatother
\renewcommand{\tablename}{Extended Data TABLE}\setcounter{table}{0}

\begin{figure*}[t!]
    \centering
    \includegraphics[]{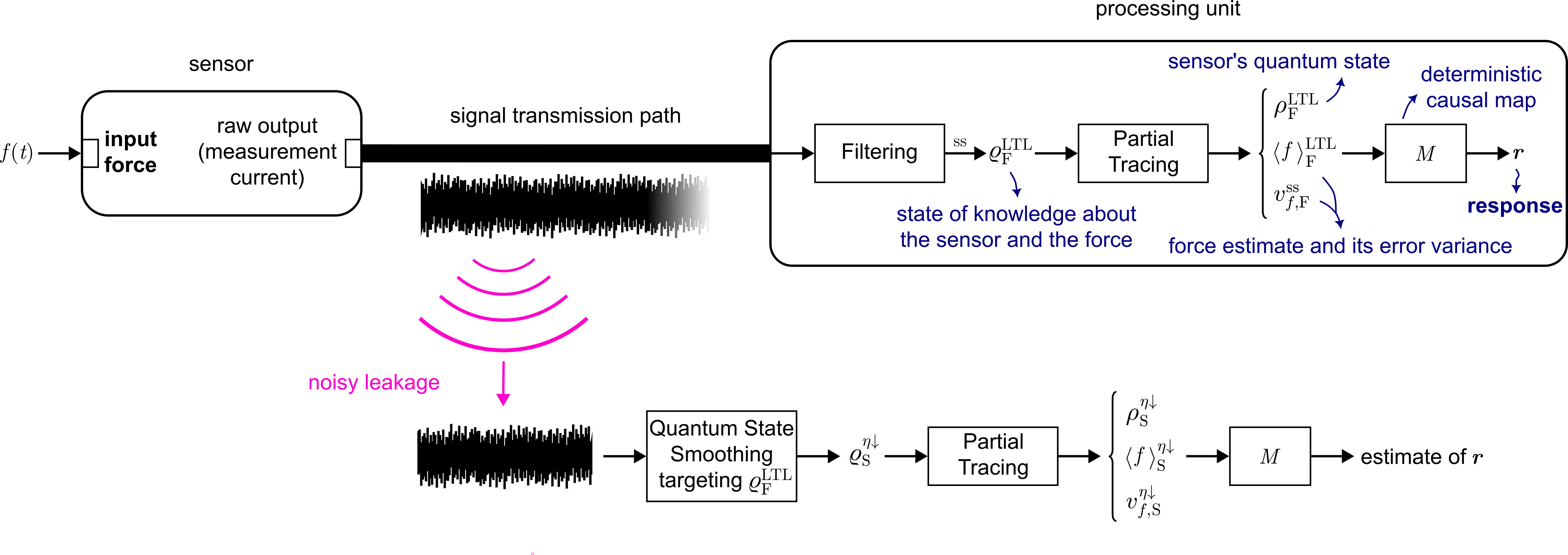}
    \caption{\textbf{Application example: estimating the response of a sensing machine.} A force sensor sends its measurement current to a separate unit for processing. The \textit{steady-state} (ss) output of the filtering process is a state of knowledge about the sensor and the force, denoted by $\varrho_\text{F}^\text{LTL}$. By tracing out the sensor part, the LTL filtered estimate of the force $(\langle f\rangle_\text{F}^\text{LTL})$ is found. This estimated value of the force is then fed to a deterministic causal map ($M$, representing a decision-making algorithm, an analogue response-generating system, or the like) to provide a final response to the input force, denoted by $\bm{r}$. Meanwhile, the leakage of the transmission path allows an external agent to record a noisy copy of a piece of the measurement current. This agent is interested in knowing $\bm{r}$, but they have no access to the processing unit. They therefore estimate the filtered value of the force using their noisy piece and feed it to $M$, which is public knowledge, to estimate $\bm{r}$. Assuming that the machine has a considerable delay before the filtering process (due to resource allocation limitations, having a long transmission path, etc.) or that the agent sacrifices time for accuracy, they apply quantum state smoothing equations to their noisy piece, with $\varrho_\text{F}^\text{LTL}$ as the target. The output, once the sensor part is traced out, gives $\langle f\rangle_\text{S}^{\eta\downarrow}$ which is the optimal post-processed estimate of $\langle f\rangle_\text{F}^\text{LTL}$. Feeding this number to $M$ yields an estimate of $\bm{r}$. If $M$ is linear, this estimate of $\bm{r}$ has the least mean squared error. See Supplementary~\ref{supp:applications}.3 for more explanation and numerical demonstration.}
    \label{SuppFig:SensingConcept}
\end{figure*}

\bibliographystyle{apsrev4-2}
\bibliography{sn-bibliography}

\section*{Methods}
\setcounter{section}{0}

\ifSubfilesClassLoaded{
    \title{Theoretical methods}
    \date{\today}
    \maketitle
}{
}

\section{Filtering and retrofiltering in LGQ systems}\label{sec:GeneralACDK}

The problem of quantum filtering and retrofiltering applies to Markovian open quantum systems, where the unconditional evolution of the system's quantum state is described by a Lindblad master equation
\beq
\hbar\frac{\dd}{\dd t}\rho(t) = -i[\hat{H},\rho(t)] + {\cal{D}}[\hat{\bf c}]\rho(t)\,.
\label{equ:LindbladMasEq}
\eeq
Here, $\hat{H}$ is the system Hamiltonian and $\hat{\bf c}$ is a vector of Lindblad operators.
An LGQ system is an open $N$-mode bosonic system, with canonical position and momentum operators for the $n$th mode $\hat{q}_n$ and $\hat{p}_n$, satisfying the canonical commutation relations $[\hat{q}_k,\hat{p}_\ell] = i\hbar\delta_{k,\ell}$. These systems arise when $\hat{H}$ is at most quadratic and $\hat{\bf c}$ is linear in the canonical operators, e.g., $\hat{H} = \bx\tp {\bf G} \bx\quad{\rm and} \quad \hat{\bf c} = {\bf B}\bx$, where $\bx = (\hat{q}_1,\hat{p}_1,...,\hat{q}_N,\hat{p}_N)\tp$ and ${\bf G}$ is Hermitian. In the Heisenberg picture, the operators evolve dynamically according to the linear Langevin equation \cite{DohJac99,wiseman2009quantum},
\beq
\dd \hat{\bf x}(t) = {\bf A}\,\bx(t)\,\dd t + {\bf E}\,\dd\hat{\bf v}_{\rm p}(t)\,,
\eeq
where $\dd\hat{\bf v}_{\rm p}(t)$ is a set of external operators that influence the evolution due to interactions with a quantum environment (i.e., quantum noise), satisfying
\begin{align}
\label{eq:WhtQNoise}
& \ex{\dd\hat{\bf v}_{\rm p}(t)} = {\bf 0}\,, \quad {\rm and} \nonumber \\ & \hspace{3.5em} {\rm Re}[\ex{\dd{\hat{\bf v}_{\rm p} (t)\,\dd\hat{\bf v}_{\rm p}\tp(t')}}] = {\bf I}\,\delta(t - t')\,\dd t^2\,,
\end{align}
where ${\bf I}$ is the identity matrix. Matrices ${\bf A}$ and ${\bf E}$ are determined by the system Hamiltonian and Lindblad operators \cite{wiseman2009quantum}. We will assume throughout that these are time-independent (as they are in our experiment), but in general they can be time-dependent without affecting any of the following.

These dynamics preserve Gaussianity, meaning that if the system is initially prepared in a Gaussian quantum state, it will remain Gaussian over its entire evolution. Gaussian states are completely described, in terms of their Wigner function, by the mean $\ex{\bx}$ and the (symmetrised) covariance matrix $V_{ij} = \frac{1}{2}\ex{\hat{x}_i \hat{x}_j + \hat{x}_j \hat{x}_i} - \ex{\hat{x}_i}\ex{\hat{x}_j}$. It is easy to show \cite{wiseman2009quantum} that these moments evolve according to 
\begin{align}
&\dd \ex{\hat{\bf x}}(t) = {\bf A}\,\ex{\bx}(t)\,\dd t\,,\\
&\frac{\dd}{\dd t}{\bf V}(t) =  \mathbf{A}\,\mathbf{V}(t) + \mathbf{V}(t)\,\mathbf{A}\tp + \mathbf{D}\,,
\end{align}
where ${\bf D} = {\bf E}{\bf E}\tp$.

We now consider a continuous-in-time measurement of the system's baths, or some fraction thereof. In order for the system to remain LGQ, it must be that the {\em conditional} dynamics are linear and preserve Gaussianity. This is the case if and only if the operator describing the measurement is linear in $\bx$~\cite{DohJac99}, i.e., in the Heisenberg picture,
\beq
\hat{\bf y}(t)\,\dd t = {\bf C}\,\bx(t)\,\dd t + \dd\hat{\bf v}_{\rm m}(t)\,,
\eeq
where $\dd \hat{\bf v}_{\rm m}$ are noise operators affecting the measurement outcome, satisfying similar conditions to Eq.~(\ref{eq:WhtQNoise}). In general, $\dd\hat{\bf v}_{\rm m}(t)$ may be correlated with $\dd\hat{\bf v}_{\rm p}(t)$, giving ${\rm Re}[\ex{{\bf E}\,\dd\hat{\bf v}_{\rm p}(t)\,\dd\hat{\bf v}_{\rm m}\tp(t)}] = {\bf \Gamma}\tp\,\dd t$. We will assume throughout that the measurement matrix ${\bf C}$ is time-independent. From now on, the explicit time arguments of other variables may be omitted for notational convenience.

Real-time conditioning (or filtering) the system on the measurement outcomes, from an initial time $t_0$ until time $t$ --- in the main text, we set $t_0 = 0$ --- causes the mean and covariance matrix to evolve according to \cite{DohJac99,wiseman2009quantum}
\begin{align}
    &\ex{\bx}\fil(t) = \ex{\bx}\fil(t-\dd t)+{\bf A}\,\ex{\bx}\fil(t-\dd t)\,\dd t \nonumber\\ & \hspace{9.5em} +{\cal K}^{+}[{\bf V}\fil(t)]\,\dd \mathbf{w}\fil(t)\;,
\label{1stMomFiltering}\\
    &{\bf V}\fil(t) = {\bf V}\fil(t-\dd t) + \mathbf{A}\,\mathbf{V}\fil(t-\dd t)\,\dd t \nonumber\\ & \hspace{5.5em} + \mathbf{V}\fil(t-\dd t)\,\mathbf{A}\tp\,\dd t + \mathbf{D}\,\dd t \nonumber\\ & \hspace{5.5em}  - \mathcal{K}^{+}[\mathbf{V}\fil(t)]\,\mathcal{K}^{+}[\mathbf{V}\fil(t)]\tp\,\dd t\,.
\label{2ndMomFiltering}
\end{align}
Here, ${\cal K}^{+}[{\bf V}] = {\bf V}{\bf C}\tp + {\bf \Gamma}\tp$ is the ``kick'' matrix and $\dd{\bf w}\fil(t) = {\bf y}(t)\,\dd t - {\bf C}\,\ex{\bx}\fil(t-\dd t)\,\dd t$ is the (filtered) vector of innovations, with statistics identical to those of an infinitesimal Wiener increment \cite{wiseman2009quantum}. The initial conditions for these equations, which are identical to the classical Kalman--Bucy filtering equations~\cite{KalBuc61}, are given by the system's unconditional state.

To describe the probability of the measurement outcomes that occurred from $t$ until the final time ${\cal T}$, one generally introduces a \textit{positive operator-valued measure} element $\hat{E}\rfil(t)$, more commonly called the retrofiltered effect. This operator encodes the probability of the \textit{future} measurement record $\overrightarrow{O}_t := \{{\bf y}(\tau):\tau\in(t,{\cal T}]\}$ occurring for any given state via $p(\overrightarrow{O}_t|\rho(t)) = {\rm Tr}[\hat{E}\rfil(t)\,\rho(t)]$ --- the \textit{past} record is denoted by $\past{O}_t:=\{{\bf y}(\tau):\tau\in(t_0,t]\}$ and the entire (\textit{past--future}) record by $\both{O}:=\{{\bf y}(\tau):\tau\in(t_0,{\cal T}]\}$~\footnote{The definitions of the past and future records here differ slightly from those in Ref.~\cite{guevara2015quantum} and those that follow: here, time $t$ is included in the past record rather than in the future record. This difference, however, does not affect the conclusions or final expressions.}. In the LGQ regime, the normalised retrofiltered effect has a Gaussian Wigner function with the mean and covariance~\cite{ZhaMol17}
\begin{align}
    & \langle\hat{\mathbf{x}}\rangle\rfil(t-dt) = \langle\hat{\mathbf{x}}\rangle\rfil(t) -\mathbf{A}\,\langle\hat{\mathbf{x}}\rangle\rfil(t)\,\dd t \nonumber\\&\hspace{9.5em} +\mathcal{K}^{-}[\mathbf{V}\rfil(t)]\,\dd\mathbf{w}\rfil(t)\,,
\label{1stmomentNREO}\\
    & \mathbf{V}\rfil(t-dt)= \mathbf{V}\rfil(t) -\mathbf{A}\,\mathbf{V}\rfil(t)\,\dd t \nonumber\\&\hspace{5.5em} - \mathbf{V}\rfil(t)\,\mathbf{A}\tp\,\dd t + \mathbf{D}\,\dd t \nonumber\\&\hspace{5.5em} - \mathcal{K}^{-}[\mathbf{V}\rfil(t)]\,\mathcal{K}^{-}[\mathbf{V}\rfil(t)]\tp\,\dd t\,,
\label{2ndmomentNREO}
\end{align}
where ${\cal K}^{-}[{\bf V}] = {\bf V}{\bf C}\tp - {\bf \Gamma}\tp$ and $\dd{\bf w}\rfil(t) = {\bf y}(t)\,\dd t - {\bf C}\,\ex{\bx}\rfil(t)\,\dd t$. These are backward-evolving equations with uninformative final conditions.
See Ref.~\cite{laverick2021quantum} for the details on how to deal with the strictly undefined final condition on the mean values.

We note that, on the right-hand sides of Eqs.~\eqref{1stMomFiltering}--\eqref{2ndmomentNREO}, only terms up to order $\dd t$ contribute and higher-order terms can be discarded (in the continuous-time limit).

Naively applying classical smoothing to the Wigner functions~\cite{LCW25}, one obtains the classically smoothed Wigner function with the mean and covariance
\begin{align}
\ex{\bx}\csm &= {\bf V}\csm\left[{\bf V}\fil\inv\,\ex{\bx}\fil + {\bf V}\rfil\inv\,\ex{\bx}\rfil\right],\\
{\bf V}\csm &= \left[{\bf V}\fil\inv + {\bf V}\rfil\inv\right]\inv\,.
\end{align}
In general, these do not correspond to a physical quantum state, with the covariance matrix violating the Heisenberg uncertainty principle~\cite{laverick2019quantum}
\beq\label{SHUP}
{\bf V} + \frac{i\hbar}{2}{\bf \Sigma} \geq 0\,,
\eeq
where ${\bf \Sigma}$ is a block-diagonal matrix with blocks $\left[\begin{smallmatrix}
0&1\\
-1&0
\end{smallmatrix}\right]$. 

For our experimental system, all covariance matrices are of the form $v{\bf I}$. The ground state variance is normalised to one (equivalent to choosing the units such that $\hbar = 2$) and Eq.~(\ref{SHUP}) reduces to $v \geq 1$.

\section{Adapting the quantum state smoothing formalism}\label{sec:QSSforLTL}

For a general quantum system, whether LGQ or not, the smoothed quantum state of Guevara and Wiseman \cite{guevara2015quantum,LWCW23} is defined as 
$\sum_{\rho\god} p(\rho\god;t|\both{O})\,\rho\god(t)\,,
$
where the true quantum state $\rho\god$ was defined by introducing a secondary observer, Bob, who performed a simultaneous measurement on any part of the system baths (or some subset thereof) that the main observer's, Alice's, measurement missed. This true state was conditioned on the past records obtained by measuring the baths observed by Alice, $\past{O}_t$, and by Bob, $\past{U}_t$. The notation $p(\bullet;t)$ should be understood as a time-dependent probability distribution of the random variable $\bullet$. This smoothed quantum state has since been shown to be the optimal post-processed Bayesian estimator of the true state \cite{CGLW21,LGW21} in terms of minimising the squared Hilbert--Schmidt distance, as well as the relative entropy ${\rm Tr}[\rho_\text{T}\,({\rm log}\,\rho_\text{T}-{\rm log}\,\rho_\text{S})]$. These cost functions are minimised by the conventional filtered state considering real-time estimation of the true state~\cite{guevara2015quantum,LGW21}. As discussed in Ref.~\cite{LWCW23}, these optimalities are independent of some specifications of Bob's measurement. Here, we exploit this independence to adapt the formalism. As a result, the cost functions being optimised will be the same.

The optimal smoothed estimate (in the Bayesian sense) of our \textit{target} state is given by
\beq\label{eq:gen-sm}
\rho\sm(t) = \sum_{\rho_{\rm tar}\in{\mathbb T}} p(\rho_{\rm tar};t|\both{O})\,\rho_{\rm tar}(t)\,,
\eeq
where $\mathbb T$ is the set of all possible realisations of the target state. When we target the LTL filtered state, the system of interest has been monitored for a long time prior to time $t_0$ by Bob. At time $t_0$, Bob stops his measurement, having obtained the measurement record $\past{U}_{t_0}:=\{{\bf y}(\tau):\tau \in (\tau_0,t_0]\,,\,\, \tau_0\ll t_0 \}$, and Alice starts to measure the system and obtains $\past{O}_t = \{{\bf y}(\tau):\tau \in (t_0,t] \}$ for $\forall t\le\mathcal{T} $. The LTL filtered state at time $t$ is the state
conditioned on $\past{U}_{t_0}\cup\past{O}_t$. As $\past{U}_{t_0}$ is not known to Alice, ${\mathbb T} = \{\rho_{\past{U}_{t_0}\cup\past{O}_t}, \forall \past{U}_{t_0}\}$. We note that Alice and Bob do not necessarily need to perform the same measurement on the same portion of the system baths, as is the case in our experiment; it is however necessary that Alice knows Bob's measurement choice as well as what portion of the baths he is measuring. When we target our ``true" state, Bob obtains $\past{U}_{t_0}$ by measuring all the system baths until $t_0$; he also obtains a $\past{U}_{t}$ by measuring any part of the baths that Alice's measurement misses (after its start at $t_0$). Our ``true" state at time $t$ is the state
conditioned on $\past{U}_{t_0}\cup\past{U}_{t}\cup\past{O}_t$ and the set ${\mathbb T}$ is given by $\{\rho_{\past{U}_{t_0}\cup\past{U}_{t}\cup\past{O}_t}, \forall \past{U}_{t_0}, \forall \past{U}_{t}\}$. We should now derive the conditional probability appearing in Eq.~\eqref{eq:gen-sm} for each element of ${\mathbb T}$ and then perform the summation.

\subsection{Quantum state smoothing for LGQ systems}\label{MT:LQG}
The derivation for LGQ systems largely follows that of the standard quantum state smoothing formalism in Ref.~\cite{laverick2019quantum}. We begin by defining the Wigner function as~\cite{wiseman2009quantum} 
\beq\label{eq:Wig}
W_\rho({\bf x}) = (2\pi)^{-2N} \int_{{\bf b} \in {\mathbb R}^{2N}} \dd^{2N} {\bf b}\, \Tr[\rho\,e^{i{\bf b}\tp (\bx - {\bf x})}]\,.
\eeq
Considering the linearity of this definition,
the Wigner function of $\rho\sm(t)$ equals
\begin{align}
W\sm({\bf x};t) &= \sum_{\rho_{\rm tar} \in {\mathbb T}}p(\rho_{\rm tar};t|\both{O})\,W_{\rm tar}({\bf x};t)\,.
\end{align}
Using Bayes' theorem, we have $p(\rho_{\rm tar};t|\both{O}) \propto p(\fut{O}_t|\rho_{\rm tar},\past{O}_t)\,p(\rho_{\rm tar}|\past{O}_t) = p(\fut{O}_t|\rho_{\rm tar})\,p(\rho_{\rm tar}|\past{O}_t)$, with the equality resulting from the Markovianity of the system. By definition of the retrofiltered effect, we know $p(\fut{O}_t|\rho_{\rm tar}(t)) = \Tr[\hat{E}\rfil(t)\,\rho_{\rm tar}(t)]$. Using the identity $\Tr[\rho\,\sigma] \propto \int\dd{\bf x}\,W_{\rho}({\bf x})\,W_{\sigma}({\bf x})$, we have 
\beq \label{1stPrt}
p(\fut{O}_t|\rho_{\rm tar}) \propto \mathcal{N}(\ex{\bx}\tar;\ex{\bx}\rfil, {\bf V}\rfil + {\bf V}\tar)\,.
\eeq
Now, all that remains is to compute the probability $p(\rho_{\rm tar}|\past{O}_t)\equiv p(\ex{\bx}_{\rm tar}|\past{O}_t)$, where the equivalence comes from the fact that, since the covariance is deterministic, the possible target states are parametrised solely by their mean $\ex{\bx}_{\rm tar}$.

When the target is the LTL filtered state, the mean satisfies  
\begin{align}\label{eq:LTL_mean}
&\dd \ex{\bx}\fil^{\rm LTL} = {\bf A} \,\ex{\bx}\fil^{\rm LTL}\,\dd t + {\cal K}^+[{\bf V}\fil^{\rm LTL}]\,\dd{\bf w}\fil^{\rm LTL}
\end{align}
where $\mathbf{V}\fil^{\rm LTL}$ obeys
\begin{align}
&\frac{\dd}{\dd t}{\bf V}\fil^{\rm LTL} = {\bf A}\mathbf{V}\fil^{\rm LTL} + \mathbf{V}\fil^{\rm LTL}\mathbf{A}\tp + \mathbf{D} \nonumber\\
&\hspace{10em}- \mathcal{K}^{+}[\mathbf{V}\fil^{\rm LTL}]\,\mathcal{K}^{+}[\mathbf{V}\fil^{\rm LTL}]\tp\,.\label{eq:LTL_Var}
\end{align} 
Eq.~\eqref{eq:LTL_mean} can be seen as a classical linear Langevin equation for $\ex{\bx}\fil^{\rm LTL}$ with the diffusion term ${\cal K}^+[{\bf V}\fil^{\rm LTL}]\,\dd{\bf w}\fil^{\rm LTL}$. Further, Alice's measurement record, from Bob's perspective, is ${\bf y}(t)\,\dd t = {\bf C}\,\ex{\bx}\fil^{\rm LTL}(t)\,\dd t + \dd{\bf w}\fil^{\rm LTL}(t)$, which can be seen as a classical linear measurement of $\ex{\bx}\fil^{\rm LTL}$ with the measurement noise increment $\dd{\bf w}\fil^{\rm LTL}$. In this representation, the correlation between the process and measurement noise equals ${\cal K}^+[{\bf V}\fil^{\rm LTL}]$. With these considerations, $p(\ex{\bx}\fil^{\rm LTL}|\past{O}_t)$ is determined by the Kalman–Bucy equations. Writing Alice's measurement record as ${\bf y}(t)\,\dd t = {\bf C}\,{\halo{\bf x}}\fil (t)\,\dd t +\dd{\halo{\bf w}\fil} (t)$, with ${\halo{\bf x}}\fil$ denoting the filtered estimate of $\ex{\bx}\fil^{\rm LTL}$, we have

\begin{align}
&\dd\halo{\bf x}\fil = {\bf A}\,\halo{\bf x}\fil\,\dd t + \halo{\cal K}^{+}[\halo{\bf V}\fil]\,\dd \halo{\bf w}\fil\,,\\
&\frac{\dd}{\dd t}\halo{\bf V}\fil = {\bf A}\halo{\bf V}\fil + \halo{\bf V}\fil{\bf A}\tp + \nonumber\\
&\hspace{4em}{\cal K}^+[{\bf V}\fil^{\rm LTL}]\,{\cal K}^+[{\bf V}\fil^{\rm LTL}]\tp -\halo{\cal K}^{+}[\halo{\bf V}\fil]\,\halo{\cal K}^{+}[\halo{\bf V}\fil]\tp\,,\label{eq:halo_Var}
\end{align}
while $\halo{\cal K}^{+}[\halo{\bf V}\fil] = \halo{\bf V}\fil {\bf C}\tp + {\cal K}^+[{\bf V}\fil^{\rm LTL}]$. Given that Eqs.~\eqref{1stMomFiltering},~\eqref{2ndMomFiltering} and~\eqref{eq:LTL_mean}--\eqref{eq:halo_Var} hold simultaneously, it is easy to show that $\halo{{\bf V}}\fil = {\bf V}\fil - {\bf V}\fil^{\rm LTL}$ and $\halo{\bf x}\fil = \ex{\bx}\fil$. Therefore, $p(\ex{\bx}\fil^{\rm LTL}|\past{O}_t) = \mathcal{N}(\ex{\bx}\fil^{\rm LTL};\ex{\bx}\fil, {\bf V}\fil - {\bf V}\fil^{\rm LTL})$.

When the target is our ``true" state, an argument similar to that in the previous paragraph but extended with the aid of Ref.~\cite{laverick2019quantum} gives $p(\ex{\bx}\god|\past{O}_t) = \mathcal{N}(\ex{\bx}\god;\ex{{\bx}}\fil, {\bf V}\fil - {\bf V}\god)$.

Considering Eq.~\eqref{1stPrt}, the two previous paragraphs and the Gaussian product identity, we have
\beq
\begin{split}
p(\ex{\bx}\tar|\both{O}) &\propto p(\fut{O}_t|\rho_{\rm tar})\,p(\rho_{\rm tar}|\past{O}_t)\\
&= \mathcal{N}(\ex{\bx}\tar;\ex{\bx}\sm, \halo{{\bf V}}\sm)\,,
\end{split}\label{eq:sm_hal_prob}
\eeq
where 
\begin{align}
\ex{\bx}\sm &= \halo{{\bf V}}\sm\left[({\bf V}\fil - {\bf V}\tar)\inv \ex{\bx}\fil + ({\bf V}\rfil + {\bf V}\tar)\inv \ex{\bx}\rfil\right],\\
\halo{{\bf V}}\sm &= \left[({\bf V}\fil - {\bf V}\tar)\inv + ({\bf V}\rfil + {\bf V}\tar)\inv\right]\inv\,.
\end{align}
This leaves us with 
\begin{align}
W\sm({\bf x}) &\propto  \int\dd\ex{\hat{\bf x}}\tar\;\mathcal{N}(\ex{\bx}\tar;\ex{\bx}\sm, \halo{{\bf V}}\sm)\,\times\nonumber\\ & \hspace{9em}\mathcal{N}({\bf x};\ex{\bx}\tar,{\bf V}\tar)\nonumber\\
&\hspace{5.5em}= \mathcal{N}({\bf x};\ex{\bx}\sm, \halo{{\bf V}}\sm + {\bf V}\tar)\,.
\end{align}
Thus, the mean and the covariance of the smoothed quantum state are
\begin{align}
\ex{\bx}\sm &= ({\bf V}\sm - {\bf V}\tar)\left[({\bf V}\fil - {\bf V}\tar)\inv \ex{\bx}\fil \right.\nonumber\\
&\hspace{9em}\left.+ ({\bf V}\rfil + {\bf V}\tar)\inv \ex{\bx}\rfil\right],\\
{\bf V}\sm &= \left[({\bf V}\fil - {\bf V}\tar)\inv + ({\bf V}\rfil + {\bf V}\tar)\inv\right]\inv + {\bf V}\tar\,.
\end{align}
\subsection{Quantum state smoothing for non-LGQ systems}

While complete characterisation of ${\mathbb T}$ was possible in the LGQ case, it is very challenging in the non-LGQ case, barring a few special examples \cite{CGLW21}. Alice should therefore settle for an approximation via a Monte Carlo-type simulation. She begins by adopting an {\em ostensible} distribution for the possible unobserved measurement records $\past{U}_{t_0}$ (and $\past{U}_{t}$). According to this distribution, she simulates a large number of $\past{U}_{t_0}$ (and $\past{U}_{t}$). Using these simulated records, Alice constructs, as we will explain shortly, an ostensible set of possible realisations of the target state, denoted by ${\mathbb T}_{\rm ost}$. With ${\mathbb T}_{\rm ost}$ in place of ${\mathbb T}$, she approximates the right-hand side of Eq.~\eqref{eq:gen-sm}. While the choice of the ostensible distribution is arbitrary, it must not prevent the later renormalisation. Different ostensible distributions will lead to different convergence rates for this Monte Carlo simulation.

With the LTL filtered state as the target, Alice takes each of the simulated  $\past{U}_{t_0}$ records and computes Bob's filtered state until time $t_0$ via
\begin{align}\label{eq:IC_LTL}
  \tilde\rho^{\rm LTL}\fil(t+\dd t) = &\,\hat{M}_{y_u(t+\dd t)}\,\hat{U}(t+\dd t, t)\,{\cal E}(\tilde\rho^{\rm LTL}\fil(t))\,\times \nonumber
  \\&\hspace{6em} \hat{U}^\dagger(t+\dd t,t) \,\hat{M}_{y_u(t+\dd t)}^\dagger\,, 
\end{align}
where $\tilde{\bullet}$ indicates that this state is unnormalised; $\hat{U}(t+\dd t,t)$ describes the unitary evolution of the system and $\hat{M}_{y_u(t)}$ is the corresponding measurement operator associated with the outcome $y_u(t)$ --- this operator depends on the type of measurement and also on the ostensible distribution \cite{GueWis20}; ${\cal E}$ is a completely positive and trace-preserving map that governs any remaining evolution of the system, either due to measurement inefficiency or couplings to other baths. By repeating this procedure for all the simulated records, Alice finds a set of possible LTL filtered states at $t_0$. She then evolves each of these initial states using her observed measurement record $\past{O}_t$ via \
\begin{align}\label{eq:nonLGQev}
\tilde\rho^{\rm LTL}\fil(t+\dd t) = &\,\hat{N}_{y_o(t+\dd t)}\,\hat{U}(t+\dd t, t)\,{\cal E}'(\tilde\rho^{\rm LTL}\fil(t))\,\times \nonumber
  \\&\hspace{6em} \hat{U}^\dagger(t+\dd t,t)\,\hat{N}_{y_o(t+\dd t)}^\dagger \,,
\end{align}
where $\hat{N}_{y_o(t)}$ and ${\cal E}'$ denote her measurement operator and the ongoing decoherence, respectively. After normalising the final output, Alice obtains a possible LTL filtered state at time $t$. As such, she constructs the ${\mathbb T}_{\rm ost}$.

Alice should next compute the conditional probability $p(\rho^{\rm LTL}\fil;t|\both{O})$, which is proportional to $p(\rho^{\rm LTL}\fil;t|\past{O}_t)\times p(\fut{O}_t|\rho^{\rm LTL}\fil;t)$. The probability $p(\rho^{\rm LTL}\fil;t|\past{O}_t)$ equals $p_{\rm ost}(\past{U}_{t_0})\times\Tr[\tilde{\rho}^{\rm LTL}\fil(t)]$, where $p_{\rm ost}(\past{U}_{t_0})$ is given by the adopted ostensible distribution~\cite{guevara2015quantum,GueWis20}. As for the other probability, $p(\fut{O}_t|\rho^{\rm LTL}\fil;t) = \Tr[\hat{E}\rfil(t)\,\rho^{\rm LTL}\fil(t)]$, where the retrofiltered effect is given by \cite{gammelmark2013past}
\begin{align}
\hat{E}\rfil(t) = &\,{\cal E}'^{\dagger}\Big( \,\hat{U}^\dagger(t+\dd t,t)\,\hat{N}_{y_o(t+\dd t)}^\dagger\hat{E}\rfil(t + \dd t)\,\times\nonumber\\&\hspace{9em}\hat{N}_{y_o(t+\dd t)}\,\hat{U}(t+\dd t,t)\Big)\,,
\end{align}
with $\hat{E}\rfil(\mathcal{T}) \propto \hat{1}$. Having these in hand, Alice can compute the summation in Eq.~\eqref{eq:gen-sm} over ${\mathbb T}_{\rm ost}$. By normalising the output, she obtains $\rho\sm^{(N_{\rm ost})}$, where $N_{\rm ost}$ is the size of ${\mathbb T}_{\rm ost}$. It is an acceptable approximation of $\rho\sm$ once $N_{\rm ost}$ is sufficiently large to satisfy the convergence criterion. A reasonable convergence criterion is ${\rm Tr}[(\,\rho\sm^{(N_{\rm ost})}-\rho\sm^{(2N_{\rm ost})}\,)^2]\le\epsilon$, where $\epsilon$ sets the convergence tolerance.

As for targeting our ``true" state, the procedure is conceptually the same. One replaces $\rho^{\rm LTL}\fil$ everywhere with $\rho_{\rm T}$; the ostensible distribution and the simulation of unobserved records are extended to include $\past U_t$; and \erf{eq:nonLGQev} changes to 
\begin{align}
  \tilde\rho\god(t+\dd t) = &\,\hat{M}'_{y_u(t+\dd t)}\,\hat{N}_{y_o(t+\dd t)}\,\hat{U}(t+\dd t, t)\,{\cal E}''(\tilde\rho\god(t))\,\times \nonumber\\&\hspace{5em}\hat{U}^\dagger(t+\dd t,t)\,\hat{N}^\dagger_{y_o(t+\dd t)}\,\hat{M}_{y_u(t+\dd t)}'^{\dagger}  \,. 
\end{align}

\ifSubfilesClassLoaded{
    \bibliographystyle{apsrev4-2}
    \bibliography{sn-bibliography}
}{
}

\end{document}

\ifSubfilesClassLoaded{
    \title{Experimental methods}
    \date{\today}
    \maketitle
}{
}

\section{Experimental setup}
 

In the following sections, we explain the experimental setup employed to characterise, stabilise and monitor our optomechanical device.
\begin{figure*}
    \centering
    \includegraphics{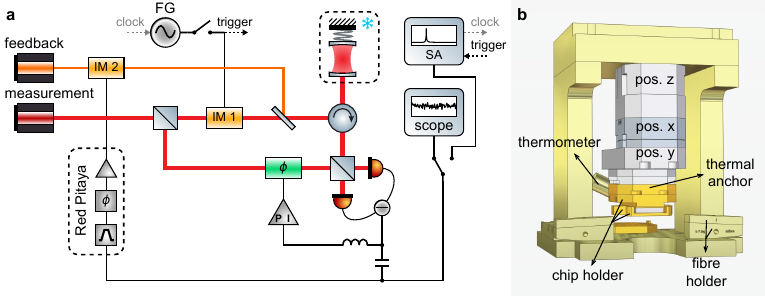}
    \caption{\textbf{Experimental setup. a,} Control and measurement circuitry. IM, intensity modulator (electro-optic). FG, function generator. SA, spectrum analyzer. Scope, oscilloscope. PI, proportional--integral controller. The $\phi$-labelled green box indicates a fibre stretcher. An on-resonance laser beam (red) interacts with the optomechanical device located inside a dilution refrigerator. The mechanical displacement is imprinted on the phase quadrature of the back-reflected beam, allowing the resonator’s motion to be monitored via homodyne detection. The measurement data is recorded by the oscilloscope. To stabilise the device, weak feedback is applied using a detuned laser (orange) and IM~2. IM~1 and the SA are used only for characterisation.
    \textbf{b,} Holding structure. The chip, with the optomechanical device fabricated on top, is placed upside-down in a holder and mounted onto a stack of three piezo-positioners (pos. x, y, z) to provide three-axis motorised control. The tapered fibre is fixed into place with adjustable angle to optimise alignment. A thermal anchor is sandwiched between chip holder and positioning stack, while a thermometer provides temperature readings. The full structure is placed on the bottom of the dilution refrigerator, over a series of windows that allow optical access for alignment.
    }
    \label{fig:methods_setup}
\end{figure*}

\subsection{Shot-noise-limited homodyne measurement}\label{SecHomoSetup}

After pumping the optomechanical cavity with an on-resonance, narrow-linewidth, tunable laser (Santec TSL-770), the back-reflected light is measured using a fibre-based homodyne interferometer (Extended Data FIG.~\ref{fig:methods_setup}\textbf{a}). Because the mechanical displacement is imprinted on the phase quadrature~\cite{bowen2015quantum}, we lock to this quadrature by feeding the low-frequency component of the balanced detector (Femto HBPR-100M-60K-IN-FC) output into an analogue servo controller (New Focus LB1005). The proportional--integral servo generates a control signal that, after amplification by a high-voltage amplifier (PiezoDrive PX200), drives a fibre stretcher (PiezoDrive FS) placed in the local oscillator arm of the interferometer. To be shot-noise-limited, the optical power at the end of this arm is about $10^4$ times higher than that at the end of the signal arm, and the laser phase noise is suppressed by matching the optical path lengths of the two arms to within a few centimetres.

\subsection{Vibration-reduced refrigeration}\label{sec:methods:refrigeration}

To minimise thermal fluctuations, the optomechanical device is placed in a dilution refrigerator (Bluefors BF-LD). While our experiments were performed with a sample stage temperature of $3.5$~K, operating in a dilution refrigerator will allow us to attain stage temperatures down to $20$~mK in the future. In normal operation, the refrigerator pulse tube, providing cooling power, induces substantial vibrations that modulate waveguide–fibre coupling, optical path length, and signal-arm polarisation. We therefore run the experiment with the pulse tube off, cooling instead via a helium ``battery" mounted on the $4$-K flange: a pre-charged vessel of liquid helium whose evaporation sustains low temperature for about three hours. As the helium battery cools only the $4$-K flange and other cryostat flanges gradually heat up during the run, the probe fibre (which is mounted on all flanges) can make a polarisation change which we regularly check and compensate.

\subsection{Stabilising feedback}
\label{sec:methods:setup_feedback}
For input optical powers $P_\text{in} > 60$~nW (injected into the long fibre going to the cryostat), 
we observe intermittent instability of the optomechanical device, characterised by high-amplitude self-oscillations (Extended Data FIG.~\ref{fig:methods_eta_check}\textbf{a}). Similar behaviour has been previously reported in these ultrahigh-Q optomechanical resonators~\cite{guo2021bringing} and attributed to coherent feedback from stray back-reflections. Alternatively, large fluctuations could result in nonlinear cavity dynamics that induce effective amplification~\cite{aspelmeyer2014cavity}. 

A minimal amount of measurement-based feedback, which reduces amplitude and broadens linewidth, stabilises the resonator. It is implemented optically by modulating a weak, slightly red-detuned laser (Santec TSL-550). 
We verify that
the feedback and measurement lasers beat only at frequencies larger than $1$ GHz, beyond the resonator’s response. This feedback laser is kept off during most characterisation.

The real-time electronic feedback signal is generated from the photocurrent by a digital signal processor (Red Pitaya STEMlab 125-14 LN). This processor implements two band-pass filters with tuneable gain and phase shift. One filter is centred at $\Omega/2\pi$ (with a $3$-dB bandwidth of $9.7$~kHz) and adjusted to stabilise the mechanical mode of interest (Methods~\ref{sec:methods:char_feedback}). The other filter is set to cool a higher-order mode that starts to lase once the fundamental mode is under control.
The generated signal is then further amplified and used to drive an electro-optic amplitude modulator (Oclaro AM1). 
As the dynamical range of the feedback setup is limited, it is only effective when engaged in the non-lasing state of the device. To ensure this, we monitor the photocurrent on the setup oscilloscope before turning on the feedback.

\subsection{Efficient optical coupling}
\label{sec:methods:optical_coupling}
To attain high $\eta$, we require an efficient interface between the cavity and the fibre-based interferometer. This is provided by a short on-chip silicon nitride waveguide that evanescently couples to the cavity with high efficiency ($\eta_\text{esc} = 82$\%, Methods~\ref{sec:methods:char_optical}) and ends on one side in a photonic crystal mirror and, on the other, in a tapered section. 
From the waveguide taper, light is transferred adiabatically into a conically tapered single-mode fibre that is brought into contact.

Maximising waveguide--fibre transfer efficiency $\eta_\text{wf}$ requires the fibre tip to have the appropriate tapering profile. Inspired by Ref.~\cite{tiecke2015efficient}, we fabricate this using a home-built heat-and-pull setup (based on a hydrogen torch) and an off-the-shelf solenoid (RS Components 177-0146, for fast pulling) \cite{khademi2024optomechanical}. Maximising $\eta_\text{wf}$ also requires a precise positional and angular alignment. This is challenging in a dilution refrigerator: Room-temperature pre-alignment is lost to thermal contraction on cool-down, while our in-situ cold alignment is hindered by limited visual access through five layers of heat shielding.
%
To overcome this challenge, the device alignment (and mounting) is first facilitated by a custom holding structure (Extended Data FIG.~\ref{fig:methods_setup}\textbf{b}). The tapered fibre is fixed with a grooved, angle-adjustable clamp, while the chip, containing several optomechanical devices, is mounted upside-down on a gold-plated, oxygen-free copper holder attached to a three-axis cryo-compatible piezo positioner stage (Attocube ANPx or ANPz). To enhance thermalisation, a thermal anchor (Attocube ATC100) is sandwiched between holder and stage, and linked to the base plate by a copper strip. A resistive thermometer (Lake Shore) monitors holder temperature.
The cold alignment is then performed using a long-working-distance zoom microscope with a separate white-light illumination source, both sitting outside the fridge. To suppress reflections from the five refrigerator windows, both microscope and illuminating beam are slightly tilted towards the chip.
With all these measures in place, we achieve $\effwf = 77\%$ (including losses in the 8-m cryostat fibre and splices). This is a particularly high efficiency for the coupling of a movable optical fibre to an on-chip waveguide inside a dilution refrigerator (compare with, e.g., 68\% reported in Ref.~\cite{Meenehan2015pulsed} for a lensed fibre and 65\% in Ref.~\cite{Pautrel2024efficient} for a conically tapered fibre).
To the best of our knowledge, the highest fibre--waveguide coupling efficiency achieved inside a dilution refrigerator is about 93\%, attained for visible light in a tapered-fibre-to-diamond-waveguide interface~\cite{Bhaskar2020experimental} using an imaging system inside the fridge.


\section{Characterisation}
\label{sec:methods:characterisation}
We here explain the methods used to characterise the system. 

\subsection{Optical cavity}
\label{sec:methods:char_optical}
To characterise the optical cavity, we inject a low power ($\Pin = 20$~nW) from the measurement laser (feedback laser off), step the laser frequency $\omgL/2\pi$ across the cavity response (with resonance frequency $\omgc/2\pi$ and decay rate $\kappa$) and measure the average intensity of the back-reflected light (Extended Data FIG.~\ref{fig:methods_optical_char}\textbf{a}). Considering the field reflection ratio
\begin{equation}
    \mathcal{R}=\dfrac{(1-2\eta_\text{esc})+2 i \Delta/\kappa}{1+2 i \Delta/\kappa}
\end{equation}
with $\Delta=\omgc-\omgL$, we fit $|\mathcal{R}|^2$ to the measured intensity response and find $\omgc/2\pi = 194.898$~THz, $\kappa/2\pi = 11.50$~GHz and $(1-2\eta_\text{esc})^2=0.40$.
\begin{figure}
    \centering
    \includegraphics{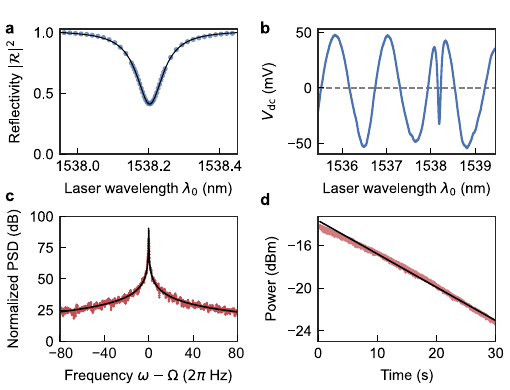}
    \caption{\textbf{Cold optical and mechanical characterisation. a, } Cavity reflectivity $|\mathcal{R}|^2$ is measured (circles) as the drive laser wavelength $\lambda_0 = 2\pi c / \omgL$ is stepped over the optical resonance. The theoretical function (line) is fit to the measurements with linewidth $\kappa/2\pi = 11.50$~GHz. \textbf{b,} Overcoupling test. The DC output $V_\text{dc} \sim \Re[\mathcal{R}\exp{(i\omgL L/c)}]$
    of the balanced detector is measured as the laser wavelength is quickly swept
    without locking the interferometer phase
    --- $L$ is the optical path difference between the two arms.
    As here $\exp{(i\omgL L/c)}\approx1$ at $\omgL=\omgc$, the negative sign of $V_\text{dc}$ at the resonance wavelength indicates that the cavity is overcoupled. 
    \textbf{c,} Mechanical spectrum around the resonance frequency $\Omega/2\pi=1.04$~MHz (with feedback off).
    \textbf{d,} Mechanical ring-down. An exponential decay is fitted with intrinsic energy decay rate $\Gamma = 2\pi\times(11.5\,$mHz).}
    \label{fig:methods_optical_char}
\end{figure}

To determine the sign in $\eta_\text{esc}=(1\pm\sqrt{0.40})/2$ (overcoupling or undercoupling of the cavity to the waveguide), we perform a test using the homodyne interferometer. While the laser wavelength is swept very quickly, across a $4$-nm interval, we record the DC output of the balanced detector (without locking on any homodyne phase). The interferometer then measures $\Re[\mathcal{R}\exp{(i\omgL L/c)}]$ where $L$ is the optical path difference between the two arms and $c$ is the speed of light in vacuum. While $L$ is almost constant during a single sweep, it has random changes each time that we do the measurement. We repeat this measurement until we find $\exp{(i\omgL L/c)} \approx 1$ at $\omgL=\omgc$, so the measurement output at this frequency approximates $\Re[\mathcal{R}]$ for $\Delta=0$. As shown in Extended Data FIG.~\ref{fig:methods_optical_char}\textbf{b}, this output is negative, indicating that the cavity is overcoupled and $\eta_\text{esc}=(1+\sqrt{0.40})/2$. In this figure, the 1.2-nm fringe spacing indicates that the optical path lengths of the two arms are different by about 2 mm.

\subsection{Mechanical resonator}
\label{sec:methods:char_mechanical}
The mechanical resonance frequency, $\Omega/2\pi = 1.04355$ MHz, is obtained by fitting the mechanical spectrum recorded with the setup spectrum analyser (Keysight EXA N9010A).
During cryostat cool-down, gas molecules condense onto the optomechanical device. To remove these, we induce a local boil-off: We repeatedly sweep the laser over the optical resonance at a slow rate ($1$~nm/s) and gradually increase the injected power. When $\Pin \approx 9~\mu$W, the cavity resonance swiftly shifts to a shorter wavelength by $0.5$~nm, indicating loss of dielectric material. We also find that $\Omega/2\pi$ has shifted up by about 600~Hz, indicating a commensurate loss of mass.

    

The mechanical decay rate $\Gamma$ is determined by a ring-down measurement, where the intensity of the measurement laser (with $\Pin = 2$~nW) is weakly modulated at frequency $\Omega/2\pi$ using an electro-optic modulator (Oclaro AM1) for $0.2$~s, after which the mechanical signal is recorded for $30$~s using the spectrum analyser's zero-span function (Extended Data FIG.~\ref{fig:methods_optical_char}\textbf{d}).
Fitting an exponential decay function yields $\Gamma/2\pi=11.5$~mHz. The modulation tone is produced by a signal generator (Siglent SDG) that also triggers acquisition and is referenced to the spectrum analyser's clock.



\subsection{Optomechanical coupling}
\label{sec:methods:char_coupling}
We evaluate the single-photon optomechanical coupling rate $g_0$ from the optical spring effect \cite{bowen2015quantum}, by sweeping the laser wavelength over the cavity response and recording the mechanical spectra.

At constant, very low injected power $\Pin = 2$~nW (Extended Data FIG.~\ref{fig:methods_spring_shift}a), we find that the effective resonance frequency $\Omega_\text{eff}$ closely follows the cavity photon number $\bar{n}_\text{cav}$ and the asymmetry around $\Delta=0$, typical of the spring effect, is almost absent. We consider this as a static photothermal effect corresponding to the expected rise in the device temperature by optical absorption (from 3.5 K towards 12.1 K, Methods~\ref{sec:methods:mode_temperature}) --- a similar shift has been reported by Refs.~\cite{usami2012optical,hauer2019dueling}. In Extended Data FIG.~\ref{fig:methods_spring_shift}\textbf{d}, $\Omega_\text{eff}$ is shown as a function of $\bar{n}_\text{cav}$ when $\Delta = 0$. While it sharply increases at small values of $\bar{n}_\text{cav}$, the thermal shift saturates after $\nc \approx 7$. Repeating this measurement with $\Pin = 60$~nW (Extended Data FIG.~\ref{fig:methods_spring_shift}\textbf{b}), we find a distinct $\Delta$-asymmetric optical spring on top of the $\Delta$-symmetric thermal shift.
\begin{figure*}
    \centering
    \includegraphics{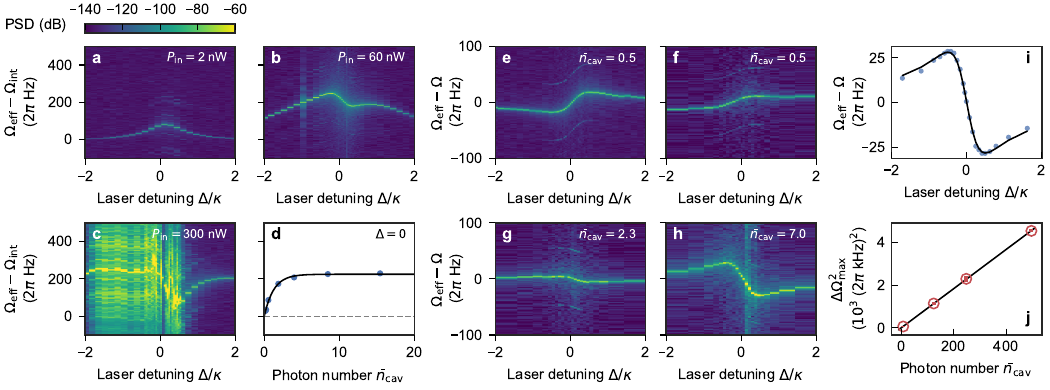}
    \caption{\textbf{Cold characterisation of single-photon optomechanical coupling rate. a-c,} Mechanical spectra of the fundamental mode ($\Omega_\text{int}$) as the laser detuning $\Delta$ is varied, with constant input power $P_\text{in}$ and feedback laser off. For low power $P_\text{in}=2$~nW (\textbf{a}), the typical detuning-asymmetric spring shift behaviour is absent. Instead, the mechanical frequency $\Omega_\text{eff}$ tracks the cavity photon number $\bar{n}_\text{cav}$, indicating an absorption-induced thermal shift. For intermediate power $P_\text{in}=60$~nW (\textbf{b}), the optical spring effect appears on top of the thermal shift. For higher power $P_\text{in}=300$~nW (\textbf{c}), high-amplitude self-oscillations occur at both positive and negative detunings. \textbf{d,} Mechanical resonance frequencies $\Omega_\text{eff}$ extracted from mechanical spectra measured on resonance ($\Delta = 0$) for a range of photon numbers $\bar{n}_\text{cav}$. A thermal shift is observed at lower cavity intensities that saturates once $\nc\gtrsim7$. An exponential saturation curve is fitted for reference. \textbf{e-h,} Mechanical spring shift spectra for constant cavity photon number $\bar{n}_\text{cav}$, measured around the thermally shifted mechanical frequency $\Omega$ with feedback laser off. At the lowest $\bar{n}_\text{cav} = 0.5$, an inverted spring shift stemming from photothermal back-action is observed. The shift is stronger before condensed gas boil-off (\textbf{e}) than after (\textbf{f}). At intermediate $\bar{n}_\text{cav} = 2.3$ (\textbf{g}), radiation pressure and photothermal back-action tend to cancel each other, leading to a small spring effect. At a higher $\bar{n}_\text{cav} = 7.0$ (\textbf{h}), where the thermal shift in panel \textbf{d} is saturated, a strong optical spring with the typical
    detuning dependence is observed. \textbf{i,} Radiation-pressure-induced spring is fitted to the mechanical frequencies extracted from panel \textbf{h}. \textbf{j,} Maximal squared spring shift $\Delta \Omega_\text{max}^2 = \max_\Delta\{|\Omega_\text{eff}^2 - \Omega^2|\}$ obtained from measuring at the three detunings $\Delta = 0, \pm \kappa/2$ with high values of $\nc$. Error bars represent the difference between the measured values of $\Delta \Omega_\text{max}^2$ with $\Delta = +\kappa/2$ and $\Delta = -\kappa/2$, while the data points are found by averaging those values. For radiation-pressure-induced spring, $\Delta \Omega_\text{max}^2=4\Omega g_0^2/\kappa\times\nc$. The reliable linear fit (with a zero y-intercept) gives the single-photon coupling rate as $g_0/2\pi=159$~kHz.}
    \label{fig:methods_spring_shift}
\end{figure*}

To isolate the optical spring effect, we do the measurement by keeping $\nc$ (and thus the thermal shift) constant across the sweep interval --- $\Pin$ is changed at each value of $\Delta$. For low $\nc = 0.5$ (Extended Data FIGs.~\ref{fig:methods_spring_shift}\textbf{e} and~\textbf{f}), we find an optical spring effect with detuning dependence opposite to a radiation-pressure-induced spring. That is, we find higher $\Omega_\text{eff}$ at $\Delta > 0$ than at $\Delta < 0$. We attribute this to the dynamical back-action of a dominating deformation force opposing the radiation pressure. This force is considered to be caused by a temperature gradient profile across the solid structure made by local absorption heating and the poor thermal conductivity of silicon nitride at low temperatures, similar to what has been reported by Ref.~\cite{hauer2019dueling}. Interestingly, this effect is stronger before the condensed gas molecules are boiled off (Extended Data FIG.~\ref{fig:methods_spring_shift}\textbf{e}), possibly indicating higher local absorption.

For a moderate $\nc = 2.3$ (Extended Data FIG.~\ref{fig:methods_spring_shift}\textbf{g}), we see the net spring effect switches sign. This is to be expected: As the device temperature rises, there should be a sharp increase in the thermal conductivity --- see, e.g., Ref.~\cite{ftouni2015thermal} --- suppressing the photothermal force and allowing the radiation pressure to dominate, similar to what has been reported by Ref.~\cite{hauer2019dueling}. With further increase of $\nc$ to 7, near the saturation limit in Extended Data FIG.~\ref{fig:methods_spring_shift}\textbf{d}, we find a strong optical spring with the typical detuning dependence (Extended Data FIG.~\ref{fig:methods_spring_shift}\textbf{h}). Fitting the radiation-pressure-induced spring~\cite{bowen2015quantum} on this data (Extended Data FIG.~\ref{fig:methods_spring_shift}\textbf{i}) gives a value of $155$~kHz for $g_0/2\pi$.

The photothermal force is then expected to be negligible for $\nc\gtrsim7$, leaving the spring effect solely driven by radiation pressure (with maximum values for $|\Omega_\text{eff}^2-\Omega^2|$ scaling linearly with $\nc$, as $g_0$ is independent of $\nc$). To verify this expectation, we need to do the measurement for higher $\nc$ where we encounter intermittent instability of the system (Extended Data FIG.~\ref{fig:methods_eta_check}\textbf{a}) and have to adapt our technique. We reduce the laser sweep to the three detunings $\Delta = 0, \pm\kappa/2$ where the spring shift (with constant $\nc$) is zero or maximal. At each detuning, we monitor the system on the oscilloscope and once it is in the non-lasing state, we record a 1-s time trace. Calculating the spectrum of the recorded trace, we extract the value of $\Omega_\text{eff}$. Performing this adapted measurement with multiple high values of $\nc$ strongly confirms the linear scaling (Extended Data FIG.~\ref{fig:methods_spring_shift}\textbf{j}) and characterises the coherent single-photon coupling rate as $g_0/2\pi=159$~kHz.

\subsection{Feedback loop}
\label{sec:methods:char_feedback}
The performance of the feedback loop is characterised and optimised by sweeping the electronic gain setting $\grp$ on the Red Pitaya and taking mechanical spectra (Extended Data FIG.~\ref{fig:methods_feedback}\textbf{a}), while the measurement and feedback laser powers are set to the values used for the final collection of monitoring data ($\Pin=154.0$~nW and $\Pfb = 0.5$~nW, respectively). By fitting on these mechanical spectra, we extract the feedback-broadened linewidth $\Gammafb$, the resonance frequency and the peak area as functions of $\grp$.
We verify that $\Gammafb$ scales linearly with $\grp$ (Extended Data FIG.~\ref{fig:methods_feedback}\textbf{b}, left), while the resonance frequency does not change (same figure, middle). This optimised performance is achieved by tuning the electronic phase shift setting on the Red Pitaya (such that $\arg[\Hfb(\Omega)] \approx \pi/2$, where $\Hfb(\omega) \propto \grp$ is the feedback loop transfer function). Importantly, we also verify that the peak area scales with $1/\Gammafb$ (Extended Data FIG.~\ref{fig:methods_feedback}\textbf{b}, right), indicating that we are not feeding excess technical noise onto the resonator.
\begin{figure}
    \centering
    \includegraphics{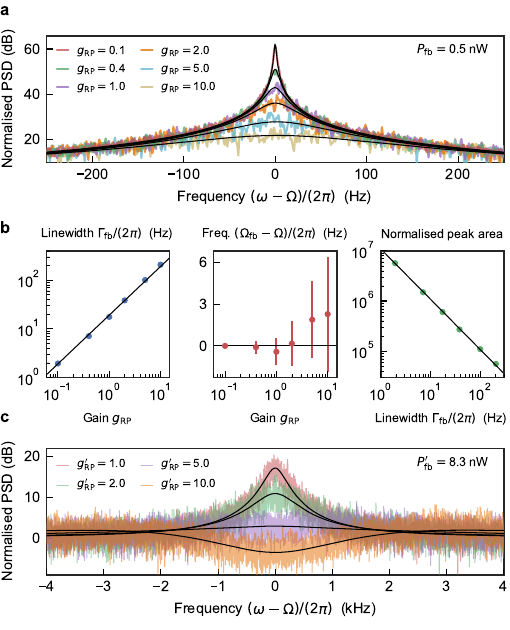}
    \caption{\textbf{Feedback loop performance. a,} Mechanical spectra, normalised to the shot noise level, for increasing feedback voltage gain $g_\text{RP}$ at low feedback laser power $P_\text{fb} = 0.5$~nW. A Lorentzian function
    is fitted (lines) with increasing linewidth and decreasing height. \textbf{b,} Estimated linewidths $\Gamma_\text{fb}$, frequencies $\Omega_\text{fb}$ and peak areas from panel \textbf{a}. A linear fit is shown for $\Gamma_\text{fb}$ versus $g_\text{RP}$, while a reciprocal fit is plotted for area versus $\Gamma_\text{fb}$. The error bars in the middle plot indicate the fit errors. \textbf{c,} Mechanical spectra for higher feedback laser power $\Pfb' = 8.3$~nW. With a strong feedback gain, noise squashing is observed.}
    \label{fig:methods_feedback}
\end{figure}

We note that with $\nc=41.4$ measurement photons (provided by the $\Pin=154$~nW input power), the average number of detuned feedback photons in the cavity is only 0.13, making the consequent dynamical back-actions and excess optical noise negligible.

To evaluate how far we are from the noise squashing regime, where there is a significant correlation between the measurement noise and the resonator process noises, we repeat this experiment with more feedback laser power $\Pfb' = 8.3$~nW. As shown in Extended Data FIG.~\ref{fig:methods_feedback}\textbf{c}, we observe noise squashing once $\grp' > 7$, equivalent to $\grp \approx \grp' \Pfb' / \Pfb > 116$ (at the main feedback power).

For the final collection of monitoring data, we set $\grp = 4$, well below the squashing limit, keeping the correlation (between the process and measurement noise) negligible. At this gain, we have $\Gammafb/2\pi = 85$~Hz.

\subsection{Resonator temperature}
\label{sec:methods:mode_temperature}
An accurate estimate of the mechanical resonator’s effective temperature is crucial for our analysis. However, because of its long intrinsic decay time, $1/\Gamma \approx 14$~s, sampling the full thermal amplitude distribution takes a long time. We assess that achieving a 10\% relative accuracy in the estimated mean energy would require monitoring the resonator in the non-lasing state for about an hour, which is impractical as the system is unstable.

We therefore estimate the average energy with the feedback on, as the system is stable and $1/\Gammafb$ is much shorter than $1/\Gamma$. In particular, we extrapolate the reciprocal fit shown in Extended Data FIG.~\ref{fig:methods_feedback}\textbf{b}, right, to find the total phonon number $n_\text{tot} = \mathcal{C} + \nth + 1/2$ at the intrinsic linewidth ($\Gammafb = \Gamma$, where $\grp = 0$). After subtracting $\mathcal{C} + 1/2$, we are left with a value for $\nth$ that corresponds to the temperature $T = 12.1$~K. This is significantly higher than the fridge temperature (3.5~K), consistent with our expectation in Methods~\ref{sec:methods:char_coupling}.

Considering the characterised values of $T$ and $g_0$, the ratio $\mathcal{C}/n_\text{th}$ with $\bar{n}_\text{cav}=41.4$ is large enough to produce a measurable level of ponderomotive squeezing of light. We verify this by changing the lock point of our homodyne setup from the phase quadrature $(\phi=90\degree)$ to $\phi=130\degree$ and recording the mechanical spectrum. We note that, with the feedback power unchanged ($\Pfb=0.5$~nW) and the Red Pitaya gain set at $g_\text{RP}^\phi=1$, the feedback-induced correlation of the measurement noise and the process noises is negligible --- the effective gain $\sin(\phi)\times g_\text{RP}^\phi\approx0.77$ is more than two orders of magnitude lower than the squashing limit. The recorded spectrum (Extended Data FIG.~\ref{fig:methods_pond_sqz}) shows ponderomotive squeezing and the corresponding asymmetry around $\Omega$, confirming the considerable ratio of back-action heating to thermal heating.
\begin{figure}[t!]
    \centering
    \includegraphics{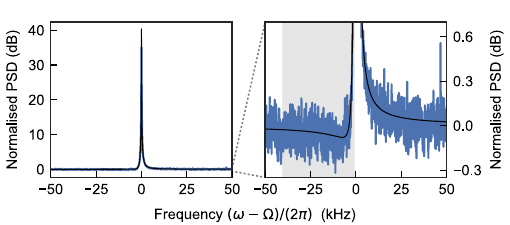}
    \caption{\textbf{Ponderomotive squeezing.} Photocurrent spectral density (in dB, collected with a 100-Hz resolution bandwidth, normalised to the shot noise level) measured with the homodyne interferometer tuned to an intermediate optical quadrature ($\phi = 130\degree$). Optical correlations mediated by the mechanical resonator give rise to ponderomotive squeezing (negative values) and render the spectrum highly asymmetric around $\Omega$. Fitting Eq.~\eqref{eq:methods:Syy_ponderomotive} to the grey-shaded region yields estimates of the single-photon optomechanical coupling $g_0$ and the resonator temperature $T$.}
    \label{fig:methods_pond_sqz}
\end{figure}

We also use this demonstration of ponderomotive squeezing to double-check the values of $g_0$ and $T$. This is possible because the recorded spectrum (once normalised to the shot noise level) approximates
\begin{align}
\begin{aligned}
    S(\omega) = &1 + 16\eta\,|\Gamma\chi(\omega)|^2  \sin^2(\phi) \times \mathcal{C}\,n_\text{tot} \\
    &+ 4 \eta\,\Re[\Gamma\chi(\omega)] \sin(2\phi)\times  \mathcal{C}
\end{aligned} \label{eq:methods:Syy_ponderomotive}
\end{align}
(with $\chi(\omega)=\Omega/(\Omega^2-\omega^2-i\omega\Gamma)$ and all the feedback replacements $\Gamma\rightarrow\Gamma_\text{fb}$ and $\nth\rightarrow\nth\times\Gamma/\Gamma_\text{fb}$ in place), where the squeezing term (the last one on the right hand side) is independent of $\nth$. By fitting Eq.~\eqref{eq:methods:Syy_ponderomotive} (Extended Data FIG.~\ref{fig:methods_pond_sqz}), we get 160.59~kHz for $g_0/2\pi$ and 12.19~K for $T$, in good agreement with previous results.



\subsection{Optical detection efficiency}
\label{sec:methods:total_losses}
In Extended Data TABLE~\ref{tab:methods:efficiency}, we quantify the various inefficiencies encountered by the probe field on its way from the optical cavity to the final phase detection.
\begin{table}[]
    \centering
    \begin{tabular}{l|c}
        efficiency element & value \\
        \hline
        cavity--waveguide coupling ($\eta_\text{esc}$) & $81.8\%$ \\
        waveguide--fibre coupling ($\effwf$, in the dilution fridge) & $76.7\%$ \\
        interferometer components' transmission & $83.2\%$ \\
        interference visibility & $99\%$ \\
        photodetector quantum efficiency & $76\%$ \\
        electronic noise suppression & $96.7\%$ \\
        \hline
        total efficiency ($\eta$) & $38\%$
    \end{tabular}
    \caption{\textbf{Contributions to optical detection inefficiency.}}
    \label{tab:methods:efficiency}
\end{table}

\subsection{Estimating optical detection efficiency from non-linear transduction}
\label{sec:methods:eta_check}
To obtain an independent estimate of the total efficiency $\eta$, we developed a new method based on the nonlinear transduction of high-amplitude mechanical motion by the optical cavity~\cite{leijssen2017nonlinear}. In our experiment, we can readily access this regime by turning the feedback off and waiting for instability to occur. Extended Data FIG.~\ref{fig:methods_eta_check}\textbf{c} shows a time trace of a photocurrent obtained under such conditions.

In this method, we use the extreme values of the phase quadrature component of the reflectivity $\Im[\mathcal{R}] = \effc \kappa \Delta/(\Delta^2 + \kappa^2/4)$ (Extended Data FIG.~\ref{fig:methods_eta_check}\textbf{b}) as a reference level. These extrema are located at $\Delta = \pm \kappa/2$, where $\Im[\mathcal{R}] = \pm \effc$, and occur in Extended Data FIG.~\ref{fig:methods_eta_check}\textbf{c} as the mechanically-shifted instantaneous detuning $\Delta = g_0 Q$ explores a large fraction of the cavity response. Here, $Q$ is the mechanical displacement normalized to its zero point fluctuations.
\begin{figure}
    \centering
    \includegraphics{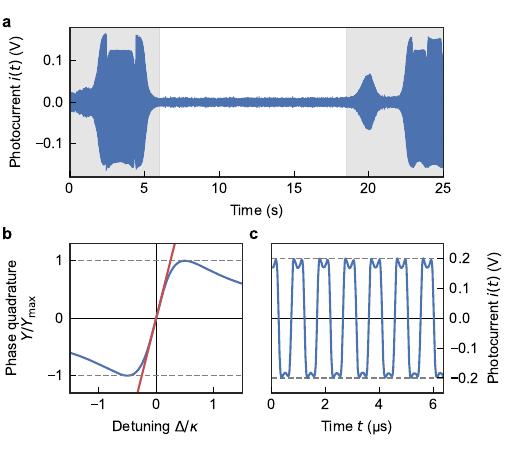}
    \caption{\textbf{Non-linear transduction of high-amplitude oscillations. a,} Intervals of intermittent instability (grey) in the mechanical amplitude are observed for higher input powers, when the feedback laser is off. These intervals get longer and more frequent with increasing laser power. \textbf{b,} Phase quadrature $Y$ of the mean intra-cavity field as a function of detuning $\Delta$ (blue). Extreme values $\pm Y_\text{max}$ are indicated (dashed lines) and occur at $\Delta = \pm \kappa/2$. For small detunings $|\Delta| \ll \kappa$, the linear approximation (red) is typically used to relate $Y$ to displacement $Q$ via $\Delta = g_0 Q$. \textbf{c,} High-amplitude self-oscillations of the resonator explore detunings beyond $|\Delta| > \kappa/2$. The extreme values in the photocurrent (dashed lines) then correspond to $\pm Y_\text{max}$ and can be used for calibration.}
    \label{fig:methods_eta_check}
\end{figure}

We proceed by switching to a semi-classical picture and writing the detected phase quadrature as
\begin{align}
    \Ydet = \Yv + \sqrt{\eta \kappa} Y \label{eq:methods:Ydet_1}\,,
\end{align}
the sum of a quantum vacuum field $\Yv$ and a classical field, where
\begin{align}
    Y = 2 \sqrt{\effc \kappa}\,\Im[\chi_a]\times\alpha_\text{in} \label{eq:methods:Ycav_1}
\end{align} 
is the mean phase quadrature of the driven intra-cavity field. Here, we consider the cavity susceptibility $\chi_a = (\kappa/2 - i \Delta)^{-1}$ and the coherent amplitude~$\alpha_\text{in}$ of the field coupling in (from the waveguide, with efficiency~$\effc$). We take $\alpha_\text{in}$ real.


Next, we express $\alpha_\text{in} = \sqrt{\nc\kappa/(4\effc)}$ in terms of the cavity photon number $\nc$ attained on resonance ($\Delta = 0$), and get
\begin{align}
    Y = \sqrt{\nc}\times\Im[\kappa \chi_a] = \sqrt{\nc}\times\frac{\kappa\Delta}{\Delta^2 + \kappa^2/4}\,,
\end{align}
where we note that $\Im[\kappa \chi_a] = \Im[\mathcal{R}]/\effc$.
As the instantaneous $\Delta$ is swept across the cavity resonance by mechanical motion, $Y$ attains a maximum value of $Y_\text{max} = \sqrt{\nc}$ at $\Delta = \kappa/2$, and $-Y_\text{max}$ at $\Delta = - \kappa/2$.

The homodyne detector produces a photocurrent $\rmi = s \Ydet$ proportional to the detected phase quadrature. The maximum photocurrent $\rmi_\text{max}$, as shown in Extended Data FIG.~\ref{fig:methods_eta_check}\textbf{c}, is obtained when $Y = Y_\text{max}$ and given by
\begin{align}
    \rmi_\text{max} = s \sqrt{\eta\kappa}\,Y_\text{max} = s \sqrt{\eta\kappa\,\nc}\,.
\end{align}
If $\eta$, $\kappa$ and $\nc$ are all known, we can use $Y_\text{max}$ as a reference to find the detector sensitivity $s = \rmi_\text{max} / \sqrt{\eta \kappa \nc}$.

If, however, $\eta$ is not known, we can determine it by comparing $\rmi_\text{max}$ to the shot-noise spectral density $S_{\rmi\rmi}^\text{SN}(\omega)$ of the photocurrent measured when the cavity is not driven, i.e., when $Y=0$ and $\Ydet = \Yv$. We then find
\begin{align}
    S_{\rmi\rmi}^\text{SN}(\omega) = s^2 S_{\Yv\Yv}(\omega) = \frac{\rmi_\text{max}^2}{\eta \kappa\,\nc} \label{eq:methods:eta_from_SN}
\end{align}
where $S_{\Yv\Yv}(\omega) = 1$ for the optical vacuum field $\Yv$.
To work out $\eta$ from Eq.~\eqref{eq:methods:eta_from_SN}, $\kappa$ needs to be entered as an angular frequency and $S_{\rmi\rmi}^\text{SN}(\omega)$ calculated as a double-sided spectrum. Note that to determine $\nc$, we still use $\effc$ and $\effwf$, which can be estimated independently and accurately.

In our experiment, we find $\rmi_\text{max} = 0.200$~V for $\nc = 41.4$, and a noise spectral density $S_{\rmi\rmi}^\text{SN}(\omega) = 3.53 \times 10^{-14}\  \mathrm{V}^2/\mathrm{Hz}$ around the mechanical resonance $\Omega$. Hence, from Eq.~\eqref{eq:methods:eta_from_SN} we find a value of 37.8\% for $\eta$, in good agreement with the previous result.

\section{Data acquisition and analysis} \label{MethodsSec:Analysis}
Time traces of the photocurrent are recorded on a low-noise oscilloscope (Tektronix MSO64). The oscilloscope produces digital samples at a rate of 5 MHz with $12$-bit vertical resolution (acquisition mode Sample). The oscilloscope's maximum record length is $6.25\times10^7$~samples. To prevent aliasing, an analogue filter is placed before the oscilloscope input.

\ifSubfilesClassLoaded{
    \bibliographystyle{apsrev4-2}
    \bibliography{sn-bibliography}
}{
}

\end{document}

\ifSubfilesClassLoaded{
    \title{Modeling and data analysis}
    \date{\today}
    \maketitle
}{
}

\subsection{Preparing the measurement records}\label{sec:RecordPrep}

The recorded data is normalised to the shot noise level and then demodulated at frequency $\Omega/2\pi$. The demodulation low-pass filter is a causal Butterworth filter, with 56.5 kHz 3-dB bandwidth and an impulse response that effectively vanishes after about $400\,\mu$s.

The filter bandwidth is chosen with two considerations. First, it is significantly smaller than $\Omega/2\pi$ and significantly larger than $\gamma_\text{th}/2\pi\approx2.79$ kHz and $\gamma_\text{opt}/2\pi\approx363$ Hz, as needed by the simplified measurement model~\cite{doherty2012quantum}. Second, it filters out all the coloured noises (including almost all of the ``measurement signal" of the other mechanical modes and the phase noise of the probe laser, Supplementary~\ref{supp:PSDs}).

Considering the filter impulse response, the first ten $400$-$\mu$s segments of the output currents are discarded. As such, we make sure that the remaining parts of the two currents are properly low-pass-filtered. These parts are then divided into shorter records. To maximise the number of records, the record length is minimised while ensuring that the steady state of filtering and smoothing is captured or closely approached.

\subsection{Analysing the records}\label{SME-ME-&Analysis}
Given that $n_\text{th}\gg1$, $\kappa\gg \Omega,g$ and $\Omega\gg\gamma_\text{th},\gamma_\text{opt}$, the effective interaction-frame stochastic master equation of the system is~\cite{doherty2012quantum,khademi2024optomechanical}
\begin{equation}
   \begin{array}{rll}
         \dd{\rho}_\text{F}(t) & = & \Gamma(n_\text{th}+1)\;\dd t\,\mathcal{D}\!\left[\hat{b}\right]\!\rho_\text{F}(t)\;+ \\&&  \Gamma n_\text{th}\;\dd t\,\mathcal{D}\!\left[\hat{b}^\dagger\right]\!\rho_\text{F}(t)\;+ \\&&
         \dfrac{1}{2}\,\Gamma \mathcal{C}\;\dd t\small\displaystyle\sum_{j=1}^{2}\mathcal{D}\!\left[\hat{X}_j\right]\!\rho_\text{F}(t)\;+ \\&& \sqrt{\dfrac{1}{2}\,\eta\Gamma \mathcal{C}\,}\;\small\displaystyle\sum_{j=1}^{2}\dd W_{\text{F},j}(t)\,\mathcal{H}\!\left[\hat{X}_j\right]\!\rho_\text{F}(t)
    \end{array}
\label{UWUQSME_RF_final} 
\end{equation}
where $\mathcal{D}[\bullet]$ and $\mathcal{H}[\bullet]$ are, respectively, the standard dissipative and diffusive measurement superoperator; $\hat{b}$ and $\hat{b}^\dagger$ are, respectively, the annihilation and creation operator of the resonator; $\hat{X}_1$ and $\hat{X}_2$ are (time-independent) interaction-frame quadrature operators and $\dd W_{\text{F},1}$ and $\dd W_{\text{F},2}$ are two independent Wiener increments, each given by the corresponding measurement record ($I_1(t)$ or $I_2(t)$):
\begin{equation}
         \dd W_{\text{F},j}(t) =\text{I}_{j}(t)\,\dd t - \sqrt{2\eta\Gamma \mathcal{C}}\;\text{Tr}\Big[\rho\fil(t)\,\hat{X}_j\Big]\,\dd t\,.
    \label{DemodulatedCurrents}
\end{equation}

Using Eq.~\eqref{UWUQSME_RF_final}, we can set the matrices employed by Eqs.~\eqref{1stMomFiltering}, \eqref{2ndMomFiltering}, \eqref{1stmomentNREO} and \eqref{2ndmomentNREO} (with $\hat{\mathbf{x}}=(\hat{X}_1,\hat{X}_2)^\top$)~\cite{laverick2021linear}:
\begin{equation}
     \mathbf{A}=-\dfrac{\Gamma}{2}\,\mathbf{I},\;\mathbf{D}=
    2\Gamma\,n_\text{tot}\,\mathbf{I},\;\mathbf{C}=\sqrt{2\eta\Gamma \mathcal{C}}\,\mathbf{I},\;\bm{\Gamma}=\mathbf{0}\,.
\label{IPmatrices}
\end{equation}
The initial conditions for Eqs.~\eqref{1stMomFiltering} and \eqref{2ndMomFiltering} are now given by the unconditional state of the resonator. At dynamical equilibrium, as in the case of our experiment, the unconditional state is Gaussian with the moments $\langle \hat{\mathbf{x}} \rangle _\text{uncon} = \mathbf{0}$ and $\mathbf{V}_\text{uncon}=2n_\text{tot}\,\mathbf{I}$ --- The Markovian master equation of the system is recovered by removing the last term on the right hand side of Eq.~\eqref{UWUQSME_RF_final}. Solving for the filtered covariance matrix then gives $\mathbf{V}\fil(t)=v\fil(t)\,\mathbf{I}$ with
\begin{equation}
    \dot{v}\fil(t)\,=-\Gamma\,v\fil(t)+2\Gamma\,n_\text{tot}-2\Gamma\,(\eta\,\mathcal{C})\,v^2\fil(t)\,,
    \label{NLODE}
\end{equation}
yielding
\begin{widetext}
\begin{equation}
    v\fil(t)=\bigg[\,\Big(\,\dfrac{1}{2n_\text{tot}-v_F^{\rm ss}}+\dfrac{2\eta \mathcal{C}}{\sqrt{1+16\eta \mathcal{C} n_\text{tot}}}\,\Big)\,e^{\Gamma t \sqrt{1+16\eta \mathcal{C} n_\text{tot}} }-\dfrac{2\eta \mathcal{C}}{\sqrt{1+16\eta \mathcal{C} n_\text{tot}} }\,\bigg]^{-1}+v_F^{\rm ss}\,,
    \label{VFT}
\end{equation}
\end{widetext}
where
\begin{equation}
    v\fil^{\rm ss}=\lim_{t\to\infty}v\fil(t)=\dfrac{\sqrt{1+16\eta \mathcal{C} n_\text{tot}}-1}{4\eta \mathcal{C}}\,.
     \label{VFSS}
\end{equation}
Having the covariance matrix, we solve
\begin{equation}
\begin{array}{rll}
    \dd\langle \hat{X}_j \rangle\fil(t) & = & -\dfrac{\Gamma}{2}\,\langle\hat{X}_\text{j} \rangle\fil(t)\,\dd t\,+ \\ && \\&&\sqrt{2\eta\Gamma \mathcal{C}}\,v\fil(t)\,\dd W_{{\rm F},j}(t)\,,
\end{array}
\label{equ:rFT}
\end{equation}
to find the stochastic first moments of the filtered state, considering Eq.~\eqref{DemodulatedCurrents}.

Solving Eq.~\eqref{2ndmomentNREO}, with an uninformative final condition, gives the covariance matrix of the effect operator as $\mathbf{V}_\text{R}(t)=v_\text{R}(t)\,\mathbf{I}$ with
\begin{equation}
\begin{array}{l}
    v\rfil(t)=v\rfil^{\rm ss}\,+\\\\\bigg[\dfrac{2\eta \mathcal{C}}{\sqrt{1+16\eta \mathcal{C} n_\text{tot}}}\,\big(e^{-\Gamma\,(t-{\cal T})\,\sqrt{1+16\eta \mathcal{C} n_\text{tot}} }-1\big)\bigg]^{-1}
\end{array}
\label{VRT}
\end{equation}
where
\begin{equation}
    v\rfil^{\rm ss}=\lim_{{\cal T}\to\infty}v\rfil(t)=\dfrac{\sqrt{1+16\eta \mathcal{C} n_\text{tot}}+1}{4\eta \mathcal{C}}\,.
    \label{VRSS}
\end{equation}
The first moments of the effect operator are then given by
\begin{equation}
    \langle \hat{X}_j \rangle\rfil(t) = v\rfil(t)\,Z_{{\rm R},j}(t)\,,\;\;\;j=1,2
\label{equ:rRT}
\end{equation}
where $Z_{{\rm R},j}(t)$ is found by solving
\begin{equation}
\begin{array}{rll}
    -\dd Z_{{\rm R},j}(t) & = & -\bigg(\,\dfrac{\Gamma}{2}+\dfrac{2\Gamma n_\text{tot}}{v\rfil(t)}\,\bigg)\,Z_{{\rm R},j}(t)\,\dd t\,+\\&&\\&&\sqrt{2\eta\Gamma \mathcal{C}}\;\text{I}_{j}(t)\,\dd t
\end{array}
\end{equation}
which goes backwards in time from the final condition $Z_{{\rm R},j}({\cal T})=0$~\cite{laverick2021linear}.

Given that the measurement records are sampled in time, the aforementioned techniques for finding the conditional mean values are accordingly adapted following the procedures explained in Ref.~\cite{stengel1994optimal}. The smoothing combination of the filtered and retro-filtered mean values is also accordingly adapted considering Ref.~\cite{Grewal2014Kalman}.


\subsection{Modelling the stabilising feedback}\label{SecFeedbackModelling}

To introduce the stabilising feedback into our analysis, we consider three points. First, the feedback can be treated as effectively Markovian~\cite{wiseman2009quantum}: With the feedback bandwidth $\mathcal{B}/2\pi=9.7$ kHz, we have $1/\mathcal{B}\ll 1/\Gamma$, so the timescale of the feedback-induced autocorrelation is much shorter than that of the dynamical autocorrelation --- the additional loop delay is less than $1~\mu$s. Second, the peak frequency of the mechanical susceptibility is unaffected by the phase-tuned feedback (Extended Data FIG.~\ref{fig:methods_feedback}\textbf{b}). Third, the effective rotating-frame dynamics remain symmetric~\cite{doherty2012quantum}, since (i) $\Omega/\Gamma_\text{fb} > 10^4$, (ii) the feedback bandwidth is centred around $\Omega$, and (iii) $\mathcal{B}\ll \Omega$.

With these considerations, the feedback can be incorporated into the (retro)filtering equations either formally or intuitively. In the formal approach, the $\mathbf{A}$, $\mathbf{D}$, and $\bm{\Gamma}$ matrices are updated by adding feedback terms, as in Ref.~\cite{wiseman2009quantum}. In the intuitive approach, one simply makes the substitutions $\Gamma\rightarrow\Gamma_\text{fb}$ and $n_\text{th}\rightarrow n_\text{th}\times\Gamma/\Gamma_\text{fb}$ in all equations. This intuitive approach performs similarly on our data because the feedback-induced correlation between the measurement noise and the process noises is negligible (Methods~\ref{sec:methods:char_feedback}).

\subsection{Noise injection and analysis}\label{sec:NI_analysis}

To perform the noise-injection experiment, we add white noise to the shot-noise-normalised data and then renormalise it to the new noise floor. The added noise is weighted such that the detection efficiency decreases from 0.38 to 0.10; the already negligible correlation between the measurement noise and the process noises also decreases by a factor of about two. The resulting noisier current is demodulated, and the outputs, excluding the initial segments, are divided into 1-ms records. Longer records are required in this experiment because the reduced detection efficiency prolongs the transients.

Each new measurement record is then processed by Eqs.~\eqref{equ:rFT} and \eqref{equ:rRT} to yield $\langle \hat{X}_j \rangle\fil^{\eta\downarrow}$ and $\langle \hat{X}_j \rangle\rfil^{\eta\downarrow}$, while $v\fil^{\eta\downarrow}$ and $v_\text{R}^{\eta\downarrow}$ are evaluated by Eqs.~\eqref{VFT} and \eqref{VRT} with the reduced detection efficiency. With $\rho\fil^{\eta\downarrow}$ in hand, the smoothed estimate of $\rho\fil^\text{LTL}$ is given by Eq.~\eqref{QSSforLTLfil} with $\langle \hat{\mathbf{x}} \rangle\fil=(\langle \hat{X}_1 \rangle\fil^{\eta\downarrow},\langle \hat{X}_2 \rangle\fil^{\eta\downarrow})^\top$, $\langle \hat{\mathbf{x}} \rangle\rfil=(\langle \hat{X}_1 \rangle\rfil^{\eta\downarrow},\langle \hat{X}_2 \rangle\rfil^{\eta\downarrow})^\top$, $\mathbf{V}\fil=v\fil^{\eta\downarrow}\mathbf{I}$, $\mathbf{V}\rfil=v\rfil^{\eta\downarrow}\mathbf{I}$ and $\mathbf{V}\fil^\text{ss}=v\fil^\text{ss}\mathbf{I}$. We note that, for the purpose of this analysis, $\mathbf{V}\fil^\text{ss}$ should not be replaced by $ v\fil^{\eta\downarrow,\text{ss}}\mathbf{I}$. When all $\mathbf{V}\fil^\text{ss}$ terms are instead replaced by $\mathbf{0}$, the new outputs of classical smoothing equations are obtained.

\section{The desiderata for a naturally occurring unravelling}\label{sec:methods_Nat_Unrav}

In order to perform true state smoothing, we make an assumption on how the environment would have naturally measured
the unobserved baths. In particular, we consider two desiderata for such measurements.

First, the resulting master equation should be invariant under the same transformations that leave the system's master equation invariant, as these are the natural symmetries the system and environment have. Master equation~\eqref{equ:LindbladMasEq} is invariant under the following transformations \cite{wiseman2009quantum}:
\beq
\hat{\bf c} \to {\bf U}\hat{\bf c}
\label{sohet}
\eeq
and 
\beq
\hat{\bf c} \to \hat{\bf c} + {\bf a}\,,\quad\,\, \hat{H} \to \hat{H} -\frac{i}{2}({\bf a}^{\dagger}\hat{\bf c} - \hat{\bf c}^{\dagger}{\bf a})\,, \label{sodyne}
\eeq
where ${\bf U}$ is an arbitrary unitary matrix and ${\bf a}$ is a vector of arbitrary complex numbers (${\bf a}^{\dagger}$ is its complex-conjugate transpose).

The second desideratum is that the measurement performed by the larger environment should be Markovian, i.e., it should be non-adaptive. This is reasonable due to the validity of the Born--Markov approximation --- an adaptive measurement would require the environment to have a memory to store (at minimum) the last measurement outcome.

Invariance under~\eqref{sodyne} limits the Markovian measurement to ``dyne''-type measurements~\cite{wiseman2009quantum}. Invariance under~\eqref{sohet} further limits this measurement to be phase-insensitive in regard to the particular Lindblad operator being detected. Heterodyne unravelling is then the unique natural one for our experimental system considering the form of its Lindblad master equation (given by Eq.~\eqref{UWUQSME_RF_final} in the limit $\eta\to0$).

\section{Full heterodyne unravelling}\label{SecFullHete}

A full heterodyne unravelling of the resonator's master equation results in the conditional covariance matrix $\mathbf{V}_\text{T}(t)=v_\text{T}(t)\,\mathbf{I}$, where $v_\text{T}(t)$ is governed by an equation similar to Eq.~\eqref{NLODE} with the last term on the right-hand side (the measurement term) being updated:
\begin{equation}
     \dot{v}_\text{T}(t) = -\Gamma\,v_\text{T}(t)+2\Gamma\,n_\text{tot}-2 \Gamma\,(\mathcal{C}+n_\text{th})\,v^2_\text{T}(t)\,.
\label{TrueRiccati}
\end{equation}
By putting $\dot{v}_\text{T}=0$, we find the steady state solution $v_\text{T}^\text{ss}=1$, yielding $\mathbf{V}_\text{T}^\text{ss}=\mathbf{I}$.

To clarify the physical meaning of this full unravelling, and to validate its consistency with the assumptions underlying the resonator’s master equation, we rewrite Eq.~\eqref{TrueRiccati} as
\begin{equation}
\begin{array}{lcl}
    \dot{v}_\text{T}(t) & = & \bigg[-\Gamma\,v_\text{T}(t)+\Gamma\,\bigg] + \\\\ && \bigg[-2 \Gamma\eta \mathcal{C}\,v^2_\text{T}(t)+2\Gamma\eta \mathcal{C}\,\bigg] + \\\\ && \bigg[ -2 \Gamma (1 - \eta) \mathcal{C}\,v^2_\text{T}(t)+2\Gamma (1 - \eta) \mathcal{C}\,\bigg] + \\\\ && \bigg[-2 \Gamma n_\text{th}\,v^2_\text{T}(t)+2\Gamma n_\text{th}\,\bigg]\;.
    \end{array}
\end{equation}
The first bracket describes the resonator in contact with a zero-temperature thermal bath. Its first term arises from dynamical drift, while the $\Gamma$ term represents the minimum diffusion required by the fluctuation–dissipation relation~\cite{wiseman2009quantum}. The second and third brackets correspond to the observed and unobserved optical baths, respectively. In each case, the first term represents heterodyne detection via the bath, while the second term is the associated diffusion, which can be interpreted as the back-action of the detection. This interpretation is consistent with the terminology ``back-action heating", used in the literature to describe the contribution of $\mathcal{C}$ to the total phonon occupancy $n_\text{tot}$ of the resonator --- we note that the terminology holds no matter what proportion $(\eta)$ of the optical bath is observed. The fourth bracket describes heterodyne detection via the hot phononic bath and its diffusive back-action. Such a measurement does not disturb the dynamical symmetry, as the rate $\Gamma n_\text{th}$ is much smaller than $\Omega$. Therefore, the resonator’s master equation has been unravelled in a manner consistent with its underlying assumptions, and the resonator can be regarded as a harmonic oscillator heated from its ground state solely by measurement back-action.

\ifSubfilesClassLoaded{
    \bibliographystyle{apsrev4-2}
    \bibliography{sn-bibliography}

\begin{thebibliography}{85}%
\makeatletter
\providecommand \@ifxundefined [1]{%
 \@ifx{#1\undefined}
}%
\providecommand \@ifnum [1]{%
 \ifnum #1\expandafter \@firstoftwo
 \else \expandafter \@secondoftwo
 \fi
}%
\providecommand \@ifx [1]{%
 \ifx #1\expandafter \@firstoftwo
 \else \expandafter \@secondoftwo
 \fi
}%
\providecommand \natexlab [1]{#1}%
\providecommand \enquote  [1]{``#1''}%
\providecommand \bibnamefont  [1]{#1}%
\providecommand \bibfnamefont [1]{#1}%
\providecommand \citenamefont [1]{#1}%
\providecommand \href@noop [0]{\@secondoftwo}%
\providecommand \href [0]{\begingroup \@sanitize@url \@href}%
\providecommand \@href[1]{\@@startlink{#1}\@@href}%
\providecommand \@@href[1]{\endgroup#1\@@endlink}%
\providecommand \@sanitize@url [0]{\catcode `\\12\catcode `\$12\catcode `\&12\catcode `\#12\catcode `\^12\catcode `\_12\catcode `\%12\relax}%
\providecommand \@@startlink[1]{}%
\providecommand \@@endlink[0]{}%
\providecommand \url  [0]{\begingroup\@sanitize@url \@url }%
\providecommand \@url [1]{\endgroup\@href {#1}{\urlprefix }}%
\providecommand \urlprefix  [0]{URL }%
\providecommand \Eprint [0]{\href }%
\providecommand \doibase [0]{https://doi.org/}%
\providecommand \selectlanguage [0]{\@gobble}%
\providecommand \bibinfo  [0]{\@secondoftwo}%
\providecommand \bibfield  [0]{\@secondoftwo}%
\providecommand \translation [1]{[#1]}%
\providecommand \BibitemOpen [0]{}%
\providecommand \bibitemStop [0]{}%
\providecommand \bibitemNoStop [0]{.\EOS\space}%
\providecommand \EOS [0]{\spacefactor3000\relax}%
\providecommand \BibitemShut  [1]{\csname bibitem#1\endcsname}%
\let\auto@bib@innerbib\@empty
\bibitem [{\citenamefont {Stengel}(1994)}]{stengel1994optimal}%
  \BibitemOpen
  \bibfield  {author} {\bibinfo {author} {\bibfnamefont {R.~F.}\ \bibnamefont {Stengel}},\ }\href@noop {} {\emph {\bibinfo {title} {Optimal control and estimation}}}\ (\bibinfo  {publisher} {Courier Corporation},\ \bibinfo {year} {1994})\BibitemShut {NoStop}%
\bibitem [{\citenamefont {Wiseman}\ and\ \citenamefont {Milburn}(2010)}]{wiseman2009quantum}%
  \BibitemOpen
  \bibfield  {author} {\bibinfo {author} {\bibfnamefont {H.~M.}\ \bibnamefont {Wiseman}}\ and\ \bibinfo {author} {\bibfnamefont {G.~J.}\ \bibnamefont {Milburn}},\ }\href@noop {} {\emph {\bibinfo {title} {Quantum measurement and control}}}\ (\bibinfo  {publisher} {Cambridge University Press},\ \bibinfo {year} {2010})\BibitemShut {NoStop}%
\bibitem [{\citenamefont {Dong}\ and\ \citenamefont {Petersen}(2022)}]{Dong2022quantum}%
  \BibitemOpen
  \bibfield  {author} {\bibinfo {author} {\bibfnamefont {D.}~\bibnamefont {Dong}}\ and\ \bibinfo {author} {\bibfnamefont {I.~R.}\ \bibnamefont {Petersen}},\ }\href {https://doi.org/10.1016/j.arcontrol.2022.04.011} {\bibfield  {journal} {\bibinfo  {journal} {Annu. Rev. Control}\ }\textbf {\bibinfo {volume} {54}},\ \bibinfo {pages} {243} (\bibinfo {year} {2022})}\BibitemShut {NoStop}%
\bibitem [{\citenamefont {Carmichael}(1993)}]{Carmichael1993open}%
  \BibitemOpen
  \bibfield  {author} {\bibinfo {author} {\bibfnamefont {H.}~\bibnamefont {Carmichael}},\ }\href@noop {} {\emph {\bibinfo {title} {An Open Systems Approach to Quantum Optics}}}\ (\bibinfo  {publisher} {Springer},\ \bibinfo {year} {1993})\BibitemShut {NoStop}%
\bibitem [{\citenamefont {Jim{\'e}nez-Mart{\'\i}nez}\ \emph {et~al.}(2018)\citenamefont {Jim{\'e}nez-Mart{\'\i}nez}, \citenamefont {Ko{\l}ody{\'n}ski}, \citenamefont {Troullinou}, \citenamefont {Lucivero}, \citenamefont {Kong},\ and\ \citenamefont {Mitchell}}]{jimenez2018signal}%
  \BibitemOpen
  \bibfield  {author} {\bibinfo {author} {\bibfnamefont {R.}~\bibnamefont {Jim{\'e}nez-Mart{\'\i}nez}}, \bibinfo {author} {\bibfnamefont {J.}~\bibnamefont {Ko{\l}ody{\'n}ski}}, \bibinfo {author} {\bibfnamefont {C.}~\bibnamefont {Troullinou}}, \bibinfo {author} {\bibfnamefont {V.~G.}\ \bibnamefont {Lucivero}}, \bibinfo {author} {\bibfnamefont {J.}~\bibnamefont {Kong}},\ and\ \bibinfo {author} {\bibfnamefont {M.~W.}\ \bibnamefont {Mitchell}},\ }\href {https://doi.org/10.1103/PhysRevLett.120.040503} {\bibfield  {journal} {\bibinfo  {journal} {Phys. Rev. Lett.}\ }\textbf {\bibinfo {volume} {120}},\ \bibinfo {pages} {040503} (\bibinfo {year} {2018})}\BibitemShut {NoStop}%
\bibitem [{\citenamefont {Duan}\ \emph {et~al.}(2025)\citenamefont {Duan}, \citenamefont {Hu}, \citenamefont {Lu}, \citenamefont {Xiao}, \citenamefont {Jia}, \citenamefont {M{\o}lmer},\ and\ \citenamefont {Xiao}}]{Duan2025concurrent}%
  \BibitemOpen
  \bibfield  {author} {\bibinfo {author} {\bibfnamefont {J.}~\bibnamefont {Duan}}, \bibinfo {author} {\bibfnamefont {Z.}~\bibnamefont {Hu}}, \bibinfo {author} {\bibfnamefont {X.}~\bibnamefont {Lu}}, \bibinfo {author} {\bibfnamefont {L.}~\bibnamefont {Xiao}}, \bibinfo {author} {\bibfnamefont {S.}~\bibnamefont {Jia}}, \bibinfo {author} {\bibfnamefont {K.}~\bibnamefont {M{\o}lmer}},\ and\ \bibinfo {author} {\bibfnamefont {Y.}~\bibnamefont {Xiao}},\ }\href {https://doi.org/10.1038/s41567-025-02855-3} {\bibfield  {journal} {\bibinfo  {journal} {Nat. Phys.}\ }\textbf {\bibinfo {volume} {21}},\ \bibinfo {pages} {909} (\bibinfo {year} {2025})}\BibitemShut {NoStop}%
\bibitem [{\citenamefont {Sayrin}\ \emph {et~al.}(2011)\citenamefont {Sayrin}, \citenamefont {Dotsenko}, \citenamefont {Zhou}, \citenamefont {Peaudecerf}, \citenamefont {Rybarczyk}, \citenamefont {Gleyzes}, \citenamefont {Rouchon}, \citenamefont {Mirrahimi}, \citenamefont {Amini}, \citenamefont {Brune}, \citenamefont {Raimond},\ and\ \citenamefont {Haroche}}]{Sayrin2011realtime}%
  \BibitemOpen
  \bibfield  {author} {\bibinfo {author} {\bibfnamefont {C.}~\bibnamefont {Sayrin}}, \bibinfo {author} {\bibfnamefont {I.}~\bibnamefont {Dotsenko}}, \bibinfo {author} {\bibfnamefont {X.}~\bibnamefont {Zhou}}, \bibinfo {author} {\bibfnamefont {B.}~\bibnamefont {Peaudecerf}}, \bibinfo {author} {\bibfnamefont {T.}~\bibnamefont {Rybarczyk}}, \bibinfo {author} {\bibfnamefont {S.}~\bibnamefont {Gleyzes}}, \bibinfo {author} {\bibfnamefont {P.}~\bibnamefont {Rouchon}}, \bibinfo {author} {\bibfnamefont {M.}~\bibnamefont {Mirrahimi}}, \bibinfo {author} {\bibfnamefont {H.}~\bibnamefont {Amini}}, \bibinfo {author} {\bibfnamefont {M.}~\bibnamefont {Brune}}, \bibinfo {author} {\bibfnamefont {J.-M.}\ \bibnamefont {Raimond}},\ and\ \bibinfo {author} {\bibfnamefont {S.}~\bibnamefont {Haroche}},\ }\href {https://doi.org/10.1038/nature10376} {\bibfield  {journal} {\bibinfo  {journal} {Nature}\ }\textbf {\bibinfo {volume} {477}},\ \bibinfo {pages} {73} (\bibinfo {year} {2011})}\BibitemShut {NoStop}%
\bibitem [{\citenamefont {Wieczorek}\ \emph {et~al.}(2015)\citenamefont {Wieczorek}, \citenamefont {Hofer}, \citenamefont {Hoelscher-Obermaier}, \citenamefont {Riedinger}, \citenamefont {Hammerer},\ and\ \citenamefont {Aspelmeyer}}]{wieczorek2015optimal}%
  \BibitemOpen
  \bibfield  {author} {\bibinfo {author} {\bibfnamefont {W.}~\bibnamefont {Wieczorek}}, \bibinfo {author} {\bibfnamefont {S.~G.}\ \bibnamefont {Hofer}}, \bibinfo {author} {\bibfnamefont {J.}~\bibnamefont {Hoelscher-Obermaier}}, \bibinfo {author} {\bibfnamefont {R.}~\bibnamefont {Riedinger}}, \bibinfo {author} {\bibfnamefont {K.}~\bibnamefont {Hammerer}},\ and\ \bibinfo {author} {\bibfnamefont {M.}~\bibnamefont {Aspelmeyer}},\ }\href {https://doi.org/10.1103/PhysRevLett.114.223601} {\bibfield  {journal} {\bibinfo  {journal} {Phys. Rev. Lett.}\ }\textbf {\bibinfo {volume} {114}},\ \bibinfo {pages} {223601} (\bibinfo {year} {2015})}\BibitemShut {NoStop}%
\bibitem [{\citenamefont {Thomas}\ \emph {et~al.}(2021)\citenamefont {Thomas}, \citenamefont {Parniak}, \citenamefont {{\O}stfeldt}, \citenamefont {M{\o}ller}, \citenamefont {B{\ae}rentsen}, \citenamefont {Tsaturyan}, \citenamefont {Schliesser}, \citenamefont {Appel}, \citenamefont {Zeuthen},\ and\ \citenamefont {Polzik}}]{thomas2021entanglement}%
  \BibitemOpen
  \bibfield  {author} {\bibinfo {author} {\bibfnamefont {R.~A.}\ \bibnamefont {Thomas}}, \bibinfo {author} {\bibfnamefont {M.}~\bibnamefont {Parniak}}, \bibinfo {author} {\bibfnamefont {C.}~\bibnamefont {{\O}stfeldt}}, \bibinfo {author} {\bibfnamefont {C.~B.}\ \bibnamefont {M{\o}ller}}, \bibinfo {author} {\bibfnamefont {C.}~\bibnamefont {B{\ae}rentsen}}, \bibinfo {author} {\bibfnamefont {Y.}~\bibnamefont {Tsaturyan}}, \bibinfo {author} {\bibfnamefont {A.}~\bibnamefont {Schliesser}}, \bibinfo {author} {\bibfnamefont {J.}~\bibnamefont {Appel}}, \bibinfo {author} {\bibfnamefont {E.}~\bibnamefont {Zeuthen}},\ and\ \bibinfo {author} {\bibfnamefont {E.~S.}\ \bibnamefont {Polzik}},\ }\href {https://doi.org/10.1038/s41567-020-1031-5} {\bibfield  {journal} {\bibinfo  {journal} {Nat. Phys.}\ }\textbf {\bibinfo {volume} {17}},\ \bibinfo {pages} {228} (\bibinfo {year} {2021})}\BibitemShut {NoStop}%
\bibitem [{\citenamefont {Magrini}\ \emph {et~al.}(2021)\citenamefont {Magrini}, \citenamefont {Rosenzweig}, \citenamefont {Bach}, \citenamefont {{Deutschmann-Olek}}, \citenamefont {Hofer}, \citenamefont {Hong}, \citenamefont {Kiesel}, \citenamefont {Kugi},\ and\ \citenamefont {Aspelmeyer}}]{Magrini2021realtime}%
  \BibitemOpen
  \bibfield  {author} {\bibinfo {author} {\bibfnamefont {L.}~\bibnamefont {Magrini}}, \bibinfo {author} {\bibfnamefont {P.}~\bibnamefont {Rosenzweig}}, \bibinfo {author} {\bibfnamefont {C.}~\bibnamefont {Bach}}, \bibinfo {author} {\bibfnamefont {A.}~\bibnamefont {{Deutschmann-Olek}}}, \bibinfo {author} {\bibfnamefont {S.~G.}\ \bibnamefont {Hofer}}, \bibinfo {author} {\bibfnamefont {S.}~\bibnamefont {Hong}}, \bibinfo {author} {\bibfnamefont {N.}~\bibnamefont {Kiesel}}, \bibinfo {author} {\bibfnamefont {A.}~\bibnamefont {Kugi}},\ and\ \bibinfo {author} {\bibfnamefont {M.}~\bibnamefont {Aspelmeyer}},\ }\href {https://doi.org/10.1038/s41586-021-03602-3} {\bibfield  {journal} {\bibinfo  {journal} {Nature}\ }\textbf {\bibinfo {volume} {595}},\ \bibinfo {pages} {373} (\bibinfo {year} {2021})}\BibitemShut {NoStop}%
\bibitem [{\citenamefont {Convy}\ \emph {et~al.}(2022)\citenamefont {Convy}, \citenamefont {Liao}, \citenamefont {Zhang}, \citenamefont {Patel}, \citenamefont {Livingston}, \citenamefont {Nguyen}, \citenamefont {Siddiqi},\ and\ \citenamefont {Whaley}}]{convy2022machine}%
  \BibitemOpen
  \bibfield  {author} {\bibinfo {author} {\bibfnamefont {I.}~\bibnamefont {Convy}}, \bibinfo {author} {\bibfnamefont {H.}~\bibnamefont {Liao}}, \bibinfo {author} {\bibfnamefont {S.}~\bibnamefont {Zhang}}, \bibinfo {author} {\bibfnamefont {S.}~\bibnamefont {Patel}}, \bibinfo {author} {\bibfnamefont {W.~P.}\ \bibnamefont {Livingston}}, \bibinfo {author} {\bibfnamefont {H.~N.}\ \bibnamefont {Nguyen}}, \bibinfo {author} {\bibfnamefont {I.}~\bibnamefont {Siddiqi}},\ and\ \bibinfo {author} {\bibfnamefont {K.~B.}\ \bibnamefont {Whaley}},\ }\href {https://doi.org/10.1088/1367-2630/ac66f9} {\bibfield  {journal} {\bibinfo  {journal} {New J. Phys.}\ }\textbf {\bibinfo {volume} {24}},\ \bibinfo {pages} {063019} (\bibinfo {year} {2022})}\BibitemShut {NoStop}%
\bibitem [{\citenamefont {Livingston}\ \emph {et~al.}(2022)\citenamefont {Livingston}, \citenamefont {Blok}, \citenamefont {Flurin}, \citenamefont {Dressel}, \citenamefont {Jordan},\ and\ \citenamefont {Siddiqi}}]{livingston2022experimental}%
  \BibitemOpen
  \bibfield  {author} {\bibinfo {author} {\bibfnamefont {W.~P.}\ \bibnamefont {Livingston}}, \bibinfo {author} {\bibfnamefont {M.~S.}\ \bibnamefont {Blok}}, \bibinfo {author} {\bibfnamefont {E.}~\bibnamefont {Flurin}}, \bibinfo {author} {\bibfnamefont {J.}~\bibnamefont {Dressel}}, \bibinfo {author} {\bibfnamefont {A.~N.}\ \bibnamefont {Jordan}},\ and\ \bibinfo {author} {\bibfnamefont {I.}~\bibnamefont {Siddiqi}},\ }\href {https://doi.org/10.1038/s41467-022-29906-0} {\bibfield  {journal} {\bibinfo  {journal} {Nat. Commun.}\ }\textbf {\bibinfo {volume} {13}},\ \bibinfo {pages} {2307} (\bibinfo {year} {2022})}\BibitemShut {NoStop}%
\bibitem [{\citenamefont {M{\"u}ller-Ebhardt}\ \emph {et~al.}(2008)\citenamefont {M{\"u}ller-Ebhardt}, \citenamefont {Rehbein}, \citenamefont {Schnabel}, \citenamefont {Danzmann},\ and\ \citenamefont {Chen}}]{muller2008entanglement}%
  \BibitemOpen
  \bibfield  {author} {\bibinfo {author} {\bibfnamefont {H.}~\bibnamefont {M{\"u}ller-Ebhardt}}, \bibinfo {author} {\bibfnamefont {H.}~\bibnamefont {Rehbein}}, \bibinfo {author} {\bibfnamefont {R.}~\bibnamefont {Schnabel}}, \bibinfo {author} {\bibfnamefont {K.}~\bibnamefont {Danzmann}},\ and\ \bibinfo {author} {\bibfnamefont {Y.}~\bibnamefont {Chen}},\ }\href {https://doi.org/10.1103/PhysRevLett.100.013601} {\bibfield  {journal} {\bibinfo  {journal} {Phys. Rev. Lett.}\ }\textbf {\bibinfo {volume} {100}},\ \bibinfo {pages} {013601} (\bibinfo {year} {2008})}\BibitemShut {NoStop}%
\bibitem [{\citenamefont {Meng}\ \emph {et~al.}(2020)\citenamefont {Meng}, \citenamefont {Brawley}, \citenamefont {Bennett}, \citenamefont {Vanner},\ and\ \citenamefont {Bowen}}]{meng2020mechanical}%
  \BibitemOpen
  \bibfield  {author} {\bibinfo {author} {\bibfnamefont {C.}~\bibnamefont {Meng}}, \bibinfo {author} {\bibfnamefont {G.~A.}\ \bibnamefont {Brawley}}, \bibinfo {author} {\bibfnamefont {J.~S.}\ \bibnamefont {Bennett}}, \bibinfo {author} {\bibfnamefont {M.~R.}\ \bibnamefont {Vanner}},\ and\ \bibinfo {author} {\bibfnamefont {W.~P.}\ \bibnamefont {Bowen}},\ }\href {https://doi.org/10.1103/PhysRevLett.125.043604} {\bibfield  {journal} {\bibinfo  {journal} {Phys. Rev. Lett.}\ }\textbf {\bibinfo {volume} {125}},\ \bibinfo {pages} {043604} (\bibinfo {year} {2020})}\BibitemShut {NoStop}%
\bibitem [{\citenamefont {Weinert}(2001)}]{Weinert01}%
  \BibitemOpen
  \bibfield  {author} {\bibinfo {author} {\bibfnamefont {H.~L.}\ \bibnamefont {Weinert}},\ }\href@noop {} {\emph {\bibinfo {title} {Fixed Interval Smoothing for State Space Models}}}\ (\bibinfo  {publisher} {Kluwer Academic},\ \bibinfo {address} {New York},\ \bibinfo {year} {2001})\BibitemShut {NoStop}%
\bibitem [{\citenamefont {Kocsis}\ \emph {et~al.}(2011)\citenamefont {Kocsis}, \citenamefont {Braverman}, \citenamefont {Ravets}, \citenamefont {Stevens}, \citenamefont {Mirin}, \citenamefont {Shalm},\ and\ \citenamefont {Steinberg}}]{Kocsis2011observing}%
  \BibitemOpen
  \bibfield  {author} {\bibinfo {author} {\bibfnamefont {S.}~\bibnamefont {Kocsis}}, \bibinfo {author} {\bibfnamefont {B.}~\bibnamefont {Braverman}}, \bibinfo {author} {\bibfnamefont {S.}~\bibnamefont {Ravets}}, \bibinfo {author} {\bibfnamefont {M.~J.}\ \bibnamefont {Stevens}}, \bibinfo {author} {\bibfnamefont {R.~P.}\ \bibnamefont {Mirin}}, \bibinfo {author} {\bibfnamefont {L.~K.}\ \bibnamefont {Shalm}},\ and\ \bibinfo {author} {\bibfnamefont {A.~M.}\ \bibnamefont {Steinberg}},\ }\href {https://doi.org/10.1126/science.1202218} {\bibfield  {journal} {\bibinfo  {journal} {Science}\ }\textbf {\bibinfo {volume} {332}},\ \bibinfo {pages} {1170} (\bibinfo {year} {2011})}\BibitemShut {NoStop}%
\bibitem [{\citenamefont {Bliokh}\ \emph {et~al.}(2013)\citenamefont {Bliokh}, \citenamefont {Bekshaev}, \citenamefont {Kofman},\ and\ \citenamefont {Nori}}]{Bliokh2013photon}%
  \BibitemOpen
  \bibfield  {author} {\bibinfo {author} {\bibfnamefont {K.~Y.}\ \bibnamefont {Bliokh}}, \bibinfo {author} {\bibfnamefont {A.~Y.}\ \bibnamefont {Bekshaev}}, \bibinfo {author} {\bibfnamefont {A.~G.}\ \bibnamefont {Kofman}},\ and\ \bibinfo {author} {\bibfnamefont {F.}~\bibnamefont {Nori}},\ }\href {https://doi.org/10.1088/1367-2630/15/7/073022} {\bibfield  {journal} {\bibinfo  {journal} {New J. Phys.}\ }\textbf {\bibinfo {volume} {15}},\ \bibinfo {pages} {073022} (\bibinfo {year} {2013})}\BibitemShut {NoStop}%
\bibitem [{\citenamefont {Aharonov}\ \emph {et~al.}(1988)\citenamefont {Aharonov}, \citenamefont {Albert},\ and\ \citenamefont {Vaidman}}]{Aharonov1988how}%
  \BibitemOpen
  \bibfield  {author} {\bibinfo {author} {\bibfnamefont {Y.}~\bibnamefont {Aharonov}}, \bibinfo {author} {\bibfnamefont {D.~Z.}\ \bibnamefont {Albert}},\ and\ \bibinfo {author} {\bibfnamefont {L.}~\bibnamefont {Vaidman}},\ }\href {https://doi.org/10.1103/PhysRevLett.60.1351} {\bibfield  {journal} {\bibinfo  {journal} {Phys. Rev. Lett.}\ }\textbf {\bibinfo {volume} {60}},\ \bibinfo {pages} {1351} (\bibinfo {year} {1988})}\BibitemShut {NoStop}%
\bibitem [{\citenamefont {Leggett}(1989)}]{leggett1989comment}%
  \BibitemOpen
  \bibfield  {author} {\bibinfo {author} {\bibfnamefont {A.~J.}\ \bibnamefont {Leggett}},\ }\href {https://doi.org/10.1103/PhysRevLett.62.2325} {\bibfield  {journal} {\bibinfo  {journal} {Phys. Rev. Lett.}\ }\textbf {\bibinfo {volume} {62}},\ \bibinfo {pages} {2325} (\bibinfo {year} {1989})}\BibitemShut {NoStop}%
\bibitem [{\citenamefont {Ferrie}\ and\ \citenamefont {Combes}(2014)}]{ferrie2014result}%
  \BibitemOpen
  \bibfield  {author} {\bibinfo {author} {\bibfnamefont {C.}~\bibnamefont {Ferrie}}\ and\ \bibinfo {author} {\bibfnamefont {J.}~\bibnamefont {Combes}},\ }\href {https://doi.org/10.1103/PhysRevLett.113.120404} {\bibfield  {journal} {\bibinfo  {journal} {Phys. Rev. Lett.}\ }\textbf {\bibinfo {volume} {113}},\ \bibinfo {pages} {120404} (\bibinfo {year} {2014})}\BibitemShut {NoStop}%
\bibitem [{\citenamefont {Gammelmark}\ \emph {et~al.}(2013)\citenamefont {Gammelmark}, \citenamefont {Julsgaard},\ and\ \citenamefont {M{\o}lmer}}]{gammelmark2013past}%
  \BibitemOpen
  \bibfield  {author} {\bibinfo {author} {\bibfnamefont {S.}~\bibnamefont {Gammelmark}}, \bibinfo {author} {\bibfnamefont {B.}~\bibnamefont {Julsgaard}},\ and\ \bibinfo {author} {\bibfnamefont {K.}~\bibnamefont {M{\o}lmer}},\ }\href {https://doi.org/10.1103/PhysRevLett.111.160401} {\bibfield  {journal} {\bibinfo  {journal} {Phys. Rev. Lett.}\ }\textbf {\bibinfo {volume} {111}},\ \bibinfo {pages} {160401} (\bibinfo {year} {2013})}\BibitemShut {NoStop}%
\bibitem [{\citenamefont {Rybarczyk}\ \emph {et~al.}(2015)\citenamefont {Rybarczyk}, \citenamefont {Peaudecerf}, \citenamefont {Penasa}, \citenamefont {Gerlich}, \citenamefont {Julsgaard}, \citenamefont {M{\o}lmer}, \citenamefont {Gleyzes}, \citenamefont {Brune}, \citenamefont {Raimond}, \citenamefont {Haroche} \emph {et~al.}}]{rybarczyk2015forward}%
  \BibitemOpen
  \bibfield  {author} {\bibinfo {author} {\bibfnamefont {T.}~\bibnamefont {Rybarczyk}}, \bibinfo {author} {\bibfnamefont {B.}~\bibnamefont {Peaudecerf}}, \bibinfo {author} {\bibfnamefont {M.}~\bibnamefont {Penasa}}, \bibinfo {author} {\bibfnamefont {S.}~\bibnamefont {Gerlich}}, \bibinfo {author} {\bibfnamefont {B.}~\bibnamefont {Julsgaard}}, \bibinfo {author} {\bibfnamefont {K.}~\bibnamefont {M{\o}lmer}}, \bibinfo {author} {\bibfnamefont {S.}~\bibnamefont {Gleyzes}}, \bibinfo {author} {\bibfnamefont {M.}~\bibnamefont {Brune}}, \bibinfo {author} {\bibfnamefont {J.~M.}\ \bibnamefont {Raimond}}, \bibinfo {author} {\bibfnamefont {S.}~\bibnamefont {Haroche}}, \emph {et~al.},\ }\href {https://doi.org/10.1103/PhysRevA.91.062116} {\bibfield  {journal} {\bibinfo  {journal} {Phys. Rev. A}\ }\textbf {\bibinfo {volume} {91}},\ \bibinfo {pages} {062116} (\bibinfo {year} {2015})}\BibitemShut {NoStop}%
\bibitem [{\citenamefont {Tan}\ \emph {et~al.}(2015)\citenamefont {Tan}, \citenamefont {Weber}, \citenamefont {Siddiqi}, \citenamefont {M{\o}lmer},\ and\ \citenamefont {Murch}}]{tan2015prediction}%
  \BibitemOpen
  \bibfield  {author} {\bibinfo {author} {\bibfnamefont {D.}~\bibnamefont {Tan}}, \bibinfo {author} {\bibfnamefont {S.~J.}\ \bibnamefont {Weber}}, \bibinfo {author} {\bibfnamefont {I.}~\bibnamefont {Siddiqi}}, \bibinfo {author} {\bibfnamefont {K.}~\bibnamefont {M{\o}lmer}},\ and\ \bibinfo {author} {\bibfnamefont {K.~W.}\ \bibnamefont {Murch}},\ }\href {https://doi.org/10.1103/PhysRevLett.114.090403} {\bibfield  {journal} {\bibinfo  {journal} {Phys. Rev. Lett.}\ }\textbf {\bibinfo {volume} {114}},\ \bibinfo {pages} {090403} (\bibinfo {year} {2015})}\BibitemShut {NoStop}%
\bibitem [{\citenamefont {Bao}\ \emph {et~al.}(2020)\citenamefont {Bao}, \citenamefont {Jin}, \citenamefont {Duan}, \citenamefont {Jia}, \citenamefont {M{\o}lmer}, \citenamefont {Shen},\ and\ \citenamefont {Xiao}}]{bao2020retrodiction}%
  \BibitemOpen
  \bibfield  {author} {\bibinfo {author} {\bibfnamefont {H.}~\bibnamefont {Bao}}, \bibinfo {author} {\bibfnamefont {S.}~\bibnamefont {Jin}}, \bibinfo {author} {\bibfnamefont {J.}~\bibnamefont {Duan}}, \bibinfo {author} {\bibfnamefont {S.}~\bibnamefont {Jia}}, \bibinfo {author} {\bibfnamefont {K.}~\bibnamefont {M{\o}lmer}}, \bibinfo {author} {\bibfnamefont {H.}~\bibnamefont {Shen}},\ and\ \bibinfo {author} {\bibfnamefont {Y.}~\bibnamefont {Xiao}},\ }\href {https://doi.org/https://doi.org/10.1038/s41467-020-19495-1} {\bibfield  {journal} {\bibinfo  {journal} {Nat. Commun.}\ }\textbf {\bibinfo {volume} {11}},\ \bibinfo {pages} {5658} (\bibinfo {year} {2020})}\BibitemShut {NoStop}%
\bibitem [{\citenamefont {Miao}\ \emph {et~al.}(2010)\citenamefont {Miao}, \citenamefont {Danilishin}, \citenamefont {M{\"u}ller-Ebhardt}, \citenamefont {Rehbein}, \citenamefont {Somiya},\ and\ \citenamefont {Chen}}]{miao2010probing}%
  \BibitemOpen
  \bibfield  {author} {\bibinfo {author} {\bibfnamefont {H.}~\bibnamefont {Miao}}, \bibinfo {author} {\bibfnamefont {S.}~\bibnamefont {Danilishin}}, \bibinfo {author} {\bibfnamefont {H.}~\bibnamefont {M{\"u}ller-Ebhardt}}, \bibinfo {author} {\bibfnamefont {H.}~\bibnamefont {Rehbein}}, \bibinfo {author} {\bibfnamefont {K.}~\bibnamefont {Somiya}},\ and\ \bibinfo {author} {\bibfnamefont {Y.}~\bibnamefont {Chen}},\ }\href {https://doi.org/10.1103/PhysRevA.81.012114} {\bibfield  {journal} {\bibinfo  {journal} {Phys. Rev. A}\ }\textbf {\bibinfo {volume} {81}},\ \bibinfo {pages} {012114} (\bibinfo {year} {2010})}\BibitemShut {NoStop}%
\bibitem [{\citenamefont {Rossi}\ \emph {et~al.}(2019)\citenamefont {Rossi}, \citenamefont {Mason}, \citenamefont {Chen},\ and\ \citenamefont {Schliesser}}]{rossi2019observing}%
  \BibitemOpen
  \bibfield  {author} {\bibinfo {author} {\bibfnamefont {M.}~\bibnamefont {Rossi}}, \bibinfo {author} {\bibfnamefont {D.}~\bibnamefont {Mason}}, \bibinfo {author} {\bibfnamefont {J.}~\bibnamefont {Chen}},\ and\ \bibinfo {author} {\bibfnamefont {A.}~\bibnamefont {Schliesser}},\ }\href {https://doi.org/10.1103/PhysRevLett.123.163601} {\bibfield  {journal} {\bibinfo  {journal} {Phys. Rev. Lett.}\ }\textbf {\bibinfo {volume} {123}},\ \bibinfo {pages} {163601} (\bibinfo {year} {2019})}\BibitemShut {NoStop}%
\bibitem [{\citenamefont {Meng}\ \emph {et~al.}(2022)\citenamefont {Meng}, \citenamefont {Brawley}, \citenamefont {Khademi}, \citenamefont {Bridge}, \citenamefont {Bennett},\ and\ \citenamefont {Bowen}}]{meng2022measurement}%
  \BibitemOpen
  \bibfield  {author} {\bibinfo {author} {\bibfnamefont {C.}~\bibnamefont {Meng}}, \bibinfo {author} {\bibfnamefont {G.~A.}\ \bibnamefont {Brawley}}, \bibinfo {author} {\bibfnamefont {S.}~\bibnamefont {Khademi}}, \bibinfo {author} {\bibfnamefont {E.~M.}\ \bibnamefont {Bridge}}, \bibinfo {author} {\bibfnamefont {J.~S.}\ \bibnamefont {Bennett}},\ and\ \bibinfo {author} {\bibfnamefont {W.~P.}\ \bibnamefont {Bowen}},\ }\href {https://doi.org/10.1126/sciadv.abm7585} {\bibfield  {journal} {\bibinfo  {journal} {Sci. Adv.}\ }\textbf {\bibinfo {volume} {8}},\ \bibinfo {pages} {eabm7585} (\bibinfo {year} {2022})}\BibitemShut {NoStop}%
\bibitem [{\citenamefont {Lammers}\ and\ \citenamefont {Hammerer}(2024)}]{lammers2024quantum}%
  \BibitemOpen
  \bibfield  {author} {\bibinfo {author} {\bibfnamefont {J.}~\bibnamefont {Lammers}}\ and\ \bibinfo {author} {\bibfnamefont {K.}~\bibnamefont {Hammerer}},\ }\href {https://doi.org/10.3389/frqst.2023.1294905} {\bibfield  {journal} {\bibinfo  {journal} {Front. Quantum Sci. Technol.}\ }\textbf {\bibinfo {volume} {2}},\ \bibinfo {pages} {1294905} (\bibinfo {year} {2024})}\BibitemShut {NoStop}%
\bibitem [{\citenamefont {Guevara}\ and\ \citenamefont {Wiseman}(2015)}]{guevara2015quantum}%
  \BibitemOpen
  \bibfield  {author} {\bibinfo {author} {\bibfnamefont {I.}~\bibnamefont {Guevara}}\ and\ \bibinfo {author} {\bibfnamefont {H.}~\bibnamefont {Wiseman}},\ }\href {https://doi.org/10.1103/PhysRevLett.115.180407} {\bibfield  {journal} {\bibinfo  {journal} {Phys. Rev. Lett.}\ }\textbf {\bibinfo {volume} {115}},\ \bibinfo {pages} {180407} (\bibinfo {year} {2015})}\BibitemShut {NoStop}%
\bibitem [{\citenamefont {Laverick}(2021)}]{laverick2021quantum}%
  \BibitemOpen
  \bibfield  {author} {\bibinfo {author} {\bibfnamefont {K.~T.}\ \bibnamefont {Laverick}},\ }\href {https://doi.org/10.1103/PhysRevResearch.3.033196} {\bibfield  {journal} {\bibinfo  {journal} {Phys. Rev. Res.}\ }\textbf {\bibinfo {volume} {3}},\ \bibinfo {pages} {033196} (\bibinfo {year} {2021})}\BibitemShut {NoStop}%
\bibitem [{\citenamefont {Campbell}\ \emph {et~al.}(2020)\citenamefont {Campbell}, \citenamefont {Shim}, \citenamefont {Kannan}, \citenamefont {Winik}, \citenamefont {Kim}, \citenamefont {Melville}, \citenamefont {Niedzielski}, \citenamefont {Yoder}, \citenamefont {Tahan}, \citenamefont {Gustavsson} \emph {et~al.}}]{campbell2020universal}%
  \BibitemOpen
  \bibfield  {author} {\bibinfo {author} {\bibfnamefont {D.~L.}\ \bibnamefont {Campbell}}, \bibinfo {author} {\bibfnamefont {Y.-P.}\ \bibnamefont {Shim}}, \bibinfo {author} {\bibfnamefont {B.}~\bibnamefont {Kannan}}, \bibinfo {author} {\bibfnamefont {R.}~\bibnamefont {Winik}}, \bibinfo {author} {\bibfnamefont {D.~K.}\ \bibnamefont {Kim}}, \bibinfo {author} {\bibfnamefont {A.}~\bibnamefont {Melville}}, \bibinfo {author} {\bibfnamefont {B.~M.}\ \bibnamefont {Niedzielski}}, \bibinfo {author} {\bibfnamefont {J.~L.}\ \bibnamefont {Yoder}}, \bibinfo {author} {\bibfnamefont {C.}~\bibnamefont {Tahan}}, \bibinfo {author} {\bibfnamefont {S.}~\bibnamefont {Gustavsson}}, \emph {et~al.},\ }\href {https://doi.org/https://doi.org/10.1103/PhysRevX.10.041051} {\bibfield  {journal} {\bibinfo  {journal} {Phys. Rev. X}\ }\textbf {\bibinfo {volume} {10}},\ \bibinfo {pages} {041051} (\bibinfo {year} {2020})}\BibitemShut {NoStop}%
\bibitem [{\citenamefont {Furusawa}\ \emph {et~al.}(1998)\citenamefont {Furusawa}, \citenamefont {S{\o}rensen}, \citenamefont {Braunstein}, \citenamefont {Fuchs}, \citenamefont {Kimble},\ and\ \citenamefont {Polzik}}]{furusawa1998unconditional}%
  \BibitemOpen
  \bibfield  {author} {\bibinfo {author} {\bibfnamefont {A.}~\bibnamefont {Furusawa}}, \bibinfo {author} {\bibfnamefont {J.~L.}\ \bibnamefont {S{\o}rensen}}, \bibinfo {author} {\bibfnamefont {S.~L.}\ \bibnamefont {Braunstein}}, \bibinfo {author} {\bibfnamefont {C.~A.}\ \bibnamefont {Fuchs}}, \bibinfo {author} {\bibfnamefont {H.~J.}\ \bibnamefont {Kimble}},\ and\ \bibinfo {author} {\bibfnamefont {E.~S.}\ \bibnamefont {Polzik}},\ }\href {https://doi.org/10.1126/science.282.5389.706} {\bibfield  {journal} {\bibinfo  {journal} {Science}\ }\textbf {\bibinfo {volume} {282}},\ \bibinfo {pages} {706} (\bibinfo {year} {1998})}\BibitemShut {NoStop}%
\bibitem [{\citenamefont {Bowen}\ \emph {et~al.}(2003)\citenamefont {Bowen}, \citenamefont {Treps}, \citenamefont {Buchler}, \citenamefont {Schnabel}, \citenamefont {Ralph}, \citenamefont {Bachor}, \citenamefont {Symul},\ and\ \citenamefont {Lam}}]{bowen2003experimental}%
  \BibitemOpen
  \bibfield  {author} {\bibinfo {author} {\bibfnamefont {W.~P.}\ \bibnamefont {Bowen}}, \bibinfo {author} {\bibfnamefont {N.}~\bibnamefont {Treps}}, \bibinfo {author} {\bibfnamefont {B.~C.}\ \bibnamefont {Buchler}}, \bibinfo {author} {\bibfnamefont {R.}~\bibnamefont {Schnabel}}, \bibinfo {author} {\bibfnamefont {T.~C.}\ \bibnamefont {Ralph}}, \bibinfo {author} {\bibfnamefont {H.-A.}\ \bibnamefont {Bachor}}, \bibinfo {author} {\bibfnamefont {T.}~\bibnamefont {Symul}},\ and\ \bibinfo {author} {\bibfnamefont {P.~K.}\ \bibnamefont {Lam}},\ }\href {https://doi.org/https://doi.org/10.1103/PhysRevA.67.032302} {\bibfield  {journal} {\bibinfo  {journal} {Phys. Rev. A}\ }\textbf {\bibinfo {volume} {67}},\ \bibinfo {pages} {032302} (\bibinfo {year} {2003})}\BibitemShut {NoStop}%
\bibitem [{\citenamefont {Wiseman}(1996)}]{wiseman1996quantum}%
  \BibitemOpen
  \bibfield  {author} {\bibinfo {author} {\bibfnamefont {H.~M.}\ \bibnamefont {Wiseman}},\ }\href {https://doi.org/10.1088/1355-5111/8/1/015} {\bibfield  {journal} {\bibinfo  {journal} {Quantum Semiclass. Opt.}\ }\textbf {\bibinfo {volume} {8}},\ \bibinfo {pages} {205} (\bibinfo {year} {1996})}\BibitemShut {NoStop}%
\bibitem [{\citenamefont {Rossi}\ \emph {et~al.}(2020)\citenamefont {Rossi}, \citenamefont {Mancino}, \citenamefont {Landi}, \citenamefont {Paternostro}, \citenamefont {Schliesser},\ and\ \citenamefont {Belenchia}}]{rossi2020experimental}%
  \BibitemOpen
  \bibfield  {author} {\bibinfo {author} {\bibfnamefont {M.}~\bibnamefont {Rossi}}, \bibinfo {author} {\bibfnamefont {L.}~\bibnamefont {Mancino}}, \bibinfo {author} {\bibfnamefont {G.~T.}\ \bibnamefont {Landi}}, \bibinfo {author} {\bibfnamefont {M.}~\bibnamefont {Paternostro}}, \bibinfo {author} {\bibfnamefont {A.}~\bibnamefont {Schliesser}},\ and\ \bibinfo {author} {\bibfnamefont {A.}~\bibnamefont {Belenchia}},\ }\href {https://doi.org/10.1103/PhysRevLett.125.080601} {\bibfield  {journal} {\bibinfo  {journal} {Phys. Rev. Lett.}\ }\textbf {\bibinfo {volume} {125}},\ \bibinfo {pages} {080601} (\bibinfo {year} {2020})}\BibitemShut {NoStop}%
\bibitem [{\citenamefont {He}\ \emph {et~al.}(2023)\citenamefont {He}, \citenamefont {Pakkiam}, \citenamefont {Gangat}, \citenamefont {Kewming}, \citenamefont {Milburn},\ and\ \citenamefont {Fedorov}}]{he2023effect}%
  \BibitemOpen
  \bibfield  {author} {\bibinfo {author} {\bibfnamefont {X.}~\bibnamefont {He}}, \bibinfo {author} {\bibfnamefont {P.}~\bibnamefont {Pakkiam}}, \bibinfo {author} {\bibfnamefont {A.~A.}\ \bibnamefont {Gangat}}, \bibinfo {author} {\bibfnamefont {M.~J.}\ \bibnamefont {Kewming}}, \bibinfo {author} {\bibfnamefont {G.~J.}\ \bibnamefont {Milburn}},\ and\ \bibinfo {author} {\bibfnamefont {A.}~\bibnamefont {Fedorov}},\ }\href {https://doi.org/10.1103/PhysRevApplied.20.034038} {\bibfield  {journal} {\bibinfo  {journal} {Phys. Rev. Appl.}\ }\textbf {\bibinfo {volume} {20}},\ \bibinfo {pages} {034038} (\bibinfo {year} {2023})}\BibitemShut {NoStop}%
\bibitem [{\citenamefont {Wiseman}\ and\ \citenamefont {Vaccaro}(2001)}]{wiseman2001inequivalence}%
  \BibitemOpen
  \bibfield  {author} {\bibinfo {author} {\bibfnamefont {H.~M.}\ \bibnamefont {Wiseman}}\ and\ \bibinfo {author} {\bibfnamefont {J.~A.}\ \bibnamefont {Vaccaro}},\ }\href {https://doi.org/10.1103/PhysRevLett.87.240402} {\bibfield  {journal} {\bibinfo  {journal} {Phys. Rev. Lett.}\ }\textbf {\bibinfo {volume} {87}},\ \bibinfo {pages} {240402} (\bibinfo {year} {2001})}\BibitemShut {NoStop}%
\bibitem [{\citenamefont {Aharonov}\ \emph {et~al.}(1964)\citenamefont {Aharonov}, \citenamefont {Bergmann},\ and\ \citenamefont {Lebowitz}}]{aharonov1964time}%
  \BibitemOpen
  \bibfield  {author} {\bibinfo {author} {\bibfnamefont {Y.}~\bibnamefont {Aharonov}}, \bibinfo {author} {\bibfnamefont {P.~G.}\ \bibnamefont {Bergmann}},\ and\ \bibinfo {author} {\bibfnamefont {J.~L.}\ \bibnamefont {Lebowitz}},\ }\href {https://doi.org/10.1103/PhysRev.134.B1410} {\bibfield  {journal} {\bibinfo  {journal} {Phys. Rev.}\ }\textbf {\bibinfo {volume} {134}},\ \bibinfo {pages} {B1410} (\bibinfo {year} {1964})}\BibitemShut {NoStop}%
\bibitem [{\citenamefont {Chantasri}\ \emph {et~al.}(2019)\citenamefont {Chantasri}, \citenamefont {Guevara},\ and\ \citenamefont {Wiseman}}]{CGW19}%
  \BibitemOpen
  \bibfield  {author} {\bibinfo {author} {\bibfnamefont {A.}~\bibnamefont {Chantasri}}, \bibinfo {author} {\bibfnamefont {I.}~\bibnamefont {Guevara}},\ and\ \bibinfo {author} {\bibfnamefont {H.~M.}\ \bibnamefont {Wiseman}},\ }\href {https://doi.org/10.1088/1367-2630/ab396e} {\bibfield  {journal} {\bibinfo  {journal} {New J. Phys.}\ }\textbf {\bibinfo {volume} {21}},\ \bibinfo {pages} {083039} (\bibinfo {year} {2019})}\BibitemShut {NoStop}%
\bibitem [{\citenamefont {Yokoyama}\ \emph {et~al.}(2025)\citenamefont {Yokoyama}, \citenamefont {Laverick}, \citenamefont {McManus}, \citenamefont {Yu}, \citenamefont {Chantasri}, \citenamefont {Asavanant}, \citenamefont {Dong}, \citenamefont {Wiseman},\ and\ \citenamefont {Yonezawa}}]{Shota25}%
  \BibitemOpen
  \bibfield  {author} {\bibinfo {author} {\bibfnamefont {S.}~\bibnamefont {Yokoyama}}, \bibinfo {author} {\bibfnamefont {K.~T.}\ \bibnamefont {Laverick}}, \bibinfo {author} {\bibfnamefont {D.}~\bibnamefont {McManus}}, \bibinfo {author} {\bibfnamefont {Q.}~\bibnamefont {Yu}}, \bibinfo {author} {\bibfnamefont {A.}~\bibnamefont {Chantasri}}, \bibinfo {author} {\bibfnamefont {W.}~\bibnamefont {Asavanant}}, \bibinfo {author} {\bibfnamefont {D.}~\bibnamefont {Dong}}, \bibinfo {author} {\bibfnamefont {H.~M.}\ \bibnamefont {Wiseman}},\ and\ \bibinfo {author} {\bibfnamefont {H.}~\bibnamefont {Yonezawa}},\ }\bibfield  {journal} {\bibinfo  {journal} {arXiv preprint}\ }\href {https://doi.org/10.48550/arXiv.2509.04754} {10.48550/arXiv.2509.04754} (\bibinfo {year} {2025})\BibitemShut {NoStop}%
\bibitem [{\citenamefont {Laverick}\ \emph {et~al.}(2019)\citenamefont {Laverick}, \citenamefont {Chantasri},\ and\ \citenamefont {Wiseman}}]{laverick2019quantum}%
  \BibitemOpen
  \bibfield  {author} {\bibinfo {author} {\bibfnamefont {K.~T.}\ \bibnamefont {Laverick}}, \bibinfo {author} {\bibfnamefont {A.}~\bibnamefont {Chantasri}},\ and\ \bibinfo {author} {\bibfnamefont {H.~M.}\ \bibnamefont {Wiseman}},\ }\href {https://doi.org/10.1103/PhysRevLett.122.190402} {\bibfield  {journal} {\bibinfo  {journal} {Phys. Rev. Lett.}\ }\textbf {\bibinfo {volume} {122}},\ \bibinfo {pages} {190402} (\bibinfo {year} {2019})}\BibitemShut {NoStop}%
\bibitem [{\citenamefont {Laverick}\ \emph {et~al.}(2021{\natexlab{a}})\citenamefont {Laverick}, \citenamefont {Chantasri},\ and\ \citenamefont {Wiseman}}]{laverick2021linear}%
  \BibitemOpen
  \bibfield  {author} {\bibinfo {author} {\bibfnamefont {K.~T.}\ \bibnamefont {Laverick}}, \bibinfo {author} {\bibfnamefont {A.}~\bibnamefont {Chantasri}},\ and\ \bibinfo {author} {\bibfnamefont {H.~M.}\ \bibnamefont {Wiseman}},\ }\href {https://doi.org/10.1103/PhysRevA.103.012213} {\bibfield  {journal} {\bibinfo  {journal} {Phys. Rev. A}\ }\textbf {\bibinfo {volume} {103}},\ \bibinfo {pages} {012213} (\bibinfo {year} {2021}{\natexlab{a}})}\BibitemShut {NoStop}%
\bibitem [{\citenamefont {Aspelmeyer}\ \emph {et~al.}(2014)\citenamefont {Aspelmeyer}, \citenamefont {Kippenberg},\ and\ \citenamefont {Marquardt}}]{aspelmeyer2014cavity}%
  \BibitemOpen
  \bibfield  {author} {\bibinfo {author} {\bibfnamefont {M.}~\bibnamefont {Aspelmeyer}}, \bibinfo {author} {\bibfnamefont {T.~J.}\ \bibnamefont {Kippenberg}},\ and\ \bibinfo {author} {\bibfnamefont {F.}~\bibnamefont {Marquardt}},\ }\href {https://doi.org/10.1103/RevModPhys.86.1391} {\bibfield  {journal} {\bibinfo  {journal} {Rev. Mod. Phys.}\ }\textbf {\bibinfo {volume} {86}},\ \bibinfo {pages} {1391} (\bibinfo {year} {2014})}\BibitemShut {NoStop}%
\bibitem [{\citenamefont {Mason}\ \emph {et~al.}(2019)\citenamefont {Mason}, \citenamefont {Chen}, \citenamefont {Rossi}, \citenamefont {Tsaturyan},\ and\ \citenamefont {Schliesser}}]{mason2019continuous}%
  \BibitemOpen
  \bibfield  {author} {\bibinfo {author} {\bibfnamefont {D.}~\bibnamefont {Mason}}, \bibinfo {author} {\bibfnamefont {J.}~\bibnamefont {Chen}}, \bibinfo {author} {\bibfnamefont {M.}~\bibnamefont {Rossi}}, \bibinfo {author} {\bibfnamefont {Y.}~\bibnamefont {Tsaturyan}},\ and\ \bibinfo {author} {\bibfnamefont {A.}~\bibnamefont {Schliesser}},\ }\href {https://doi.org/10.1038/s41567-019-0533-5} {\bibfield  {journal} {\bibinfo  {journal} {Nat. Phys.}\ }\textbf {\bibinfo {volume} {15}},\ \bibinfo {pages} {745} (\bibinfo {year} {2019})}\BibitemShut {NoStop}%
\bibitem [{\citenamefont {Bowen}\ and\ \citenamefont {Milburn}(2015)}]{bowen2015quantum}%
  \BibitemOpen
  \bibfield  {author} {\bibinfo {author} {\bibfnamefont {W.~P.}\ \bibnamefont {Bowen}}\ and\ \bibinfo {author} {\bibfnamefont {G.~J.}\ \bibnamefont {Milburn}},\ }\href@noop {} {\emph {\bibinfo {title} {Quantum optomechanics}}}\ (\bibinfo  {publisher} {CRC press},\ \bibinfo {year} {2015})\BibitemShut {NoStop}%
\bibitem [{\citenamefont {Guo}\ \emph {et~al.}(2019)\citenamefont {Guo}, \citenamefont {Norte},\ and\ \citenamefont {Gr{\"o}blacher}}]{guo2019feedback}%
  \BibitemOpen
  \bibfield  {author} {\bibinfo {author} {\bibfnamefont {J.}~\bibnamefont {Guo}}, \bibinfo {author} {\bibfnamefont {R.}~\bibnamefont {Norte}},\ and\ \bibinfo {author} {\bibfnamefont {S.}~\bibnamefont {Gr{\"o}blacher}},\ }\href {https://doi.org/10.1103/PhysRevLett.123.223602} {\bibfield  {journal} {\bibinfo  {journal} {Phys. Rev. Lett.}\ }\textbf {\bibinfo {volume} {123}},\ \bibinfo {pages} {223602} (\bibinfo {year} {2019})}\BibitemShut {NoStop}%
\bibitem [{\citenamefont {Hirschel}\ \emph {et~al.}(2024)\citenamefont {Hirschel}, \citenamefont {Vadakkumbatt}, \citenamefont {Baker}, \citenamefont {Schweizer}, \citenamefont {Sankey}, \citenamefont {Singh},\ and\ \citenamefont {Davis}}]{hirschel2024superfluid}%
  \BibitemOpen
  \bibfield  {author} {\bibinfo {author} {\bibfnamefont {M.}~\bibnamefont {Hirschel}}, \bibinfo {author} {\bibfnamefont {V.}~\bibnamefont {Vadakkumbatt}}, \bibinfo {author} {\bibfnamefont {N.~P.}\ \bibnamefont {Baker}}, \bibinfo {author} {\bibfnamefont {F.~M.}\ \bibnamefont {Schweizer}}, \bibinfo {author} {\bibfnamefont {J.~C.}\ \bibnamefont {Sankey}}, \bibinfo {author} {\bibfnamefont {S.}~\bibnamefont {Singh}},\ and\ \bibinfo {author} {\bibfnamefont {J.~P.}\ \bibnamefont {Davis}},\ }\href {https://doi.org/10.1103/PhysRevD.109.095011} {\bibfield  {journal} {\bibinfo  {journal} {Phys. Rev. D}\ }\textbf {\bibinfo {volume} {109}},\ \bibinfo {pages} {095011} (\bibinfo {year} {2024})}\BibitemShut {NoStop}%
\bibitem [{\citenamefont {Baker}\ \emph {et~al.}(2024)\citenamefont {Baker}, \citenamefont {Bowen}, \citenamefont {Cox}, \citenamefont {Dolan}, \citenamefont {Goryachev},\ and\ \citenamefont {Harris}}]{baker2024optomechanical}%
  \BibitemOpen
  \bibfield  {author} {\bibinfo {author} {\bibfnamefont {C.~G.}\ \bibnamefont {Baker}}, \bibinfo {author} {\bibfnamefont {W.~P.}\ \bibnamefont {Bowen}}, \bibinfo {author} {\bibfnamefont {P.}~\bibnamefont {Cox}}, \bibinfo {author} {\bibfnamefont {M.~J.}\ \bibnamefont {Dolan}}, \bibinfo {author} {\bibfnamefont {M.}~\bibnamefont {Goryachev}},\ and\ \bibinfo {author} {\bibfnamefont {G.}~\bibnamefont {Harris}},\ }\href {https://doi.org/10.1103/PhysRevD.110.043005} {\bibfield  {journal} {\bibinfo  {journal} {Phys. Rev. D}\ }\textbf {\bibinfo {volume} {110}},\ \bibinfo {pages} {043005} (\bibinfo {year} {2024})}\BibitemShut {NoStop}%
\bibitem [{\citenamefont {Girdhar}\ and\ \citenamefont {Doherty}(2020)}]{girdhar2020testing}%
  \BibitemOpen
  \bibfield  {author} {\bibinfo {author} {\bibfnamefont {P.}~\bibnamefont {Girdhar}}\ and\ \bibinfo {author} {\bibfnamefont {A.~C.}\ \bibnamefont {Doherty}},\ }\href {https://doi.org/10.1088/1367-2630/abb43c} {\bibfield  {journal} {\bibinfo  {journal} {New J. Phys.}\ }\textbf {\bibinfo {volume} {22}},\ \bibinfo {pages} {093073} (\bibinfo {year} {2020})}\BibitemShut {NoStop}%
\bibitem [{\citenamefont {Guo}\ and\ \citenamefont {Gr{\"o}blacher}(2022)}]{guo2022integrated}%
  \BibitemOpen
  \bibfield  {author} {\bibinfo {author} {\bibfnamefont {J.}~\bibnamefont {Guo}}\ and\ \bibinfo {author} {\bibfnamefont {S.}~\bibnamefont {Gr{\"o}blacher}},\ }\href {https://doi.org/10.1038/s41377-022-00966-7} {\bibfield  {journal} {\bibinfo  {journal} {Light Sci. Appl.}\ }\textbf {\bibinfo {volume} {11}},\ \bibinfo {pages} {282} (\bibinfo {year} {2022})}\BibitemShut {NoStop}%
\bibitem [{\citenamefont {Guo}\ \emph {et~al.}(2023)\citenamefont {Guo}, \citenamefont {Chang}, \citenamefont {Yao},\ and\ \citenamefont {Gr{\"o}blacher}}]{guo2023active}%
  \BibitemOpen
  \bibfield  {author} {\bibinfo {author} {\bibfnamefont {J.}~\bibnamefont {Guo}}, \bibinfo {author} {\bibfnamefont {J.}~\bibnamefont {Chang}}, \bibinfo {author} {\bibfnamefont {X.}~\bibnamefont {Yao}},\ and\ \bibinfo {author} {\bibfnamefont {S.}~\bibnamefont {Gr{\"o}blacher}},\ }\href {https://doi.org/10.1038/s41467-023-40442-3} {\bibfield  {journal} {\bibinfo  {journal} {Nat. Commun.}\ }\textbf {\bibinfo {volume} {14}},\ \bibinfo {pages} {4721} (\bibinfo {year} {2023})}\BibitemShut {NoStop}%
\bibitem [{\citenamefont {Fedorov}\ \emph {et~al.}(2020)\citenamefont {Fedorov}, \citenamefont {Beccari}, \citenamefont {Engelsen},\ and\ \citenamefont {Kippenberg}}]{fedorov2020fractal}%
  \BibitemOpen
  \bibfield  {author} {\bibinfo {author} {\bibfnamefont {S.~A.}\ \bibnamefont {Fedorov}}, \bibinfo {author} {\bibfnamefont {A.}~\bibnamefont {Beccari}}, \bibinfo {author} {\bibfnamefont {N.~J.}\ \bibnamefont {Engelsen}},\ and\ \bibinfo {author} {\bibfnamefont {T.~J.}\ \bibnamefont {Kippenberg}},\ }\href {https://doi.org/10.1103/PhysRevLett.124.025502} {\bibfield  {journal} {\bibinfo  {journal} {Phys. Rev. Lett.}\ }\textbf {\bibinfo {volume} {124}},\ \bibinfo {pages} {025502} (\bibinfo {year} {2020})}\BibitemShut {NoStop}%
\bibitem [{\citenamefont {Doherty}\ \emph {et~al.}(2012)\citenamefont {Doherty}, \citenamefont {Szorkovszky}, \citenamefont {Harris},\ and\ \citenamefont {Bowen}}]{doherty2012quantum}%
  \BibitemOpen
  \bibfield  {author} {\bibinfo {author} {\bibfnamefont {A.~C.}\ \bibnamefont {Doherty}}, \bibinfo {author} {\bibfnamefont {A.}~\bibnamefont {Szorkovszky}}, \bibinfo {author} {\bibfnamefont {G.~I.}\ \bibnamefont {Harris}},\ and\ \bibinfo {author} {\bibfnamefont {W.~P.}\ \bibnamefont {Bowen}},\ }\href {https://doi.org/10.1098/rsta.2011.0531} {\bibfield  {journal} {\bibinfo  {journal} {Phil. Trans. R. Soc. A.}\ }\textbf {\bibinfo {volume} {370}},\ \bibinfo {pages} {5338} (\bibinfo {year} {2012})}\BibitemShut {NoStop}%
\bibitem [{\citenamefont {Serafini}(2017)}]{serafini2017quantum}%
  \BibitemOpen
  \bibfield  {author} {\bibinfo {author} {\bibfnamefont {A.}~\bibnamefont {Serafini}},\ }\href@noop {} {\emph {\bibinfo {title} {Quantum continuous variables: a primer of theoretical methods}}}\ (\bibinfo  {publisher} {CRC press},\ \bibinfo {year} {2017})\BibitemShut {NoStop}%
\bibitem [{\citenamefont {Guevara}\ and\ \citenamefont {Wiseman}(2020)}]{GueWis20}%
  \BibitemOpen
  \bibfield  {author} {\bibinfo {author} {\bibfnamefont {I.}~\bibnamefont {Guevara}}\ and\ \bibinfo {author} {\bibfnamefont {H.~M.}\ \bibnamefont {Wiseman}},\ }\href {https://doi.org/10.1103/PhysRevA.102.052217} {\bibfield  {journal} {\bibinfo  {journal} {Phys. Rev. A}\ }\textbf {\bibinfo {volume} {102}},\ \bibinfo {pages} {052217} (\bibinfo {year} {2020})}\BibitemShut {NoStop}%
\bibitem [{\citenamefont {Chantasri}\ \emph {et~al.}(2021)\citenamefont {Chantasri}, \citenamefont {Guevara}, \citenamefont {Laverick},\ and\ \citenamefont {Wiseman}}]{CGLW21}%
  \BibitemOpen
  \bibfield  {author} {\bibinfo {author} {\bibfnamefont {A.}~\bibnamefont {Chantasri}}, \bibinfo {author} {\bibfnamefont {I.}~\bibnamefont {Guevara}}, \bibinfo {author} {\bibfnamefont {K.~T.}\ \bibnamefont {Laverick}},\ and\ \bibinfo {author} {\bibfnamefont {H.~M.}\ \bibnamefont {Wiseman}},\ }\href {https://doi.org/10.1016/j.physrep.2021.07.003} {\bibfield  {journal} {\bibinfo  {journal} {Phys. Rep.}\ }\textbf {\bibinfo {volume} {930}},\ \bibinfo {pages} {1} (\bibinfo {year} {2021})}\BibitemShut {NoStop}%
\bibitem [{\citenamefont {Laverick}\ \emph {et~al.}(2021{\natexlab{b}})\citenamefont {Laverick}, \citenamefont {Guevara},\ and\ \citenamefont {Wiseman}}]{LGW21}%
  \BibitemOpen
  \bibfield  {author} {\bibinfo {author} {\bibfnamefont {K.~T.}\ \bibnamefont {Laverick}}, \bibinfo {author} {\bibfnamefont {I.}~\bibnamefont {Guevara}},\ and\ \bibinfo {author} {\bibfnamefont {H.~M.}\ \bibnamefont {Wiseman}},\ }\href {https://doi.org/10.1103/PhysRevA.104.032213} {\bibfield  {journal} {\bibinfo  {journal} {Phys. Rev. A}\ }\textbf {\bibinfo {volume} {104}},\ \bibinfo {pages} {032213} (\bibinfo {year} {2021}{\natexlab{b}})}\BibitemShut {NoStop}%
\bibitem [{\citenamefont {Gisin}\ and\ \citenamefont {Percival}(1992{\natexlab{a}})}]{gisin1992quantum}%
  \BibitemOpen
  \bibfield  {author} {\bibinfo {author} {\bibfnamefont {N.}~\bibnamefont {Gisin}}\ and\ \bibinfo {author} {\bibfnamefont {I.~C.}\ \bibnamefont {Percival}},\ }\href {https://doi.org/10.1088/0305-4470/25/21/023} {\bibfield  {journal} {\bibinfo  {journal} {J. Phys. A Math. Gen.}\ }\textbf {\bibinfo {volume} {25}},\ \bibinfo {pages} {5677} (\bibinfo {year} {1992}{\natexlab{a}})}\BibitemShut {NoStop}%
\bibitem [{\citenamefont {Gisin}\ and\ \citenamefont {Percival}(1992{\natexlab{b}})}]{gisin1992wave}%
  \BibitemOpen
  \bibfield  {author} {\bibinfo {author} {\bibfnamefont {N.}~\bibnamefont {Gisin}}\ and\ \bibinfo {author} {\bibfnamefont {I.~C.}\ \bibnamefont {Percival}},\ }\href {https://doi.org/10.1016/0375-9601(92)90264-M} {\bibfield  {journal} {\bibinfo  {journal} {Phys. Lett. A}\ }\textbf {\bibinfo {volume} {167}},\ \bibinfo {pages} {315} (\bibinfo {year} {1992}{\natexlab{b}})}\BibitemShut {NoStop}%
\bibitem [{\citenamefont {Wiseman}\ and\ \citenamefont {Di{\'o}si}(2001)}]{WisDio01}%
  \BibitemOpen
  \bibfield  {author} {\bibinfo {author} {\bibfnamefont {H.~M.}\ \bibnamefont {Wiseman}}\ and\ \bibinfo {author} {\bibfnamefont {L.}~\bibnamefont {Di{\'o}si}},\ }\href {https://doi.org/10.1016/S0301-0104(01)00296-8} {\bibfield  {journal} {\bibinfo  {journal} {Chem. Phys.}\ }\textbf {\bibinfo {volume} {268}},\ \bibinfo {pages} {91} (\bibinfo {year} {2001})},\ \bibinfo {note} {erratum {\em ibid.} {\bf 271}, 227 (2001)}\BibitemShut {NoStop}%
\bibitem [{\citenamefont {Fraser}\ and\ \citenamefont {Potter}(1969)}]{fraser1969optimum}%
  \BibitemOpen
  \bibfield  {author} {\bibinfo {author} {\bibfnamefont {D.}~\bibnamefont {Fraser}}\ and\ \bibinfo {author} {\bibfnamefont {J.}~\bibnamefont {Potter}},\ }\href {https://doi.org/10.1109/TAC.1969.1099196} {\bibfield  {journal} {\bibinfo  {journal} {IEEE Trans. Automat. Contr.}\ }\textbf {\bibinfo {volume} {14}},\ \bibinfo {pages} {387} (\bibinfo {year} {1969})}\BibitemShut {NoStop}%
\bibitem [{\citenamefont {Laverick}\ \emph {et~al.}(2025)\citenamefont {Laverick}, \citenamefont {Chantasri},\ and\ \citenamefont {Wiseman}}]{LCW25}%
  \BibitemOpen
  \bibfield  {author} {\bibinfo {author} {\bibfnamefont {K.~T.}\ \bibnamefont {Laverick}}, \bibinfo {author} {\bibfnamefont {A.}~\bibnamefont {Chantasri}},\ and\ \bibinfo {author} {\bibfnamefont {H.~M.}\ \bibnamefont {Wiseman}},\ }\href {https://doi.org/10.1103/j71g-pnmb} {\bibfield  {journal} {\bibinfo  {journal} {Phys. Rev. A}\ }\textbf {\bibinfo {volume} {112}},\ \bibinfo {pages} {022411} (\bibinfo {year} {2025})}\BibitemShut {NoStop}%
\bibitem [{\citenamefont {Tsang}(2009)}]{tsang2009optimal}%
  \BibitemOpen
  \bibfield  {author} {\bibinfo {author} {\bibfnamefont {M.}~\bibnamefont {Tsang}},\ }\href {https://doi.org/10.1103/PhysRevA.80.033840} {\bibfield  {journal} {\bibinfo  {journal} {Phys. Rev. A}\ }\textbf {\bibinfo {volume} {80}},\ \bibinfo {pages} {033840} (\bibinfo {year} {2009})}\BibitemShut {NoStop}%
\bibitem [{\citenamefont {Brawley}\ \emph {et~al.}(2016)\citenamefont {Brawley}, \citenamefont {Vanner}, \citenamefont {Larsen}, \citenamefont {Schmid}, \citenamefont {Boisen},\ and\ \citenamefont {Bowen}}]{brawley2016nonlinear}%
  \BibitemOpen
  \bibfield  {author} {\bibinfo {author} {\bibfnamefont {G.~A.}\ \bibnamefont {Brawley}}, \bibinfo {author} {\bibfnamefont {M.~R.}\ \bibnamefont {Vanner}}, \bibinfo {author} {\bibfnamefont {P.~E.}\ \bibnamefont {Larsen}}, \bibinfo {author} {\bibfnamefont {S.}~\bibnamefont {Schmid}}, \bibinfo {author} {\bibfnamefont {A.}~\bibnamefont {Boisen}},\ and\ \bibinfo {author} {\bibfnamefont {W.~P.}\ \bibnamefont {Bowen}},\ }\href {https://doi.org/10.1038/ncomms10988} {\bibfield  {journal} {\bibinfo  {journal} {Nat. Commun.}\ }\textbf {\bibinfo {volume} {7}},\ \bibinfo {pages} {10988} (\bibinfo {year} {2016})}\BibitemShut {NoStop}%
\bibitem [{\citenamefont {Deleglise}\ \emph {et~al.}(2008)\citenamefont {Deleglise}, \citenamefont {Dotsenko}, \citenamefont {Sayrin}, \citenamefont {Bernu}, \citenamefont {Brune}, \citenamefont {Raimond},\ and\ \citenamefont {Haroche}}]{deleglise2008reconstruction}%
  \BibitemOpen
  \bibfield  {author} {\bibinfo {author} {\bibfnamefont {S.}~\bibnamefont {Deleglise}}, \bibinfo {author} {\bibfnamefont {I.}~\bibnamefont {Dotsenko}}, \bibinfo {author} {\bibfnamefont {C.}~\bibnamefont {Sayrin}}, \bibinfo {author} {\bibfnamefont {J.}~\bibnamefont {Bernu}}, \bibinfo {author} {\bibfnamefont {M.}~\bibnamefont {Brune}}, \bibinfo {author} {\bibfnamefont {J.-M.}\ \bibnamefont {Raimond}},\ and\ \bibinfo {author} {\bibfnamefont {S.}~\bibnamefont {Haroche}},\ }\href {https://doi.org/10.1038/nature07288} {\bibfield  {journal} {\bibinfo  {journal} {Nature}\ }\textbf {\bibinfo {volume} {455}},\ \bibinfo {pages} {510} (\bibinfo {year} {2008})}\BibitemShut {NoStop}%
\bibitem [{\citenamefont {Khademi}(2024)}]{khademi2024optomechanical}%
  \BibitemOpen
  \bibfield  {author} {\bibinfo {author} {\bibfnamefont {S.}~\bibnamefont {Khademi}},\ }\emph {\bibinfo {title} {Optomechanical monitoring of quantum Brownian motion and the challenge of Heisenberg}},\ \href {https://doi.org/10.14264/193275c} {Ph.D. thesis},\ \bibinfo  {school} {The University of Queensland} (\bibinfo {year} {2024})\BibitemShut {NoStop}%
\bibitem [{\citenamefont {Doherty}\ and\ \citenamefont {Jacobs}(1999)}]{DohJac99}%
  \BibitemOpen
  \bibfield  {author} {\bibinfo {author} {\bibfnamefont {A.~C.}\ \bibnamefont {Doherty}}\ and\ \bibinfo {author} {\bibfnamefont {K.}~\bibnamefont {Jacobs}},\ }\href {https://doi.org/10.1103/PhysRevA.60.2700} {\bibfield  {journal} {\bibinfo  {journal} {Phys. Rev. A}\ }\textbf {\bibinfo {volume} {60}},\ \bibinfo {pages} {2700} (\bibinfo {year} {1999})}\BibitemShut {NoStop}%
\bibitem [{\citenamefont {Kalman}\ and\ \citenamefont {Bucy}(1961)}]{KalBuc61}%
  \BibitemOpen
  \bibfield  {author} {\bibinfo {author} {\bibfnamefont {R.~E.}\ \bibnamefont {Kalman}}\ and\ \bibinfo {author} {\bibfnamefont {R.~S.}\ \bibnamefont {Bucy}},\ }\href {https://doi.org/10.1115/1.3658902} {\bibfield  {journal} {\bibinfo  {journal} {J. Basic Eng.}\ }\textbf {\bibinfo {volume} {83}},\ \bibinfo {pages} {95} (\bibinfo {year} {1961})}\BibitemShut {NoStop}%
\bibitem [{Note1()}]{Note1}%
  \BibitemOpen
  \bibinfo {note} {The definitions of the past and future records here differ slightly from those in Ref.~\cite {guevara2015quantum} and those that follow: here, time $t$ is included in the past record rather than in the future record. This difference, however, does not affect the conclusions or final expressions.}\BibitemShut {Stop}%
\bibitem [{\citenamefont {Zhang}\ and\ \citenamefont {M\o{}lmer}(2017)}]{ZhaMol17}%
  \BibitemOpen
  \bibfield  {author} {\bibinfo {author} {\bibfnamefont {J.}~\bibnamefont {Zhang}}\ and\ \bibinfo {author} {\bibfnamefont {K.}~\bibnamefont {M\o{}lmer}},\ }\href {https://doi.org/10.1103/PhysRevA.96.062131} {\bibfield  {journal} {\bibinfo  {journal} {Phys. Rev. A}\ }\textbf {\bibinfo {volume} {96}},\ \bibinfo {pages} {062131} (\bibinfo {year} {2017})}\BibitemShut {NoStop}%
\bibitem [{\citenamefont {Laverick}\ \emph {et~al.}(2023)\citenamefont {Laverick}, \citenamefont {Warszawski}, \citenamefont {Chantasri},\ and\ \citenamefont {Wiseman}}]{LWCW23}%
  \BibitemOpen
  \bibfield  {author} {\bibinfo {author} {\bibfnamefont {K.~T.}\ \bibnamefont {Laverick}}, \bibinfo {author} {\bibfnamefont {P.}~\bibnamefont {Warszawski}}, \bibinfo {author} {\bibfnamefont {A.}~\bibnamefont {Chantasri}},\ and\ \bibinfo {author} {\bibfnamefont {H.~M.}\ \bibnamefont {Wiseman}},\ }\href {https://doi.org/10.1103/PRXQuantum.4.040340} {\bibfield  {journal} {\bibinfo  {journal} {PRX Quantum}\ }\textbf {\bibinfo {volume} {4}},\ \bibinfo {pages} {040340} (\bibinfo {year} {2023})}\BibitemShut {NoStop}%
\bibitem [{\citenamefont {Guo}(2021)}]{guo2021bringing}%
  \BibitemOpen
  \bibfield  {author} {\bibinfo {author} {\bibfnamefont {J.}~\bibnamefont {Guo}},\ }\emph {\bibinfo {title} {Bringing Classical Mechanical Resonators towards the Quantum Regime}},\ \href {https://doi.org/10.4233/uuid:3d240b0f-cf5b-4a63-86a1-c67bc24fe9b5} {Ph.D. thesis},\ \bibinfo  {school} {Delft University of Technology} (\bibinfo {year} {2021})\BibitemShut {NoStop}%
\bibitem [{\citenamefont {Tiecke}\ \emph {et~al.}(2015)\citenamefont {Tiecke}, \citenamefont {Nayak}, \citenamefont {Thompson}, \citenamefont {Peyronel}, \citenamefont {de~Leon}, \citenamefont {Vuleti{\'c}},\ and\ \citenamefont {Lukin}}]{tiecke2015efficient}%
  \BibitemOpen
  \bibfield  {author} {\bibinfo {author} {\bibfnamefont {T.~G.}\ \bibnamefont {Tiecke}}, \bibinfo {author} {\bibfnamefont {K.~P.}\ \bibnamefont {Nayak}}, \bibinfo {author} {\bibfnamefont {J.~D.}\ \bibnamefont {Thompson}}, \bibinfo {author} {\bibfnamefont {T.}~\bibnamefont {Peyronel}}, \bibinfo {author} {\bibfnamefont {N.~P.}\ \bibnamefont {de~Leon}}, \bibinfo {author} {\bibfnamefont {V.}~\bibnamefont {Vuleti{\'c}}},\ and\ \bibinfo {author} {\bibfnamefont {M.~D.}\ \bibnamefont {Lukin}},\ }\href {https://doi.org/10.1364/OPTICA.2.000070} {\bibfield  {journal} {\bibinfo  {journal} {Optica}\ }\textbf {\bibinfo {volume} {2}},\ \bibinfo {pages} {70} (\bibinfo {year} {2015})}\BibitemShut {NoStop}%
\bibitem [{\citenamefont {Meenehan}\ \emph {et~al.}(2015)\citenamefont {Meenehan}, \citenamefont {Cohen}, \citenamefont {MacCabe}, \citenamefont {Marsili}, \citenamefont {Shaw},\ and\ \citenamefont {Painter}}]{Meenehan2015pulsed}%
  \BibitemOpen
  \bibfield  {author} {\bibinfo {author} {\bibfnamefont {S.~M.}\ \bibnamefont {Meenehan}}, \bibinfo {author} {\bibfnamefont {J.~D.}\ \bibnamefont {Cohen}}, \bibinfo {author} {\bibfnamefont {G.~S.}\ \bibnamefont {MacCabe}}, \bibinfo {author} {\bibfnamefont {F.}~\bibnamefont {Marsili}}, \bibinfo {author} {\bibfnamefont {M.~D.}\ \bibnamefont {Shaw}},\ and\ \bibinfo {author} {\bibfnamefont {O.}~\bibnamefont {Painter}},\ }\href {https://doi.org/10.1103/PhysRevX.5.041002} {\bibfield  {journal} {\bibinfo  {journal} {Phys. Rev. X}\ }\textbf {\bibinfo {volume} {5}},\ \bibinfo {pages} {041002} (\bibinfo {year} {2015})}\BibitemShut {NoStop}%
\bibitem [{\citenamefont {Pautrel}\ \emph {et~al.}(2024)\citenamefont {Pautrel}, \citenamefont {Malabat}, \citenamefont {Waquier}, \citenamefont {Colombano}, \citenamefont {Morassi}, \citenamefont {Lemaître},\ and\ \citenamefont {Favero}}]{Pautrel2024efficient}%
  \BibitemOpen
  \bibfield  {author} {\bibinfo {author} {\bibfnamefont {S.}~\bibnamefont {Pautrel}}, \bibinfo {author} {\bibfnamefont {F.}~\bibnamefont {Malabat}}, \bibinfo {author} {\bibfnamefont {L.}~\bibnamefont {Waquier}}, \bibinfo {author} {\bibfnamefont {M.}~\bibnamefont {Colombano}}, \bibinfo {author} {\bibfnamefont {M.}~\bibnamefont {Morassi}}, \bibinfo {author} {\bibfnamefont {A.}~\bibnamefont {Lemaître}},\ and\ \bibinfo {author} {\bibfnamefont {I.}~\bibnamefont {Favero}},\ }\href {https://doi.org/10.1364/OE.521697} {\bibfield  {journal} {\bibinfo  {journal} {Opt. Express}\ }\textbf {\bibinfo {volume} {32}},\ \bibinfo {pages} {26954} (\bibinfo {year} {2024})}\BibitemShut {NoStop}%
\bibitem [{\citenamefont {Bhaskar}\ \emph {et~al.}(2020)\citenamefont {Bhaskar}, \citenamefont {Riedinger}, \citenamefont {Machielse}, \citenamefont {Levonian}, \citenamefont {Nguyen}, \citenamefont {Knall}, \citenamefont {Park}, \citenamefont {Englund}, \citenamefont {Lončar}, \citenamefont {Sukachev},\ and\ \citenamefont {Lukin}}]{Bhaskar2020experimental}%
  \BibitemOpen
  \bibfield  {author} {\bibinfo {author} {\bibfnamefont {M.~K.}\ \bibnamefont {Bhaskar}}, \bibinfo {author} {\bibfnamefont {R.}~\bibnamefont {Riedinger}}, \bibinfo {author} {\bibfnamefont {B.}~\bibnamefont {Machielse}}, \bibinfo {author} {\bibfnamefont {D.~S.}\ \bibnamefont {Levonian}}, \bibinfo {author} {\bibfnamefont {C.~T.}\ \bibnamefont {Nguyen}}, \bibinfo {author} {\bibfnamefont {E.~N.}\ \bibnamefont {Knall}}, \bibinfo {author} {\bibfnamefont {H.}~\bibnamefont {Park}}, \bibinfo {author} {\bibfnamefont {D.}~\bibnamefont {Englund}}, \bibinfo {author} {\bibfnamefont {M.}~\bibnamefont {Lončar}}, \bibinfo {author} {\bibfnamefont {D.~D.}\ \bibnamefont {Sukachev}},\ and\ \bibinfo {author} {\bibfnamefont {M.~D.}\ \bibnamefont {Lukin}},\ }\href {https://doi.org/10.1038/s41586-020-2103-5} {\bibfield  {journal} {\bibinfo  {journal} {Nature}\ }\textbf {\bibinfo {volume} {580}},\ \bibinfo {pages} {60} (\bibinfo {year} {2020})}\BibitemShut {NoStop}%
\bibitem [{\citenamefont {Usami}\ \emph {et~al.}(2012)\citenamefont {Usami}, \citenamefont {Naesby}, \citenamefont {Bagci}, \citenamefont {Melholt~Nielsen}, \citenamefont {Liu}, \citenamefont {Stobbe}, \citenamefont {Lodahl},\ and\ \citenamefont {Polzik}}]{usami2012optical}%
  \BibitemOpen
  \bibfield  {author} {\bibinfo {author} {\bibfnamefont {K.}~\bibnamefont {Usami}}, \bibinfo {author} {\bibfnamefont {A.}~\bibnamefont {Naesby}}, \bibinfo {author} {\bibfnamefont {T.}~\bibnamefont {Bagci}}, \bibinfo {author} {\bibfnamefont {B.}~\bibnamefont {Melholt~Nielsen}}, \bibinfo {author} {\bibfnamefont {J.}~\bibnamefont {Liu}}, \bibinfo {author} {\bibfnamefont {S.}~\bibnamefont {Stobbe}}, \bibinfo {author} {\bibfnamefont {P.}~\bibnamefont {Lodahl}},\ and\ \bibinfo {author} {\bibfnamefont {E.~S.}\ \bibnamefont {Polzik}},\ }\href {https://doi.org/10.1038/nphys2196} {\bibfield  {journal} {\bibinfo  {journal} {Nat. Phys.}\ }\textbf {\bibinfo {volume} {8}},\ \bibinfo {pages} {168} (\bibinfo {year} {2012})}\BibitemShut {NoStop}%
\bibitem [{\citenamefont {Hauer}\ \emph {et~al.}(2019)\citenamefont {Hauer}, \citenamefont {Clark}, \citenamefont {Kim}, \citenamefont {Doolin},\ and\ \citenamefont {Davis}}]{hauer2019dueling}%
  \BibitemOpen
  \bibfield  {author} {\bibinfo {author} {\bibfnamefont {B.~D.}\ \bibnamefont {Hauer}}, \bibinfo {author} {\bibfnamefont {T.~J.}\ \bibnamefont {Clark}}, \bibinfo {author} {\bibfnamefont {P.~H.}\ \bibnamefont {Kim}}, \bibinfo {author} {\bibfnamefont {C.}~\bibnamefont {Doolin}},\ and\ \bibinfo {author} {\bibfnamefont {J.~P.}\ \bibnamefont {Davis}},\ }\href {https://doi.org/10.1103/PhysRevA.99.053803} {\bibfield  {journal} {\bibinfo  {journal} {Phys. Rev. A}\ }\textbf {\bibinfo {volume} {99}},\ \bibinfo {pages} {053803} (\bibinfo {year} {2019})}\BibitemShut {NoStop}%
\bibitem [{\citenamefont {Ftouni}\ \emph {et~al.}(2015)\citenamefont {Ftouni}, \citenamefont {Blanc}, \citenamefont {Tainoff}, \citenamefont {Fefferman}, \citenamefont {Defoort}, \citenamefont {Lulla}, \citenamefont {Richard}, \citenamefont {Collin},\ and\ \citenamefont {Bourgeois}}]{ftouni2015thermal}%
  \BibitemOpen
  \bibfield  {author} {\bibinfo {author} {\bibfnamefont {H.}~\bibnamefont {Ftouni}}, \bibinfo {author} {\bibfnamefont {C.}~\bibnamefont {Blanc}}, \bibinfo {author} {\bibfnamefont {D.}~\bibnamefont {Tainoff}}, \bibinfo {author} {\bibfnamefont {A.~D.}\ \bibnamefont {Fefferman}}, \bibinfo {author} {\bibfnamefont {M.}~\bibnamefont {Defoort}}, \bibinfo {author} {\bibfnamefont {K.~J.}\ \bibnamefont {Lulla}}, \bibinfo {author} {\bibfnamefont {J.}~\bibnamefont {Richard}}, \bibinfo {author} {\bibfnamefont {E.}~\bibnamefont {Collin}},\ and\ \bibinfo {author} {\bibfnamefont {O.}~\bibnamefont {Bourgeois}},\ }\href {https://doi.org/10.1103/PhysRevB.92.125439} {\bibfield  {journal} {\bibinfo  {journal} {Phys. Rev. B}\ }\textbf {\bibinfo {volume} {92}},\ \bibinfo {pages} {125439} (\bibinfo {year} {2015})}\BibitemShut {NoStop}%
\bibitem [{\citenamefont {Leijssen}\ \emph {et~al.}(2017)\citenamefont {Leijssen}, \citenamefont {La~Gala}, \citenamefont {Freisem}, \citenamefont {Muhonen},\ and\ \citenamefont {Verhagen}}]{leijssen2017nonlinear}%
  \BibitemOpen
  \bibfield  {author} {\bibinfo {author} {\bibfnamefont {R.}~\bibnamefont {Leijssen}}, \bibinfo {author} {\bibfnamefont {G.~R.}\ \bibnamefont {La~Gala}}, \bibinfo {author} {\bibfnamefont {L.}~\bibnamefont {Freisem}}, \bibinfo {author} {\bibfnamefont {J.~T.}\ \bibnamefont {Muhonen}},\ and\ \bibinfo {author} {\bibfnamefont {E.}~\bibnamefont {Verhagen}},\ }\href {https://doi.org/10.1038/ncomms16024} {\bibfield  {journal} {\bibinfo  {journal} {Nat. Commun.}\ }\textbf {\bibinfo {volume} {8}},\ \bibinfo {pages} {16024} (\bibinfo {year} {2017})}\BibitemShut {NoStop}%
\bibitem [{\citenamefont {Grewal}\ and\ \citenamefont {Andrews}(2014)}]{Grewal2014Kalman}%
  \BibitemOpen
  \bibfield  {author} {\bibinfo {author} {\bibfnamefont {M.~S.}\ \bibnamefont {Grewal}}\ and\ \bibinfo {author} {\bibfnamefont {A.~P.}\ \bibnamefont {Andrews}},\ }\bibinfo {title} {Optimal smoothers},\ in\ \href {https://doi.org/https://doi.org/10.1002/9781118984987.ch6} {\emph {\bibinfo {booktitle} {Kalman Filtering}}}\ (\bibinfo  {publisher} {John Wiley \& Sons, Ltd},\ \bibinfo {year} {2014})\ Chap.~\bibinfo {chapter} {6}, pp.\ \bibinfo {pages} {239--279}\BibitemShut {NoStop}%
\bibitem [{\citenamefont {Bayley}\ and\ \citenamefont {Hammersley}(1946)}]{bayley1946effective}%
  \BibitemOpen
  \bibfield  {author} {\bibinfo {author} {\bibfnamefont {G.~V.}\ \bibnamefont {Bayley}}\ and\ \bibinfo {author} {\bibfnamefont {J.~M.}\ \bibnamefont {Hammersley}},\ }\href {https://doi.org/10.2307/2983560} {\bibfield  {journal} {\bibinfo  {journal} {J. R. Stat. Soc. Suppl.}\ }\textbf {\bibinfo {volume} {8}},\ \bibinfo {pages} {184} (\bibinfo {year} {1946})}\BibitemShut {NoStop}%
\bibitem [{\citenamefont {Hofer}\ \emph {et~al.}(2013)\citenamefont {Hofer}, \citenamefont {Vasilyev}, \citenamefont {Aspelmeyer},\ and\ \citenamefont {Hammerer}}]{hofer2013time}%
  \BibitemOpen
  \bibfield  {author} {\bibinfo {author} {\bibfnamefont {S.~G.}\ \bibnamefont {Hofer}}, \bibinfo {author} {\bibfnamefont {D.~V.}\ \bibnamefont {Vasilyev}}, \bibinfo {author} {\bibfnamefont {M.}~\bibnamefont {Aspelmeyer}},\ and\ \bibinfo {author} {\bibfnamefont {K.}~\bibnamefont {Hammerer}},\ }\href {https://doi.org/https://doi.org/10.1103/PhysRevLett.111.170404} {\bibfield  {journal} {\bibinfo  {journal} {Phys. Rev. Lett.}\ }\textbf {\bibinfo {volume} {111}},\ \bibinfo {pages} {170404} (\bibinfo {year} {2013})}\BibitemShut {NoStop}%
\bibitem [{\citenamefont {Chen}\ \emph {et~al.}(2020)\citenamefont {Chen}, \citenamefont {Rossi}, \citenamefont {Mason},\ and\ \citenamefont {Schliesser}}]{chen2020entanglement}%
  \BibitemOpen
  \bibfield  {author} {\bibinfo {author} {\bibfnamefont {J.}~\bibnamefont {Chen}}, \bibinfo {author} {\bibfnamefont {M.}~\bibnamefont {Rossi}}, \bibinfo {author} {\bibfnamefont {D.}~\bibnamefont {Mason}},\ and\ \bibinfo {author} {\bibfnamefont {A.}~\bibnamefont {Schliesser}},\ }\href {https://doi.org/https://doi.org/10.1038/s41467-020-14768-1} {\bibfield  {journal} {\bibinfo  {journal} {Nat. Commun.}\ }\textbf {\bibinfo {volume} {11}},\ \bibinfo {pages} {943} (\bibinfo {year} {2020})}\BibitemShut {NoStop}%
\bibitem [{\citenamefont {Harris}\ \emph {et~al.}(2013)\citenamefont {Harris}, \citenamefont {McAuslan}, \citenamefont {Stace}, \citenamefont {Doherty},\ and\ \citenamefont {Bowen}}]{harris2013minimum}%
  \BibitemOpen
  \bibfield  {author} {\bibinfo {author} {\bibfnamefont {G.~I.}\ \bibnamefont {Harris}}, \bibinfo {author} {\bibfnamefont {D.~L.}\ \bibnamefont {McAuslan}}, \bibinfo {author} {\bibfnamefont {T.~M.}\ \bibnamefont {Stace}}, \bibinfo {author} {\bibfnamefont {A.~C.}\ \bibnamefont {Doherty}},\ and\ \bibinfo {author} {\bibfnamefont {W.~P.}\ \bibnamefont {Bowen}},\ }\href {https://doi.org/https://doi.org/10.1103/PhysRevLett.111.103603} {\bibfield  {journal} {\bibinfo  {journal} {Phys. Rev. Lett.}\ }\textbf {\bibinfo {volume} {111}},\ \bibinfo {pages} {103603} (\bibinfo {year} {2013})}\BibitemShut {NoStop}%
\end{thebibliography}%
}{
}

\end{document}

\onecolumngrid
\newpage
\section*{Supplementary Information for\\Post-processed estimation of quantum state trajectories}
\begin{center}
  Soroush Khademi, Jesse J. Slim, Kiarn T. Laverick, Jin Chang, Jingkun Guo,\\Simon Gröblacher, Howard M. Wiseman, and Warwick P. Bowen
\end{center}
\renewcommand{\thesection}{S\arabic{section}}\setcounter{section}{0}
\renewcommand{\thesubsection}{S\arabic{section}.\arabic{subsection}}
\renewcommand{\theequation}{S\arabic{equation}}\setcounter{equation}{0}
\renewcommand{\figurename}{FIG.}
\renewcommand{\thefigure}{S\arabic{figure}}\setcounter{figure}{0}
\makeatletter
\renewcommand{\theHfigure}{S.\arabic{figure}}
\makeatother
\renewcommand{\tablename}{TABLE}
\renewcommand{\thetable}{S\arabic{table}}\setcounter{table}{0}

\title{Supplementary Information for \\ Post-processed estimation of quantum state trajectories}

\author{Soroush Khademi}

\author{Jesse J. Slim}

\author{Kiarn T. Laverick}

\author{Jin Chang}

\author{Jingkun Guo}

\author{Simon Gröblacher}

\author{Howard M. Wiseman}

\author{Warwick P. Bowen}

\maketitle

\section{Sample traces of \texorpdfstring{$\langle\hat{X}_1\rangle$}{<X_1>}}\label{sec:DataForX1}

The values of $\langle\hat{X}_1\rangle$, for sample measurement records, are shown in FIG.~\ref{SuppFig:X1&errorbar}\textbf{a} (main experiment) and FIG.~\ref{SuppFig:X1&errorbar}\textbf{c} (noise-injection experiment). They are qualitatively similar to the sample traces of $\langle\hat{X}_2\rangle$ shown in FIG.~\ref{FigData1}\textbf{b} and FIG.~\ref{FigData3}\textbf{b}, as $\hat{X}_1$ and $\hat{X}_2$ have identical statistics due to dynamical symmetry. 

\section{Extended data for $\delta_\text{C}$}

\begin{figure*}[t!]
    \centering
    \includegraphics[]{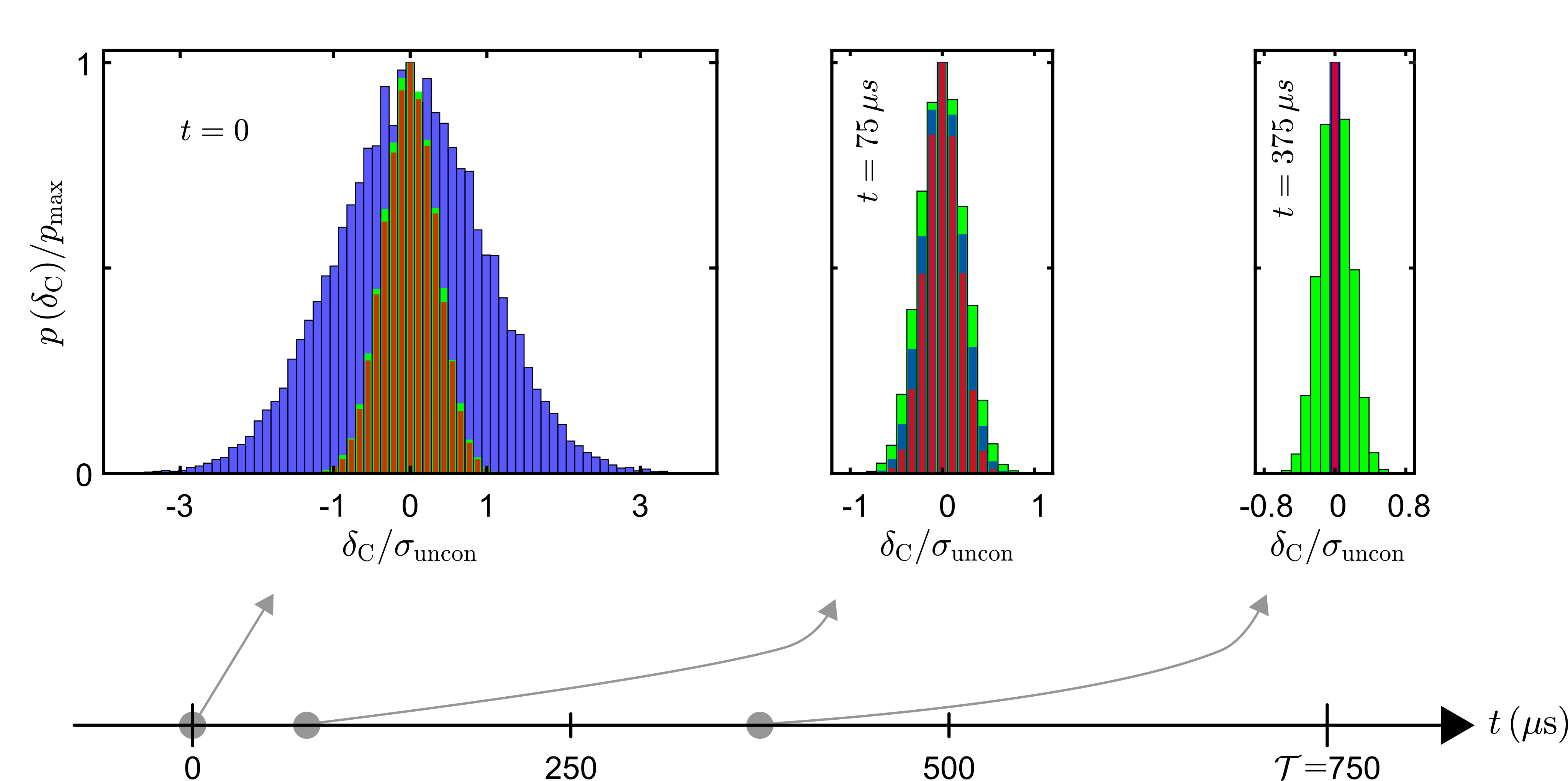}
    \caption{\textbf{Distributions of} $\delta_\text{C}(t)=\langle\hat{X}_j\rangle_\text{F}^\text{LTL}(t)-\langle\hat{X}_j\rangle_\text{C}(t),\;j=1,2$\textbf{.} With the main experiment data, the blue histograms are for C = F (filtering), the red ones for C = S (quantum smoothing) and the green ones for C = cS (classical smoothing). $p(\delta_\text{C})$ is the probability distribution of the realised $\delta_\text{C}$ values and $p_\text{max}$ is the maximum bar height of the corresponding histogram.}
    \label{SuppFig:delta_C}
\end{figure*}

The distributions of $\delta_\text{C}(t)=\langle\hat{X}_j\rangle_\text{F}^\text{LTL}(t)-\langle\hat{X}_j\rangle_\text{C}(t),\;j=1,2$ for three sample times are shown in FIG.~\ref{SuppFig:delta_C}. They all have zero means. That is, filtering (C = F) and quantum state smoothing (C = S) are unbiased estimations of the target mean values. Interestingly, the classical smoothing (C = cS) distributions are also unbiased. The standard deviations of all distributions decrease over time, as also shown by FIG.~\ref{FigData1}\textbf{c}, with the quantum state smoothing ones remaining the smallest.

\section{Inference self-consistency relations}\label{SecConsistency}

In this supplementary section, we examine the equality $\mathbb{V}\text{ar}_{\text{ens}}\big[\langle\hat{X}_j\rangle_\text{C}(t)\big]=\sigma^2_\text{uncon}-v_\text{C}(t)$ for an infinitely large ensemble of measurement records. The effect of the finite size of the experimental ensemble is discussed in the next supplementary section.

For filtering ($\text{C}=\text{F}$), Eq.~\eqref{equ:rFT} gives
\beq
\ex{\hat{X}_j}\fil(t) = e^{-\Gamma t/2}\ex{\hat{X}_j}\fil(0) + \sqrt{2\eta\Gamma {\cal C}}\int_{0}^{t} e^{-\frac{\Gamma}{2}(t - s)} v\fil(s)\dd W_{{\rm F},j}(s)\,.
\eeq
Considering $\mathbb{V}\text{ar}_{\text{ens}}\big[\langle\hat{X}_j\rangle_\text{F}(t)\big]={\mathbb E}_{\text{ens}}\big[\ex{\hat{X}_j}_\text{F}^2(t)\big] - {\mathbb E}^2_{\text{ens}}\big[\ex{\hat{X}_j}\fil(t)\big]$, where ${\mathbb E}_{\text{ens}}[\bullet]$ denotes averaging, ${\mathbb E}_{\text{ens}}\big[\dd W_{{\rm F},j}(s)\big] = 0$ and ${\mathbb E}_{\text{ens}}\big[\dd W_{{\rm F},j}(s)\,\dd W_{{\rm F},j}(s')\big] = \delta(s-s')\,\dd s^2$, we find
\beq
\mathbb{V}\text{ar}_{\text{ens}}\big[\langle\hat{X}_j\rangle_\text{F}(t)\big] = e^{-\Gamma t}\,\mathbb{V}\text{ar}_{\text{ens}}\big[\langle\hat{X}_j\rangle_\text{F}(0)\big] + 2\eta\Gamma \mathcal{C}\int_{0}^{t} \dd s\,e^{-\Gamma (t - s)}v^2_\text{F}(s)\,. 
\eeq
As $\ex{\hat X_j}\fil(0)$ is fixed (at zero), $\mathbb{V}\text{ar}_{\text{ens}}\big[\langle\hat{X}_j\rangle_\text{F}(0)\big] = 0$ and 
\beq
\mathbb{V}\text{ar}_{\text{ens}}\big[\langle\hat{X}_j\rangle_\text{F}(t)\big] = 2\eta\Gamma \mathcal{C}\int_{0}^{t} \dd s\,e^{-\Gamma (t - s)}v^2_\text{F}(s)\,.
\label{eq:intv2}
\eeq
Let us now also consider the quantity $\ell(t) = \sigma^2_\text{uncon} - v\fil(t)$. Considering Eq.~(\ref{NLODE}), we have
\beq
\frac{\dd}{\dd t}\ell(t) = -\Gamma \ell(t) + 2\eta\Gamma \mathcal{C} v^2_\text{F}(t)\,,
\eeq
which has the solution
\beq
\ell(t) = e^{-\Gamma t}\ell(0) + 2\eta\Gamma \mathcal{C} \int_{0}^{t}\dd s\,e^{-\Gamma (t - s)}v^2_\text{F}(s)\,.
\eeq
As $v\fil(0) = \sigma^2_\text{uncon}$, $\ell(0) = 0$ and 
\beq
\ell(t) = 2\eta\Gamma \mathcal{C} \int_{0}^{t}\dd s\,e^{-\Gamma (t - s)}v^2_\text{F}(s)\,.
\label{eq:lt}
\eeq
We can therefore conclude from Eqs.~\eqref{eq:intv2} and \eqref{eq:lt} that
\beq
\mathbb{V}\text{ar}_{\text{ens}}\big[\langle\hat{X}_j\rangle_\text{F}(t)\big] = \ell(t) = \sigma^2_\text{uncon} - v\fil(t)\,.
\label{equ:Inf_Consistency_F}
\eeq

\begin{figure*}[t!]
    \centering
    \includegraphics[]{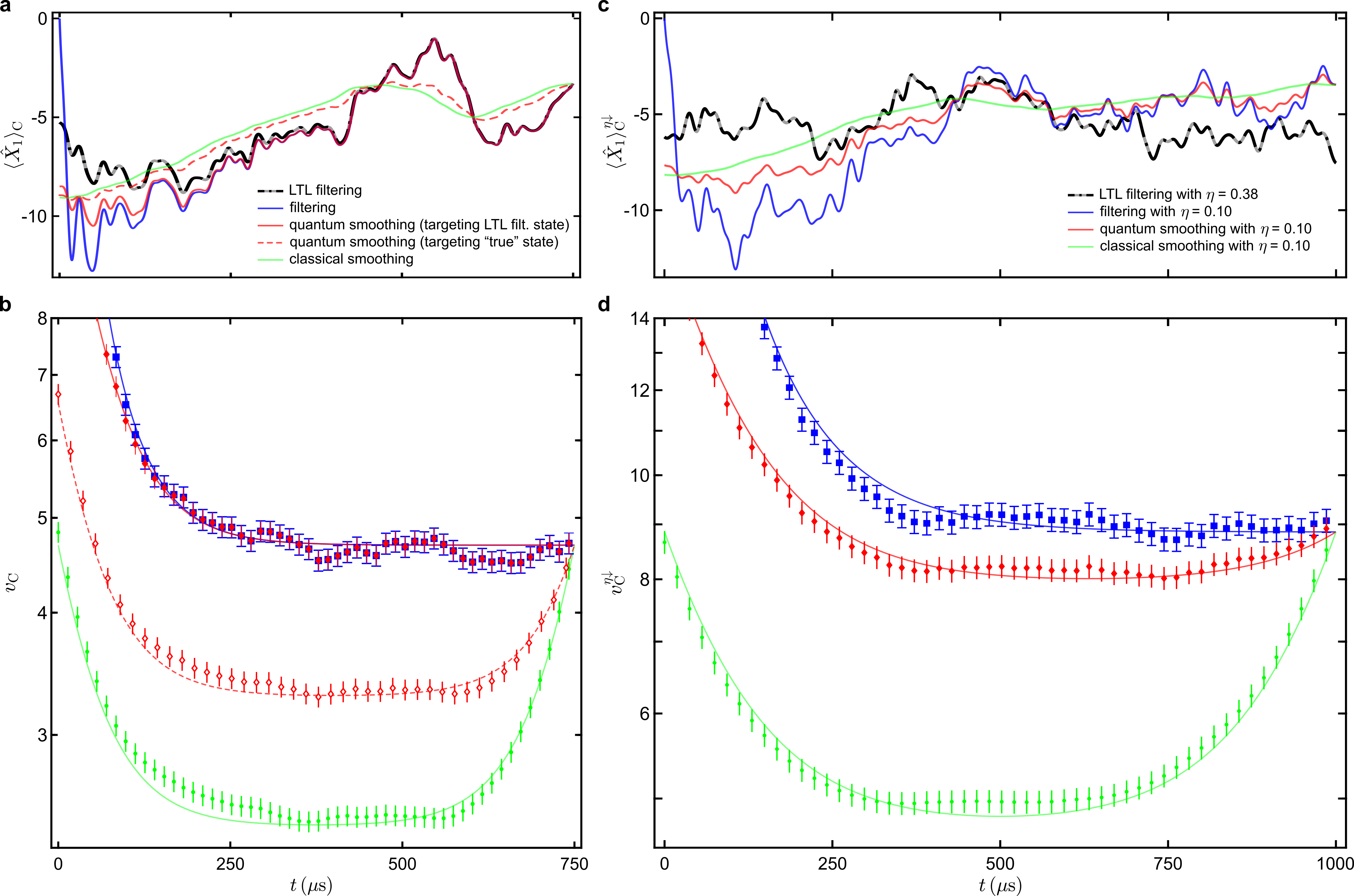}
    \caption{\textbf{Estimation analysis extended data.} \textbf{a-b,} The main experiment data.
    \textbf{a,} Stochastic traces for the quadrature mean $\langle\hat{X}_1\rangle(t)$ for a sample measurement record. \textbf{b,} Deterministic conditional variances ($v_\text{C}(t)$, curves) and the corresponding value of $\sigma^2_\text{uncon}-\mathbb{V}\mathrm{ar}_{\text{ens}}[\langle\hat{X}_j\rangle_\text{C}(t)]$ (data points) with error bars. \textbf{c-d,} The noise-injection ($\eta$-reduced) experiment data.
    \textbf{c,} Stochastic traces for the quadrature mean $\langle\hat{X}_1\rangle(t)$ for a sample measurement record. \textbf{d,} Deterministic conditional variances ($v_\text{C}^{\eta\downarrow}(t)$, curves) and the corresponding value of $\sigma^2_\text{uncon}-\mathbb{V}\mathrm{ar}_{\text{ens}}[\langle\hat{X}_j\rangle_\text{C}^{\eta\downarrow}(t)]$ (data points) with error bars. In \textbf{b} and \textbf{d}, the vertical axes are limited to near steady-state values for visual clarity, and each error bar spans two standard deviations of the error (caused by the finite size of the ensemble of measurement records).}
    \label{SuppFig:X1&errorbar}
\end{figure*}

To examine the case of quantum state smoothing ($\text{C}=\text{S}$), we note that
\begin{equation}
\begin{array}{rcl}
    \mathbb{V}\text{ar}_{\text{ens}}\big[\langle\hat{X}_j\rangle_\text{S}(t)\big] & = & \left(\dfrac{v\sm-v\tar}{v\fil-v\tar}\right)^2\;\mathbb{V}\text{ar}_{\text{ens}}\big[\langle\hat{X}_j\rangle_\text{F}(t)\big]+\\&&\left(\dfrac{v\sm-v\tar}{v\rfil+v\tar}\right)^2\;\mathbb{V}\text{ar}_{\text{ens}}\big[\langle\hat{X}_j\rangle_\text{R}(t)\big]+\\&&
    2\left(\dfrac{v\sm-v\tar}{v\fil-v\tar}\right)\left(\dfrac{v\sm-v\tar}{v\rfil+v\tar}\right)\;
    \mathbb{C}\text{ov}_{\text{ens}}\big[\langle\hat{X}_j\rangle_\text{F}(t),\langle\hat{X}_j\rangle_\text{R}(t)\big]
    \label{equ:VarX_S}
\end{array}
\end{equation}
where $v\tar$ is the variance of the target state. To evaluate this expression, we first investigate the value of $\mathbb{V}\text{ar}_{\text{ens}}\big[\langle\hat{X}_j\rangle_\text{R}(t)\big]$.
Considering the definition of the retrofiltered effect operator, we have $p(\fut{O}_t|\rho_{\rm uncon}) \propto \Tr[\hat E\rfil\,\rho_{\rm uncon}]$. Given that both the effect operator and the unconditional state are Gaussian, and that $\Tr[\rho\,\sigma] = 4\pi\int\dd{\bf x} \,W_{\rho}({\bf x})\,W_{\sigma}({\bf x})$, we find 
\beq
p(\fut{O}_t|\rho_{\rm uncon}) \propto \mathcal{N}(\ex{\bx}\rfil;\ex{\bx}_{\rm uncon}, {\bf V}\rfil + {\bf V}_{\rm uncon})\equiv p(\ex{\bx}\rfil;t)
\eeq
where the final equivalence (after marginalising over records yielding the same $\ex{\bx}\rfil(t)$) comes from the fact that $\ex{\bx}\rfil(t)$ is a deterministic function of $\fut{O}_t$, with the other variables in the Gaussian being independent of the measurement records. The standard probability theory then yields
\beq\label{equ:Inf_Consistency_R}
\mathbb{V}\text{ar}_{\rm ens}\big[\langle\hat{X}_j\rangle_\text{R}(t)\big] = \sigma_{\rm uncon}^2 + v\rfil(t)\,.
\eeq
The next step to evaluate the right-hand side of Eq.~\eqref{equ:VarX_S} is to consider the fact that $\mathbb{V}\text{ar}_{\text{ens}}\big[\langle\hat{X}_j\rangle_\text{F}(t)-\langle\hat{X}_j\rangle_\text{R}(t)\big]=v\fil(t)+v\rfil(t)$, which alongside Eqs.~\eqref{equ:Inf_Consistency_F} and~\eqref{equ:Inf_Consistency_R} gives
\begin{equation}
    \mathbb{C}\text{ov}_{\text{ens}}\big[\langle\hat{X}_j\rangle_\text{F}(t),\langle\hat{X}_j\rangle_\text{R}(t)\big]=\sigma^2_\text{uncon} - v\fil(t)\,.
    \label{equ:Consistency_Cov}
\end{equation}
It is then straightforward to conclude from Eqs.~\eqref{equ:Inf_Consistency_F}, \eqref{equ:Inf_Consistency_R} and \eqref{equ:Consistency_Cov} that
\begin{equation}
    \mathbb{V}\text{ar}_{\text{ens}}\big[\langle\hat{X}_j\rangle_\text{S}(t)\big] = \sigma^2_\text{uncon} - v\sm(t)\,.
\end{equation}

Finally, for classical smoothing ($\text{C}=\text{cS}$), we have
\begin{equation}
\begin{array}{rcl}
    \mathbb{V}\text{ar}_{\text{ens}}\big[\langle\hat{X}_j\rangle_\text{cS}(t)\big] & = & \left(\dfrac{v\csm}{v\fil}\right)^2\;\mathbb{V}\text{ar}_{\text{ens}}\big[\langle\hat{X}_j\rangle_\text{F}(t)\big]+\\&&\left(\dfrac{v\csm}{v\rfil}\right)^2\;\mathbb{V}\text{ar}_{\text{ens}}\big[\langle\hat{X}_j\rangle_\text{R}(t)\big]+\\&&
    2\,\left(\dfrac{v\csm^2}{v\fil\,v\rfil}\right)\;
    \mathbb{C}\text{ov}_{\text{ens}}\big[\langle\hat{X}_j\rangle_\text{F}(t),\langle\hat{X}_j\rangle_\text{R}(t)\big]
    \label{equ:VarX_cS}
\end{array}
\end{equation}
which, considering Eqs.~\eqref{equ:Inf_Consistency_F}, \eqref{equ:Inf_Consistency_R} and \eqref{equ:Consistency_Cov}, results in
\begin{equation}
    \mathbb{V}\text{ar}_{\text{ens}}\big[\langle\hat{X}_j\rangle_\text{cS}(t)\big] = \sigma^2_\text{uncon} - v\csm(t)\,.
\end{equation}

\section{Error bars for the self-consistency tests}\label{sec:ErrorBars_forSC}

There is always a statistical error involved in the inference self-consistency tests because, in practice, the number of available measurement records is finite. As the quantity of interest is the variance of a Gaussian random variable estimated from a finite number of realisations, the standard deviation of this error equals the \textit{standard error of variance} (SEV)~\cite{bayley1946effective}. 

To calculate this, one may use the formula
\begin{equation} \label{sigmaSEV}
    \sigma_\text{SEV,C}(t) = \sqrt{\dfrac{4}{2N_\text{eff}}}\,\times\,\left(\sigma^2_\text{uncon}-\mathbb{V}\mathrm{ar}_{\text{ens}}\left[\langle\hat{X}_j\rangle_\text{C}(t)\right]\right)
\end{equation}
where
\begin{equation}
    N_\text{eff} = N_{\mathrm{ens}}\times\dfrac{1}{1+2\displaystyle\sum_{k=1}^{N_{\mathrm{ens}}-1}\left(1-\frac{k}{N_{\mathrm{ens}}}\right)\left(e^{-(\Gamma_\text{fb}/2)\mathcal{T}k}\right)^2}\,
\end{equation}
is the product of the large size of the ensemble ($N_{\mathrm{ens}}=16653$) and a temporal-correlation factor smaller than one (where $\mathcal{T}=750\,\mu$s) --- the replacement $\Gamma\to\Gamma_\text{fb}$ has been performed in the denominator as the stabilising feedback is on. For the noise-injection experiment, $N_{\mathrm{ens}}=12489$ and $\mathcal{T}=1000\,\mu$s. For our data, the error bars (spanning $2\,\sigma_\text{SEV,C}(t)$) obtained through this method do not include the deterministic values of $v_\text{C}(t)$ about one-third of the time --- see FIG.~\ref{SuppFig:X1&errorbar}\textbf{b} (for the main experiment) and FIG.~\ref{SuppFig:X1&errorbar}\textbf{d} (for the noise-injection experiment). This is a plausible rate.

There is, however, an unexpected output from Eq.~\eqref{sigmaSEV} at time zero for $\text{C}=\text{F}$. The equation yields a nonzero value where there is no error. A numerical approach is therefore employed. We \textit{resample} the ensemble of measurement records --- where each \textit{sample} is a block of 48 consecutive measurement records (about 10 correlation lengths), drawn with replacement --- to make 99 new ensembles having the same size as the original one. We then use a readjusted $n_\text{th}$ for analysing each ensemble. Finally, the error's standard deviation is calculated across the 100 ensembles that we have in hand. This gives us better error bars. They vanish at time zero for $\text{C}=\text{F}$ and do not include the deterministic curves about one-third of the time.

\section{Squared Hilbert--Schmidt distance --- general theory}\label{supp:HSc-Dis-Squ}
In this supplementary section, we theoretically derive the ensemble-averaged value of the estimation cost function that equals
\beq
\overline{\Delta^2_{\rm HS}}[\rho_{\rm tar},\rho_{\rm C}] = \sum_{\rho_{\rm tar}\in\mathbb{T}} \int \dd\both{O} p(\rho_{\rm tar},\both{O})\,{\rm Tr}[(\rho_{\rm tar}-\rho_{\rm C})^2]\,. 
\eeq 

As a straightforward generalisation of a result in Ref.~\cite{CGW19}, we have
\beq
\overline{\Delta^2_{\rm HS}}[\rho_{\rm tar},\rho_{\rm C}] = \sum_{\rho_{\rm tar}\in\mathbb{T}} \int \dd\both{O} p(\rho_{\rm tar},\both{O})\left(\text{P}[\rho_{\rm tar}] - \text{P}[\rho_{\rm C}]\right)\,,\;\;\;\;\text{for\;\,C\,=\,F,\,S}
\label{eq:HS_int_Purity}
\eeq
where $\text{P}[\rho]$ denotes the purity of $\rho$. In LGQ systems, since the purity equals $1/\sqrt{\det[{\bf V}]}$ --- where we have chosen $\hbar$ such that this equality holds --- and  ${\bf V}$ is deterministic with the same ${\bf V}_{\rm tar}$ for $\forall \rho_{\rm tar}\in \mathbb{T}$, Eq.~\eqref{eq:HS_int_Purity} gives
\beq
\overline{\Delta^2_{\rm HS}}[\rho_{\rm tar},\rho_{\rm C}] = 1/\sqrt{\det[{\bf V_{\rm tar}}]} - 1/\sqrt{\det[{\bf V_{\rm C}}]}\,,\;\;\;\;\text{for\;\,C\,=\,F,\,S}.
\eeq
As all covariance matrices are proportional to ${\bf I}$ for our experimental system, this expression is simplified to
\beq
\overline{\Delta^2_{\rm HS}}[\rho_{\rm tar},\rho_{\rm C}] = v_{\rm tar}^{-1} - v_{\rm C}^{-1}\,,\;\;\;\;\text{for\;\,C\,=\,F,\,S}.
\eeq

Doing the calculation for classical smoothing is slightly more complicated. Using $p(\rho_{\rm tar},\both{O}) = p(\rho_{\rm tar}|\both{O})\,p(\both{O})$, expanding the trace term and using the definition of the smoothed quantum state, Eq.~\eqref{eq:gen-sm}, one obtains
\beq
\overline{\Delta^2_{\rm HS}}[\rho_{\rm tar},\rho_{\rm cS}] = \int \dd\both{O} p(\both{O}) \left({\rm P}[\rho_{\rm cS}] - 2\Tr[\rho_{\rm cS}\,\rho\sm]+  \sum_{\rho_{\rm tar}} p(\rho_{\rm tar}|\both{O})\,{\rm P}[\rho_{\rm tar}]\right)\,.
\eeq
Regarding the second term of the integrand, it is known~\cite{wiseman2009quantum} that $\Tr[\rho\,\sigma] = 4\pi\int\dd{\bf x} \,W_{\rho}({\bf x})\,W_{\sigma}({\bf x})$, where $W_{\rho}({\bf x})$ denotes the Wigner function of $\rho$ --- the $4\pi$ factor is necessary for ${\rm P}[\rho]=1/\sqrt{\det[{\bf V}]}$ to hold. Regarding the argument of the summation above, $p(\rho_{\rm tar}|\both{O})\,{\rm P}[\rho_{\rm tar}]$, because the target state is completely described by the mean $\ex{\bx}_{\rm tar}$ and covariance ${\bf V}_{\rm tar}$, the latter of which is deterministic, we have $p(\rho_{\rm tar}|\both{O})\equiv p(\ex{\bx}_{\rm tar}|\both{O})$. Now, since the purity is independent of the measurement records $\both{O}$ and the mean $\ex{\bx}_{\rm tar}$, we have
\beq
\overline{\Delta^2_{\rm HS}}[\rho_{\rm tar},\rho_{\rm cS}] =  1/\sqrt{\det[{\bf V_{\rm cS}}]} + 1/\sqrt{\det[{\bf V_{\rm tar}}]} - 8\pi \int \dd\both{O} p(\both{O}) \int\dd{\bf x}\,W_{\rm cS}({\bf x})\,W_{\sm}({\bf x}) \,.
\label{equ:D2HScal}
\eeq
Focusing on the last term on the right-hand side, we compute the overlap of the two Gaussian Wigner functions and find $\int\dd{\bf x}\,W_{\rm cS}({\bf x})\,W_{\sm}({\bf x}) = \mathcal{N}(\ex{\bx}\sm;\ex{\bx}_{\rm cS},{\bf V}\sm + {\bf V}_{\rm cS})$; the probability of the records can be determined via
\beq
\begin{split}
p(\both{O}) &\propto \Tr[{\hat E}\rfil\,\tilde{\rho}\fil]\\
&\propto \int\dd{\bf x}\,W\fil({\bf x})\,W\rfil({\bf x})\\
&= \mathcal{N}(\ex{\bx}\fil;\ex{\bx}\rfil,{\bf V}\rfil + {\bf V}\fil)\,,\\
\end{split}
\eeq
where $\tilde{\rho}\fil$ is the unnormalised filtered state introduced in Ref.~\cite{guevara2015quantum} and the final equality follows from normalising the distribution. Importantly, we see that this distribution depends only on the difference ${\bf r}_{\rm d} = \ex{\bx}\fil - \ex{\bx}\rfil$. Now, we simplify the difference $\ex{\bx}\csm - \ex{\bx}\sm$ as follows: 
\beq
\begin{split}
\ex{\bx}\csm - \ex{\bx}\sm &= {\bf V}\csm({\bf V}\fil\inv \ex{\bx}\fil + {\bf V}\rfil\inv \ex{\bx}\rfil) - ({\bf V}\sm - {\bf V}\tar)(({\bf V}\fil - {\bf V}\tar)\inv \ex{\bx}\fil + ({\bf V}\rfil + {\bf V}\tar)\inv \ex{\bx}\rfil)\\
& = {\bf Z}\,(\ex{\bx}\rfil - \ex{\bx}\fil)\\
& = -{\bf Z}\,{\bf r}_{\rm d} 
\end{split}\label{eq:csm_sm_diff}
\eeq
where ${\bf Z} = {\bf V}_{\rm cS} {\bf V}\rfil^{-1} - ({\bf V}\sm - {\bf V}\tar)({\bf V}\rfil + {\bf V}\tar)^{-1}$, and we have used ${\bf V}\csm\inv = {\bf V}\fil\inv + {\bf V}\rfil\inv$ and $({\bf V}\sm - {\bf V}\tar)\inv = ({\bf V}\fil - {\bf V}\tar)\inv + ({\bf V}\rfil + {\bf V}\tar)\inv$. Thus, $\mathcal{N}(\ex{\bx}\sm;\ex{\bx}_{\rm cS},{\bf V}\sm + {\bf V}_{\rm cS}) = \mathcal{N}({\bf Z}\,{\bf r}_{\rm d};0,{\bf V}\sm + {\bf V}_{\rm cS})$. The last term on the right-hand side of Eq.~\eqref{equ:D2HScal} then equals
\beq
\begin{split}
-8\pi\int \dd\both{O} p(\both{O}) \int\dd{\bf x}\,W_{\rm cS}({\bf x})\,W_{\sm}({\bf x})&\equiv -8\pi\int \dd{\bf r}_{\rm d}\;\mathcal{N}({\bf r}_{\rm d};0,{\bf V}\rfil + {\bf V}\fil)\;\mathcal{N}({\bf Z}\,{\bf r}_{\rm d};0,{\bf V}\sm + {\bf V}_{\rm cS}) \\
&= -4\,\frac{\sqrt{\det\Big[\left(\,({\bf V}\fil+{\bf V}\rfil)\inv + {\bf Z}\tp ({\bf V}\sm + {\bf V}\csm)\inv {\bf Z}\,\right)\inv\Big]}}{\sqrt{\det[{\bf V}\fil+{\bf V}\rfil]}\,\sqrt{\det[{\bf V}\sm+{\bf V}\csm]}}\,.
\end{split}
\eeq
Therefore
\beq
\overline{\Delta^2_{\rm HS}}[\rho_{\rm tar},\rho_{\rm cS}] =  1/\sqrt{\det[{\bf V_{\rm cS}}]} + 1/\sqrt{\det[{\bf V_{\rm tar}}]} - 4\,\frac{\sqrt{\det\Big[\left(\,({\bf V}\fil+{\bf V}\rfil)\inv + {\bf Z}\tp ({\bf V}\sm + {\bf V}\csm)\inv {\bf Z}\,\right)\inv\Big]}}{\sqrt{\det[{\bf V}\fil+{\bf V}\rfil]}\,\sqrt{\det[{\bf V}\sm+{\bf V}\csm]}}\,.
\eeq

As all covariance matrices are proportional to ${\bf I}$ for our experimental system, the expression is simplified to
\beq
\overline{\Delta^2_{\rm HS}}[\rho_{\rm tar},\rho_{\rm cS}] =  v_{\rm cS}^{-1} + v_{\rm tar}^{-1} - 4\left((v\sm+v\csm) + z^2 (v\fil+v\rfil)\right)\inv
\eeq
where 
\beq
z = \frac{v_{\rm cS}}{v\rfil} - \frac{v\sm - v\tar}{v\rfil + v\tar}\,.
\eeq

\section{Distance between the mean values --- theory}\label{supp:delta_C}

We here derive the theory curves for the standard deviation of the distance $\delta_\text{C} = \ex{\hat{X}_j}_{\rm tar}-\ex{\hat{X}_j}_{\rm C}$. It equals
\beq
{\rm Std}(\delta_\text{C}) = \sqrt{\iint \dd \ex{\hat{X}_j}_{\rm tar}\,\dd\both{O}\,p(\ex{\hat{X}_j}_{\rm tar},\both{O})\,\delta_\text{C}^2\,}\,.
\label{equ:Std-deltaC}
\eeq

For $\text{C}=\text{F}$, $\delta_{\rm C}$ does not depend explicitly on the 
future measurement record, since $\ex{\hat{X}_j}\tar$ depends only on the past record. This makes averaging over the future record on the right-hand side of Eq.~\eqref{equ:Std-deltaC} trivial, reducing the standard deviation to ${\rm Std}(\delta_{\text{F}}) = \sqrt{\iint \dd \ex{\hat{X}_j}\tar\,\dd\past{O} \,p(\ex{\hat{X}_j}\tar,\past{O})\,\delta_{\text{F}}^2\,}$. 
Now, performing the average over the target mean first, using $p(\ex{\hat{X}_j}\tar,\past{O}) = p(\ex{\hat{X}_j}\tar|\past{O})\,p(\past{O})$, and $p(\ex{\hat{X}_j}\tar|\past{O}) = \mathcal{N}(\ex{\hat{X}_j}\tar;\ex{\hat{X}_j}\fil,v\fil - v\tar)$ (see Methods~\ref{MT:LQG}, just prior to Eq.~\eqref{eq:sm_hal_prob}), we get $\int\dd\ex{\hat{X}_j}\tar\,p(\ex{\hat{X}_j}\tar|\past{O}) \,(\ex{\hat{X}_j}\tar-\ex{\hat{X}_j}_{\rm F})^2 = v\fil - v\tar$, which is independent of the measurement record $\past{O}$. We therefore have ${\rm Std}(\delta_{\text{F}}) = \sqrt{v\fil - v\tar}$.

For the smoothed estimate of the target ($\text{C}=\text{S}$), the average over the future record cannot be performed trivially as it was the case for filtering. Nevertheless, following a similar logic, one needs to compute (using Eq.~\eqref{eq:sm_hal_prob}) $\int\dd\ex{\hat{X}_j}\tar\, p(\ex{\hat{X}_j}\tar|\both{O})\,(\ex{\hat{X}_j}\tar-\ex{\hat{X}_j}_{\rm S})^2 = \halo{v}\sm = v\sm - v\tar$, which is independent of the particular record $\both{O}$. This makes the final averaging trivial. Thus, ${\rm Std}(\delta_{\text{S}}) = \sqrt{v\sm - v_{\rm tar}}$.

For classical smoothing ($\text{C}=\text{cS}$), we first consider the term $\int \dd \ex{\hat{X}_j}\tar\, p(\ex{\hat{X}_j}\tar|\both{O})\,(\ex{\hat{X}_j}\tar-\ex{\hat{X}_j}_{\rm cS})^2$. Since $p(\ex{\hat{X}_j}\tar|\both{O}) = \mathcal{N}(\ex{\hat{X}_j}\tar;\ex{\hat{X}_j}\sm,\halo{v}\sm)$ (from Eq.~\eqref{eq:sm_hal_prob}), one finds
\beq
\int \dd \ex{\hat{X}_j}\tar\, p(\ex{\hat{X}_j}\tar|\both{O})\,(\ex{\hat{X}_j}_{\rm cS} - \ex{\hat{X}_j}\tar)^2 = \halo{v}\sm + (\ex{\hat{X}_j}_{\rm cS} - \ex{\hat{X}_j}_{\rm S})^2\,.
\eeq
This results in
\beq
{\rm Std}(\delta_{\text{cS}}) = \sqrt{\halo{v}\sm + \int \dd\both{O} p(\both{O})\,(\ex{\hat{X}_j}_{\rm cS} - \ex{\hat{X}_j}_{\rm S})^2}\;,\\
\eeq
where we have used the fact that $\halo{v}\sm$ is independent of the measurement record. Using Eq.~\eqref{eq:csm_sm_diff}, we have $\ex{\hat{X}_j}_{\rm cS} - \ex{\hat{X}_j}_{\rm S} = -z\,r_{\rm d}$. Considering $p(\both{O}) \equiv \mathcal{N}(r_{\rm d};0, v\fil + v\rfil)$, we conclude
\begin{align}
{\rm Std}(\delta_{\text{cS}}) &= \sqrt{\halo{v}\sm + \int \dd r_{\rm d} \;\mathcal{N}(r_{\rm d};0,v\fil + v\rfil)\;(z\,r_{\rm d})^2}\\
&=\sqrt{\halo{v}\sm + z^2(v\fil + v\rfil)}\,.
\end{align}

\section{Application note}\label{supp:applications}

In this section, we present example applications of the inference techniques discussed in this work. After outlining potential applications in quantum computation and quantum communication, we analyse a sensing application in detail and simulate it using an optomechanical force sensor.

\subsection{Qubit leakage detection}\label{supp:applications_sub1}

Ref.~\cite{campbell2020universal} reports a superconducting composite qubit in which the logical qubit is encoded in a two-dimensional subspace of a larger physical Hilbert space. The physical state may then ``leak out'' of the logical subspace. With a nearby coupled resonator, the authors show that, while the resonator cannot distinguish between the states that are inside the qubit subspace, it can distinguish whether a physical state lies inside or outside that subspace. As the authors note, this capability makes it possible, in principle, to monitor for leakage through the resonator, enabling postselection of leakage-contaminated computation sequences (or error correction on the leakage channel). Our quantum state smoothing techniques could improve leakage detection in such a postselection setting by making more efficient use of the finite-duration continuous measurement record, especially in the presence of excess technical noise.

\subsection{Continuous-in-time quantum teleportation}\label{supp:applications_sub2}

Ref.~\cite{hofer2013time} discusses continuous-in-time quantum teleportation of a quantum field to a stationary mechanical resonator. In this protocol, the stationary system is entangled with the output field of an interacting cavity. This output field and the field to be teleported then undergo a continuous-in-time Bell measurement, and the result is fed back to the stationary system with negligible delay --- therefore, post-processing of the measurement record cannot be used to determine the feedback. Inspired by this work, we consider continuous-in-time teleportation of a quantum field $\mathcal{A}$ to a travelling field $\mathcal{B}_1$ at a distance. While this problem has similarities to early optical continuous-variable teleportation experiments~\cite{furusawa1998unconditional,bowen2003experimental}, the entanglement source here is taken to be a narrow-band dynamical system, such as an optomechanically coupled mechanical resonator~\cite{chen2020entanglement}, so that the output continuous-wave fields $\mathcal{B}_1$ and $\mathcal{B}_2$ have temporal correlations governed by the source dynamics. A continuous-in-time Bell measurement is then performed on $\mathcal{A}$ and $\mathcal{B}_2$, and the outcome is used to displace $\mathcal{B}_1$ at the destination, thereby teleporting within the entanglement bandwidth. Suppose now that the finite-duration Bell measurement record contains excess technical noise, yielding an inaccurate estimate of the displacement to be applied to $\mathcal{B}_1$, and hence degrading the performance of our teleportation scheme. To partially compensate for this degradation, the destination system could be sent through an optical delay line, since it is a travelling field. Such a delay would allow the Bell measurement record to be post-processed using our quantum state smoothing techniques, taking into account the quantum dynamics of the entanglement source. This could yield an estimate of the required displacement that is more accurate than that obtained by real-time processing.

\subsection{Estimating the response of a sensing machine}\label{supp:applications_sub3}

An LGQ dynamical system, like our experimental resonator, can be engineered to become susceptible to an external force and thereby turned into a sensor. Under some plausible conditions on the force and its coupling, it will be an LGQ sensor, such that both the quantum state of the sensor and the value of the input force can be inferred in real time by applying filtering equations to the measurement current~\cite{harris2013minimum}. A better estimate of the force value can be obtained in post-processing, with the classical smoothing equations being optimal~\cite{tsang2009optimal}. The \textit{force's value}, however, is not always the quantity of interest. If a machine is using the sensor to respond to the input force, the quantity of interest is the \textit{force's filtered estimate} --- see Extended Data FIG.~\ref{SuppFig:SensingConcept}. In what follows, we elaborate on this scenario with two goals in mind: (i) to showcase the usefulness of our quantum state smoothing techniques in sensing setups, and (ii) to differentiate between the optimal smoothed estimate of the \textit{force's value}, and the optimal smoothed estimate of the \textit{force's filtered estimate}, with the latter case relying on our quantum state smoothing techniques.  
The significance of the latter goal becomes clearer when we recall that quantum mechanics can impose a limit on the sensor's real-time precision, preventing the \textit{force's filtered estimate} from arbitrarily approaching the \textit{force's value}.

In Extended Data FIG.~\ref{SuppFig:SensingConcept}, a force sensor sends its measurement current to a distant unit for processing. First, $\varrho_\text{F}^\text{LTL}$, the post-transient filtered state of knowledge about the sensor and the force, is calculated. Next, by tracing out the sensor part, the LTL filtered estimate of the force ($\langle f\rangle_\text{F}^\text{LTL}$) and its error variance ($v_{f,\text{F}}^\text{ss}$) are obtained. Subsequently, the estimated value of the force is  mapped to the machine's response. This causal mapping, denoted by $M$, is deterministic and represents, for example, the machine's decision-making algorithm or its analogue response-generating system. Meanwhile, the signal transmission path between the sensor and the centre is leaking and, consequently, an external agent obtains a noisy copy of a piece of the measurement current. This agent is interested in knowing the machine's response, but does not have any access to the processing unit. As $M$ is assumed to be public knowledge, the agent uses their data to estimate $\langle f\rangle_\text{F}^\text{LTL}$ and feeds the result to $M$ to estimate the machine's response.

In case there is no considerable delay before the filtering of the measurement record by the processing unit, and the agent needs to know the machine's response without a time lag, the optimal strategy for them is to apply filtering equations to their noisy piece. This yields $\varrho_\text{F}^{\eta\downarrow}$ and subsequently an estimate of the force, $\langle f\rangle_\text{F}^{\eta\downarrow}$, alongside its error variance $v_{f,\text{F}}^{\eta\downarrow}$. $\langle f\rangle_\text{F}^{\eta\downarrow}$ is the optimal estimate of $\langle f\rangle_\text{F}^\text{LTL}$ using the past measurement data with the error variance $v_{f,\text{F}}^{\eta\downarrow}-v_{f,\text{F}}^\text{ss}$. Feeding $\langle f\rangle_\text{F}^{\eta\downarrow}$ to $M$ results in an estimate for the response which has the least mean squared error (using only the past measurement data) if $M$ is linear.

In case the delay before the filtering by the processing unit is considerable (because the transmission path is long, there exist resource allocation limitations, etc.), or the agent can sacrifice time for knowing the machine's response more accurately, the optimal strategy is to apply quantum state smoothing equations to their noisy piece with $\varrho_\text{F}^\text{LTL}$ as the target. The output is $\varrho_\text{S}^{\eta\downarrow}$, a quantum-smoothed state of knowledge about the sensor and the force. Tracing out the sensor part yields a non-optimal estimate of the force, $\langle f\rangle_\text{S}^{\eta\downarrow}$, alongside the variance $v_{f,\text{S}}^{\eta\downarrow}$. $\langle f\rangle_\text{S}^{\eta\downarrow}$ is the optimal smoothed estimate of $\langle f\rangle_\text{F}^\text{LTL}$. The variance $v_{f,\text{S}}^{\eta\downarrow}$ indicates two quantities: (i) the ensemble distribution of $\langle f\rangle_\text{S}^{\eta\downarrow}$, where $\mathbb{V}\text{ar}_\text{ens}\big[\langle f\rangle_\text{S}^{\eta\downarrow}\big]+v_{f,\text{S}}^{\eta\downarrow}$ equals the force variance, and (ii) the error variance in the optimal estimation of $\langle f\rangle_\text{F}^\text{LTL}$, which equals $v_{f,\text{S}}^{\eta\downarrow}-v_{f,\text{F}}^\text{ss}$.
The agent then estimates the machine's response by feeding $\langle f\rangle_\text{S}^{\eta\downarrow}$ to $M$. This final estimate is better than that found by filtering, and has the least mean squared error (using both the past and future measurement data) if $M$ is linear. 

It is important to note that the covariance matrix of $\varrho_\text{F}^\text{LTL}$, which serves as the target of our smoothing technique above, has nonzero off-diagonal blocks. Therefore, the pair $\big(\langle f\rangle_\text{S}^{\eta\downarrow}, v_{f,\text{S}}^{\eta\downarrow}\big)$ is inferred hand in hand with the sensor's smoothed quantum state $\rho_\text{S}^{\eta\downarrow}$, and is not calculated independently.

In the rest of this subsection, we simulate the above scenario with an optomechanical force sensor to demonstrate numerically the advantage of our quantum state smoothing techniques. We begin by comparing a classical model of the sensor with the quantum one, showing that quantum mechanics imposes a fundamental limit on the sensor’s real-time precision. This implies that $\langle f\rangle_\text{F}^\text{LTL}\neq f$ no matter how ideal the modelled sensor is, and that the optimal estimation of $\langle f\rangle_\text{F}^\text{LTL}$ is always a different task from the optimal estimation of $f$. We then analyse the full scenario at this quantum limit of the sensor.

\subsubsection{An optomechanical force sensor}

The classical equations of motion for an optomechanical force sensor operating deep inside the fast-cavity regime are
\begin{equation}
\begin{array}{rcl}
    \dd q &= & \Omega\,p\,\dd t\;,\\
    \dd p &= & -\Omega\,q\,\dd t -\Gamma\,p\,\dd t +\sqrt{2\Gamma}\,\dd W_\text{th}-2\sqrt{\Gamma\mathcal{C_\text{ba}}\,}\dd W_\text{ba} + g_{f}\,f\,\dd t\; .
    \end{array}
\end{equation}
Here, $\dd W_\text{th}$ and $\dd W_\text{ba}$ are Wiener increments that, respectively, represent the thermal noise and a possible back-action noise:
\begin{equation}
\begin{array}{rcl}
    \langle\dd W_\text{th}(t)\,\dd W_\text{th}(t')\rangle & = & 2\,\zeta_\text{th}\,\delta(t-t')\,\dd t^2\,,\vspace{1ex}\\
    \langle\dd W_\text{ba}(t)\,\dd W_\text{ba}(t')\rangle & = & \delta(t-t')\,\dd t^2
\end{array}
\end{equation}
with $\zeta_\text{th}$ computed as $(e^{\hbar\Omega/(k_\text{B}T)}-1)^{-1}$, and $\mathcal{C}_\text{ba}$ weighting the back-action noise; $f$ is the time-varying force, scaled to be dimensionless and of order unity, and $g_f$ is its coupling rate. The word ``back-action" is used here in the general sense, i.e., the dynamical effect of having a probe optical beam.

Now, assuming that the force is narrow-band and centred around $\Omega/2\pi$, and that all the dynamical rates of our resonator (including the one that involves $g_f$) are much smaller than $\Omega$, we can switch to a frame rotating at frequency $\Omega/2\pi$ and write the effective equations of motion of the mechanical quadratures as
\begin{equation}
\begin{array}{rcl}
    \dd X_j & = & -\dfrac{\Gamma}{2}\,X_j\,\dd t+\sqrt{\Gamma}\,\dd W_{\text{th},j}-\sqrt{2\Gamma\mathcal{C_\text{ba}}\,}\dd W_{\text{ba},j} + \dfrac{g_{f}}{2}\,f_j\,\dd t\;, \label{eq:directlybefore}
    \end{array}
\end{equation}
where $f_1$ and $f_2$ describe the slowly varying envelope of $f$ [i.e., $f=-\sin(\Omega t)\,f_1+\cos(\Omega t)\,f_2$] and 
\begin{equation}
\begin{array}{rcl}
    \langle\dd W_{\text{th},k}(t)\,\dd W_{\text{th},l}(t')\rangle & = & 2\,\zeta_\text{th}\,\delta(t-t')\,\dd t^2\,\delta_{k,l}\;,\vspace{1ex}\\
    \langle\dd W_{\text{ba},k}(t)\,\dd W_{\text{ba},l}(t')\rangle & = & \delta(t-t')\,\dd t^2\,\delta_{k,l}\;.
\end{array}
\end{equation}
Measuring and demodulating the phase of the back-reflected laser beam then gives
\begin{equation}
    y_j\,\dd t = \sqrt{2\eta\Gamma\mathcal{C}\,}\,X_j\,\dd t+\dd W_{{\rm m},j}\,,\;\;\;\;\;\;j=1,2
\end{equation}
where $\eta\in[0,1]$ is the efficiency of the detection (introduced in this classical model for convenience), and 
\begin{equation}
\begin{array}{rcl}
    \langle\dd W_{{\rm m},k}(t)\,\dd W_{{\rm m},l}(t')\rangle & = & \,\delta(t-t')\,\dd t^2\,\delta_{k,l}\;.
\end{array}
\end{equation}
We model the slowly varying force quadratures, $f_1$ and $f_2$, as zero-mean Ornstein--Uhlenbeck processes, 
\begin{equation}
\begin{array}{rcl}
    \dd f_j & = & -\dfrac{\Gamma_f}{2}\,f_j\,\dd t+\sqrt{\Gamma_f\,}\dd W_{f,j}\,,\;\;\;\;\;\;j=1,2
    \end{array}
    \label{force-quads}
\end{equation}
where $\Gamma_f$ quantifies the autocorrelation of the force, and
\begin{equation}
\begin{array}{rcl}
    \langle\dd W_{f,k}(t)\,\dd W_{f,l}(t')\rangle & = & \delta(t-t')\,\dd t^2\,\delta_{k,l}\;.
\end{array}
\end{equation}

Although this classical system can then be simulated with the vector of \textit{state variables} $\mathbf{x}=(X_1,X_2,f_1,f_2)^\top$, it is equally informative to do the simulation only for $\mathbf{x}'=(X_1,f_1)^\top$ because the pairs $(X_1,f_1)$ and $(X_2,f_2)$ have decoupled and statistically identical dynamics. Crucially, the results of such a simulation are also applicable to a real (quantum) optomechanical sensor as long as we impose the quantum mechanical constraint $\mathcal{C}_\text{ba}\ge\mathcal{C}$, and recompute $\zeta_\text{th}$ as $0.5+(e^{\hbar\Omega/(k_\text{B}T)}-1)^{-1}$.

\subsubsection{Limits on the precision of real-time sensing}

Starting from no measurement ($\mathcal{C}=0$), increasing $\mathcal{C}$ yields a more accurate real-time estimate of the force, i.e., a decrease in the error variance at the \textit{steady state} of the filtering process ($v_{f,\text{F}}^\text{ss}$). Nonetheless, even as $\mathcal{C}\rightarrow\infty$, thermal and back-action noises may impose a non-zero lower bound on $v_{f,\text{F}}^\text{ss}$. 
Here we consider the fundamental limits of this lower bound. 

\begin{figure*}[]
    \centering
    \includegraphics[]{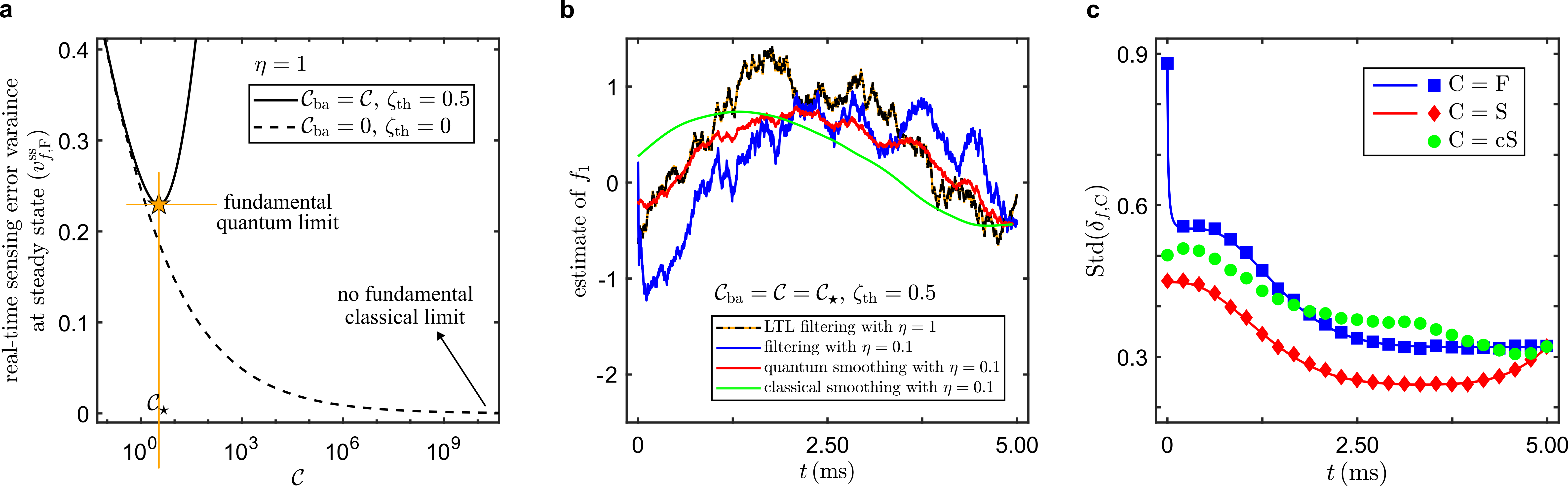}
    \caption{\textbf{Simulation of the response estimation scenario with an optomechanical force sensor.} This is the only simulation-based figure of this work. \textbf{a,} Perfect optical detection ($\eta=1$) and minimum thermal and back-action noises (weighted by $\zeta_{\rm th}$ and $\mathcal{C}_{\rm ba}$, respectively) are assumed for an optomechanical force sensor. While there is no fundamental classical limit on the steady-state error variance of real-time sensing, there exists a fundamental quantum limit. This limit is achieved at $\mathcal{C}_\star$, where we run the response estimation setup with an \textit{identity map} as $M$. \textbf{b,} The machine's filtered estimate of the force component $f_1$ is shown in a 5-ms interval, alongside the real-time and post-processed estimates found by the external agent (whose effective measurement efficiency has dropped to 0.1 because of the added noise). \textbf{c,} Markers show the standard deviation of $\delta_{f,\rm C}=\langle f_1\rangle_{\rm F}^{{\rm LTL}}-\langle f_1\rangle_{\rm C}^{\eta\downarrow}$ over 5,000 simulated records. The blue and red lines represent the deterministic values of $(v_{f,\text{F}}^{\eta\downarrow}-v_{f,\text{F}}^\text{ss})^{1/2}$ and $(v_{f,\text{S}}^{\eta\downarrow}-v_{f,\text{F}}^\text{ss})^{1/2}$, respectively. In this simulation, the force variance equals one; its coupling rate is $g_f=2\pi\times1.77$~kHz; its autocorrelation constant is $\Gamma_f =2\pi\times50$~Hz; and the mechanical decay rate equals $\Gamma_f$.}
    \label{SuppFig:SensingStandard}
\end{figure*}

First, we set $\eta=1$ (perfect optical detection). Next, to reduce the thermal noise contribution, we consider a GHz resonator with a quality factor of about $10^8$, cooled to mK temperatures in a dilution refrigerator. Assuming that our thermal bath model remains effective, we can then neglect the term $(e^{\hbar\Omega/(k_\text{B}T)}-1)^{-1}$ and set $\zeta_\text{th}$ to $0$ in the hypothetical classical model and to $0.5$ in the quantum model. To reduce the back-action noise contribution, we set $\mathcal{C}_\text{ba}=0$ in the classical model and $\mathcal{C}_\text{ba}=\mathcal{C}$ in the quantum model. Subsequently, we simulate the sensor under these two models, while the mechanical decay rate $\Gamma$ and the force-related parameters ($g_f$ and $\Gamma_f$) are fixed to the values listed in the caption of FIG.~\ref{SuppFig:SensingStandard}. Panel \textbf{a} of this figure shows that, in the classical model, $v_{f,\text{F}}^\text{ss}$ approaches zero with increasing values of $\mathcal{C}$, so that no fundamental limit exists for its lower bound. In the quantum model, however, a nonzero lower bound --- similar to the more familiar \textit{standard quantum limit}~\cite{bowen2015quantum} --- does appear at an optimal value of $\mathcal{C}$. Such dependence of $v_{f,\text{F}}^\text{ss}$ on $\mathcal{C}$ arises because, in the quantum model, increasing $\mathcal{C}$ reduces the imprecision noise but also increases the back-action noise. At this fundamental quantum limit, we simulate the full response estimation scenario.

\subsubsection{Response estimation at the fundamental quantum limit}

We simulate the full scenario shown in Extended Data FIG.~\ref{SuppFig:SensingConcept} using the optomechanical force sensor operating at the fundamental quantum limit, with $M$ taken to be the \textit{identity map} and the effective detection efficiency for the external agent set to 0.1 --- reduced from 1.0 by the leakage noise. The results are plotted in
FIG.~\ref{SuppFig:SensingStandard}\textbf{b-c} for the force quadrature $f_1$ ---  as discussed before, the other quadrature, $f_2$, has identical statistics. In \textbf{b}, example time traces are shown. The agent's $\langle f_1\rangle_\text{S}^{\eta\downarrow}$, the red line, typically follows the machine's $\langle f_1\rangle_\text{F}^\text{LTL}$, the dash-dotted line, more closely than $\langle f_1\rangle_\text{F}^{\eta\downarrow}$ and $\langle f_1\rangle_\text{cS}^{\eta\downarrow}$. In \textbf{c}, the standard deviation of $\delta_{f,\rm C}=\langle f_1\rangle_{\rm F}^{{\rm LTL}}-\langle f_1\rangle_{\rm C}^{\eta\downarrow}$, calculated over 5,000 simulated records, is plotted for $\text{C}=\text{S}$, F and cS. It shows that our quantum state smoothing technique outperforms filtering and classical smoothing. Furthermore, the calculated values for the standard deviation of $\delta_{f,\rm S}$ are in close agreement with $(v_{f,\text{S}}^{\eta\downarrow}-v_{f,\text{F}}^\text{ss})^{1/2}$. The agreement demonstrates one of the roles played by $v_{f,\text{S}}^{\eta\downarrow}$, the quantity output by our post-processing technique as a variance.
These results clarify the applicability of our quantum state smoothing techniques to some sensing setups.

With the above conclusion, our Application Note ends. \textbf{The rest of this Supplementary Information} concerns the theoretical frameworks and experiments discussed in the main text.

\section{Conditional variances in the steady state}\label{sec:SSvariances}
Assuming $\eta\,\mathcal{C}\gg\tfrac{1}{2}$, Eqs.~\eqref{VFSS} and \eqref{VRSS} give
\begin{equation}
v_\text{F}^\text{ss}\approx v_\text{R}^\text{ss}\approx \sqrt{\frac{n_\text{tot}}{\eta\,\mathcal{C}}}
\end{equation}
paving the way to derive approximate expressions for the steady-state smoothing variances (at $0\ll t\ll\mathcal{T}$).

Once we take the ``true'' state of the resonator to be a displaced ground state, the variance of its smoothed estimate is
\begin{equation}
    v_\text{S}^\text{ss}\approx 1+\frac{1}{2}\!\left(\sqrt{\frac{n_\text{tot}}{\eta\,\mathcal{C}}}-\sqrt{\frac{\eta\,\mathcal{C}}{n_\text{tot}}}\right).
\end{equation}
This is never smaller than one, the ground state variance, as required for a valid quantum state.

For classical smoothing,
\begin{equation}
v_\text{cS}^\text{ss}\approx \frac{1}{2}\sqrt{\frac{n_\text{tot}}{\eta\,\mathcal{C}}}\,.
\end{equation}
This becomes smaller than one when $\eta>0.25$ and $\mathcal{C}/n_\text{th}>1/(4\eta-1)$, indicating that the Gaussian function produced by the classical smoothing equations does not always correspond to a physical quantum state as it can violate the Heisenberg uncertainty principle. In our experiment only the first condition holds.

Finally, the difference
\begin{equation}
v_\text{S}^\text{ss}-v_\text{cS}^\text{ss}\approx 1-\frac{1}{2v_\text{F}^\text{ss}}
\end{equation}
is always positive. For this difference to be appreciable, say larger than $v_\text{cS}^\text{ss}/10$, so that the benefits of quantum state smoothing are most pronounced, one needs $v_\text{F}^\text{ss}\lesssim 20$.

\section{Power spectral densities of the currents}\label{supp:PSDs}

\begin{figure*}[]
    \centering
    \includegraphics[]{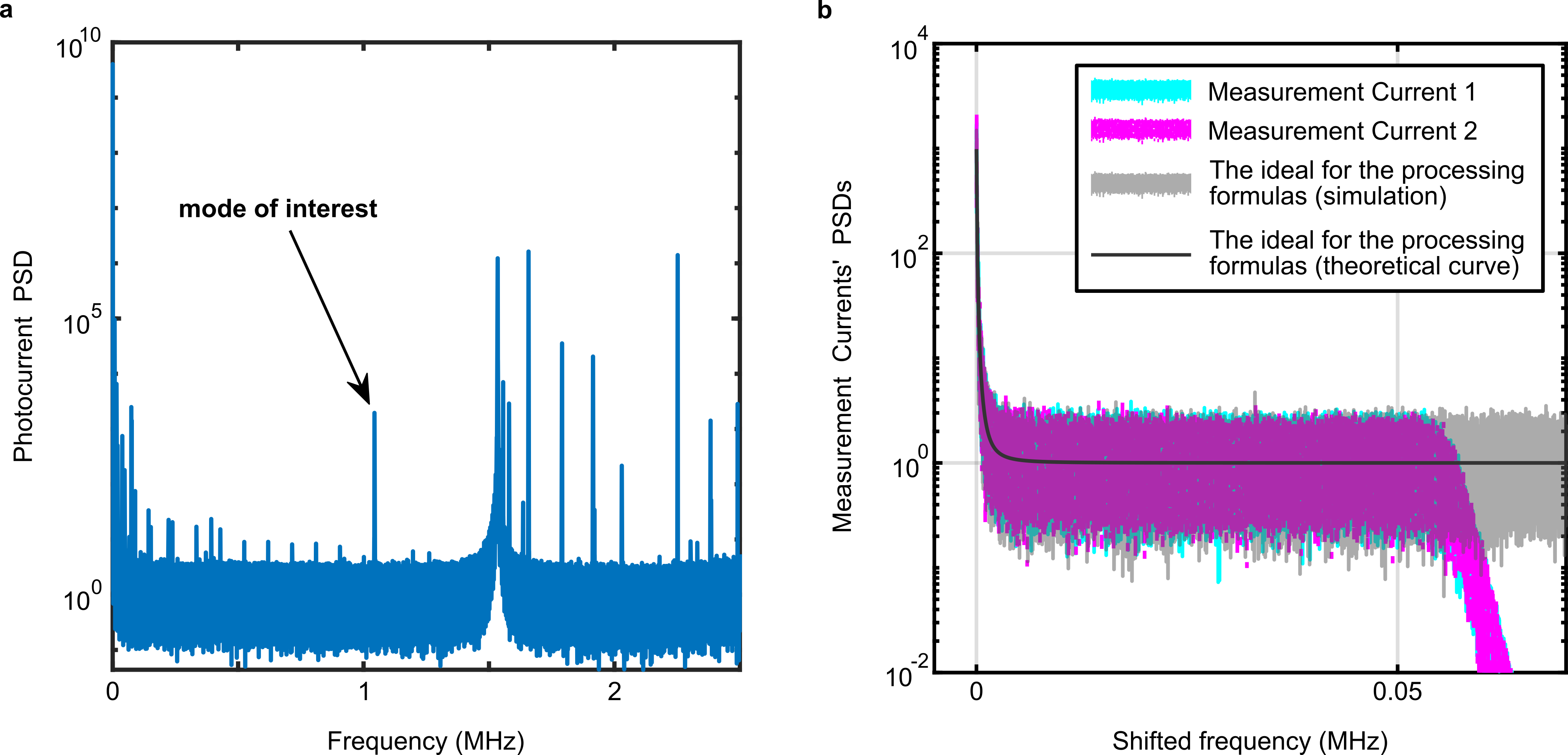}
    \caption{\textbf{Data spectra.} The shot-noise-normalised power spectral densities (PSDs) of (\textbf{a}) the photocurrent and (\textbf{b}) the final measurement currents. The latter are obtained by demodulating the photocurrent at $\Omega/2\pi=1.04355~\text{MHz}$ and low-pass filtering the output. In \textbf{b}, we also show (only for reference) the ideal case of a perfectly white noise floor, as assumed by the processing formulas.}
    \label{SuppFig:PSDs}
\end{figure*}

The spectra of the captured photocurrent (with the feedback on, at final data collection) and the measurement currents (used by our measurement model) are shown in FIG.~\ref{SuppFig:PSDs}. They are normalised to the shot-noise level, which is measured independently by blocking the signal arm of the homodyne setup --- it thus includes electronic noise. The noise floor around our ``signal of interest'' is flat and has a value of unity, indicating that \textit{higher-order mode contamination}, discussed in Ref.~\cite{khademi2024optomechanical}, is negligible. If more laser power were injected, this desirable situation would be compromised by the broad peak centred at about 1.53~MHz.

\section{Fidelity comparison}\label{supp:Fidelity}

\begin{figure*}[]
    \centering
    \includegraphics[]{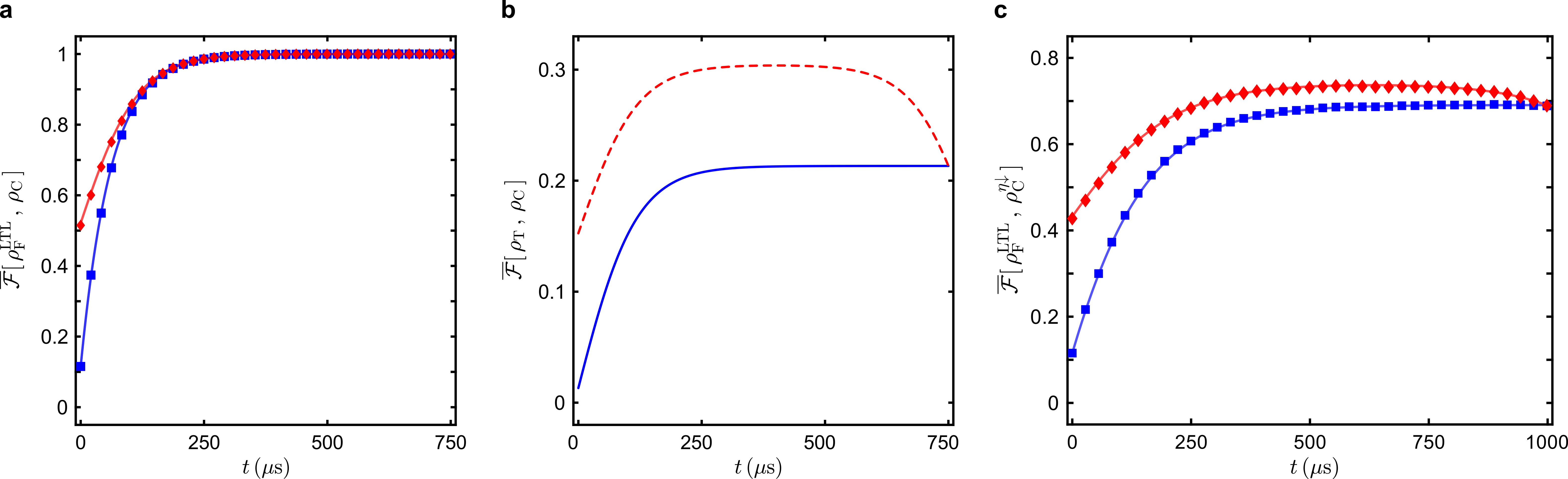}
    \caption{\textbf{Fidelity comparison.} Quantum fidelity between the target state and its estimator states, averaged over the ensemble of measurement records. It is shown in blue for the real-time estimator, i.e., the filtered state ($\text{C}=\text{F}$), and in red for the post-processed estimator, i.e., the smoothed state ($\text{C}=\text{S}$). \textbf{a} and \textbf{b} correspond to the main experiment, whereas \textbf{c} corresponds to the noise-injection ($\eta$-reduced) experiment. Data points and curves show the experimental values and theoretical predictions, respectively.}
    \label{SuppFig:Fidelity}
\end{figure*}

Here we compare the quantum fidelity between the target state and the filtered state with the quantum fidelity  between the target state and its smoothed estimate. FIG.~\ref{SuppFig:Fidelity} shows that the latter is greater than or equal to the former. Although this is consistent with the expectation that future measurement data should help increase the fidelity between the estimate and the target state, we note that neither the smoothed state nor the filtered state is the optimal estimator of the target state in terms of fidelity, whether post-processed or real-time. That is, quantum fidelity is not among the cost functions optimised by either our novel or conventional inference protocol. In fact, we do not adopt quantum fidelity as a cost function for post-processed state estimation because the filtered state, as the standard conditional state in the literature, is not the optimal real-time estimator for any of our targets in terms of fidelity.

FIG.~\ref{SuppFig:Fidelity} does not show the fidelity between the target state and the output of the classical smoothing equations. As discussed in Sec.~\ref{Sec:traj}, those equations do not, in general, output a valid density matrix. As a result, such a fidelity calculation does not necessarily yield a meaningful value, i.e., a real number in the interval $[0,1]$.

\newpage

\section{Glossary}\label{supp:Glossary}

\begin{tabular}{ll}
 \hline
	abbreviation & {} \\
 \hline
 LGQ & linear-Gaussian quantum \\
  LTL & long-time-limit \\
  Std & standard deviation  \\
VACF & velocity autocorrelation function \\
PSD & power spectral density \\
 \hline
\end{tabular}

\vspace{6pt}
\begin{center}
	\begin{tabular}{ll}
 \hline
	conditioning &\\notation & {} \\
 \hline
 $\bullet_\text{uncon}$ & (the moment of) the unconditional state \\
  $\bullet_\text{C}$ & (the moment of) the conditional state/effect operator/distribution \\
  $\bullet_\text{F}$ & (the moment of) the filtered state \vspace{3pt} \\
  $\bullet_\text{F}^\text{LTL}$ & (the moment of) the long-time-limit filtered state \vspace{3pt} \\
  $\bullet_\text{T}$ & (the moment of) the ``true" state \\
  $\bullet_\text{tar}$ & (the moment of) the estimated target state \\
  $\bullet_\text{R}$ & (the moment of) the retrofiltered effect operator \\
  $\bullet_\text{S}$ & (the moment of) the smoothed state --- the estimated target should be inferred from the context\\
  $\bullet_\text{cS}$ & (the moment of) the Gaussian distribution found by applying the classical smoothing equations \vspace{3pt}\\
  $\bullet_\text{C}^\text{ss}$ & (the moment of) the conditional state/effect operator/distribution at \textit{steady state of the conditioning} \\
  & \textit{process} \\ $\bullet_\text{C}^{\eta\downarrow}$ & (the moment of) the state/effect operator/distribution conditioned on the noise-added data \vspace{3pt} \\
 \hline
	\end{tabular}
\end{center}

\vspace{6pt}
{\renewcommand{\arraystretch}{1.4}
\begin{tabular}{ll}
 \hline
	operator/function symbol & {} \\
 \hline
 $\text{Tr}[\bullet]$ & trace of $\bullet$ \\
  $\overline{\Delta^2_\text{HS}}[\bullet,\star]$ & squared Hilbert--Schmidt distance between $\bullet$ and $\star$, ensemble-averaged over measure-\, \\ & ment records \\
    $p(\bullet)$ & probability of $\bullet$ \\
  $\text{det}[\bullet]$ & determinant of $\bullet$\\
  $\mathbb{V}\mathrm{ar}_{\text{ens}}[\bullet]$ & ensemble variance of $\bullet$\\
  $\mathcal{N}(\bullet;\star,\blacksquare)$ & Gaussian distribution of $\bullet$ with mean $\star$ and variance $\blacksquare$ \\
  $\text{P}[\bullet]$ & purity of $\bullet$\\
  $\overline{\mathcal{F}}[\bullet,\star]$ & fidelity between $\bullet$ and $\star$, ensemble-averaged over measurement records \\
  \hline
\end{tabular}}

\vspace{16pt}
\begin{tabular}{m{248pt} m{36pt} m{200pt}}
 \hline
	system parameter & symbol & experimental value \\
 \hline
 mechanical resonance frequency & $\Omega/2\pi$ & 1.04355~MHz\\
 mechanical energy decay rate & $\Gamma$ & $2\pi\times11.5$~mHz \\
 optical resonance wavelength (in vacuum) & - & 1538.2 nm \\
 optical energy decay rate & $\kappa$ & $2\pi\times11.5$~GHz \\
 single-photon optomechanical coupling rate & $g_0$ & $2\pi\times159$~kHz\\
  measurement laser detuning &$\Delta$& 0 at final data collection\\
  mean intracavity measurement photon number & $\bar{n}_\text{cav}$ & 41.4 at final data collection \\
 (boosted) optomechanical coupling rate & $g$ & $2\pi\times1.02$~MHz at final data collection\\ optomechanical cooperativity & $\mathcal{C}$ & 
 $3.16\times10^4$ at final data collection \\
 mechanical mode temperature & $T$ & 12.1~K at final data collection laser power \\
 thermal phonon number & $n_\text{th}$ & $2.45\times10^5$ at final data collection laser power \\
 cavity--waveguide coupling efficiency & $\eta_\text{esc}$ & 0.82 \\
 waveguide--fibre coupling efficiency (in the dilution fridge) & $\eta_\text{wf}$ & $0.77$ \\
 optical detection efficiency & $\eta$ & 0.38 \\
 effective mechanical decay rate after feedback & $\Gamma_\text{fb}$ & $2\pi\times85$~Hz at final data collection \\
  \hline
\end{tabular}


\end{document}